\documentclass[phdthesis,12pt,final]{wuthesis}

\usepackage{ifluatex}
\ifluatex
\usepackage{fontspec} 

\usepackage{polyglossia}
\setmainlanguage[variant=american]{english}
\else 
\usepackage[utf8]{inputenc}
\usepackage[T1]{fontenc}

\usepackage[USenglish]{babel}
\fi

\usepackage{microtype} 
\usepackage{csquotes}
\usepackage[vskip=0pt,begintext=\textooquote,endtext=\textcoquote]{quoting}
\SetBlockEnvironment{quoting}

\usepackage{amsmath}
\usepackage{amsfonts}

\usepackage[section]{placeins} 
\usepackage[inline, shortlabels]{enumitem}
\usepackage{threeparttable}
\usepackage{booktabs}
\usepackage{array}
\usepackage{siunitx}
\usepackage{makecell}

\usepackage{mdframed}

\usepackage[hidelinks,backref=page]{hyperref}
\renewcommand*{\backref}[1]{}
\renewcommand*{\backrefalt}[4]{%
	\ifcase #1 (Not cited.)%
	\or        (Cited on page~#2.)%
	\else      (Cited on pages~#2.)%
	\fi}
\usepackage[noabbrev]{cleveref}

\usepackage{caption}

\usepackage{braket}
\usepackage{color}
\usepackage{mathtools}
\usepackage{float}

\newcommand*{\brakett}[2]{\langle #1 \lvert #2 \rangle}

\newcommand{\um}{\ensuremath{\mu}{\rm m}}

\newcommand{\tw}{\ensuremath{\theta_A}}

\newcommand{\dt}[1]{ \frac{\partial#1}{\partial t}}
\newcommand{\T}{\ensuremath{\Theta}}
\newcommand{\TD}{\ensuremath{\Theta^\dagger}}
\newcommand{\M}{\ensuremath{K_j}}
\newcommand{\MD}{\ensuremath{K_j}^\dagger}
\newcommand{\Mtr}{\ensuremath{\tilde{K}_j}}
\newcommand{\MDtr}{\ensuremath{\tilde{K}_j^\dagger}}
\newcommand{\Vg}{\ensuremath{V_{\mathrm{gnd}}}}
\newcommand{\Ve}{\ensuremath{V_{\mathrm{ex}}}}
\newcommand{\SX}{\ensuremath{\langle \sigma_x\rangle}}
\newcommand{\SY}{\ensuremath{\langle \sigma_y\rangle}}
\newcommand{\SZ}{\ensuremath{\langle \sigma_z\rangle}}
\newcommand{\zf}{\ensuremath{\tilde{Z}_t}}
\DeclareMathOperator{\Tr}{Tr}

\renewcommand{\thesisauthor}{Jonathan Tyler Monroe}

\renewcommand{\thesisdepartment}{Department of Physics}
\renewcommand{\thesisfield}{Physics}

\renewcommand{\thesismonth}{May}
\renewcommand{\thesisyear}{2021}

\renewcommand{\thesistitle}{Partial Measurements of Quantum Systems}

\renewcommand{\thesissupervisor}{Professor Kater Murch, Chair}
\renewcommand{\thesiscommittee}{Kater Murch, Chair\\
James Buckley \\
Erik Henriksen  \\
Zohar Nussinov \\
Jung-Tsung Shen}

\renewcommand{\thesisdedication}{Dedicated to E. Monroe}

\hypersetup{
	pdftitle={\thesistitle},
	pdfauthor={\thesisauthor},
	pdfpagemode=UseOutlines,
	bookmarksnumbered=true,
	bookmarksopen=true,
	bookmarksopenlevel=1
}
\IfFileExists{upquote.sty}{\usepackage{upquote}}{}

\begin{document}

\frontmatter


\begin{thesistitlepage}
\end{thesistitlepage}

\begin{thesiscopyrightpage}
\end{thesiscopyrightpage}

\begin{singlespace}
\setcounter{page}{2}
\renewcommand*\contentsname{Table of Contents}
\tableofcontents

\cleardoublepage
\phantomsection
\addcontentsline{toc}{chapter}{\listfigurename}
\listoffigures

\cleardoublepage
\phantomsection
\addcontentsline{toc}{chapter}{\listtablename}
\listoftables
\end{singlespace}

\begin{thesisacknowledgments}
I am delighted to thank the many people who have guided  and supported me during my graduate studies. 
The bountiful support of those close to me filled my experience with joy.

I begin with the central driver of my success: Kater Murch.
I count myself exceptionally fortunate to have been mentored by such a personable and passionate advisor as Kater.
Kater's example has instilled the virtue of clear explanations.

I feel honored to have worked among such great colleagues in the Murch lab.
I thank Dian Tan for his patient tutelage of fabrication techniques and Mahdi Naghiloo for his willing explanation of every aspect of our lab, especially as codified in his dissertation.
Patrick Harrington has been an inspiring figure in my life. I admire his profound insights, and I feel deep appreciation to have worked and played with him for four years.
I appreciate the varied insights and passions of Maryam Abassi, Taeho Lee, Xingrui Song, Yungzhao Wang, and Weijian Chen.
Working closely with Daria Kowsari, Chandrashekhar Gaikwad, and Kaiwen Zheng has engendered passion and devotion.
Thank you all for being such excellent scientists and sharing your experiences.

I am grateful for my committee's helpful guidance and feedback. 
I particularly thank Henric Krawczynski and Zohar Nussinov for their continued mentorship throughout my graduate tenure.
I thank Jim Buckley, Erik Henricksen, and JT Shen for their eager willingness to read my dissertation.

I thankfully acknowledge the critical roles of Todd Hardt, Linda Trowler, and Sarah Akin played in supporting my graduate work. Thank you for maintaining infrastructure that makes the ship run smoothly.
I thank Rahul Gupta for training me on equipment and working hard to maintain high cleanroom standards.

I am grateful for a cohort of colleagues who have enriched life during grad school.
Kiandohkt Amiri, Jesse Balgley, Nat\'alia Calleya, Furqan Dar, and Kainen Utt have been dear friends to commensurate and celebrate with.

Finally, I would not have thrived during my graduate career without the endless love and support of Esther.
Thank you for keeping me happy and healthy; I am elated for our future. 

Thank you all.
\end{thesisacknowledgments}

\begin{thesisdedicationpage}\label{dedication}
\end{thesisdedicationpage}

\cleardoublepage
\phantomsection
\begin{thesisabstract}
Projective measurement is a commonly used assumption in quantum mechanics.
However, advances in quantum measurement techniques allow for partial measurements, which accurately estimate state information while keeping the wavefunction intact.
We employ partial measurements to study two phenomena. 
First, we investigate an uncertainty relation---in the style of Heisenberg's 1929 thought experiment---which includes partial measurements in addition to projective measurements.
We find that a weak partial measurement can decrease the uncertainty between two incompatible (non-commuting) observables.
In the second study, we investigate the foundation of irreversible dynamics resulting from partial measurements. 
We do so by comparing the forward and time-reversed probabilities of measurement outcomes resulting from post-selected feedback protocols with both causal and reversed-causal order.
We find that the statistics of partial measurements produce entropy in accordance with generalized second laws of thermodynamics.

We perform these experiments using superconducting qubits. We describe the fabrication process for these devices and detail a novel fabrication technique that allows fast, single-step lithography of Josephson-junction-based superconducting circuits.
The technique simplifies processing by utilizing a direct-write photolithography system, in contrast to traditional electron-beam lithography. 
Despite their large lithographic area, Josephson junctions made with this method have low critical currents and high coherence times.
\end{thesisabstract}


\mainmatter
\chapter{Introduction}
\section{Motivation}

Quantum mechanics offers a strange view of reality when compared to the classical world.
Particles can occupy distinct states simultaneously, and measuring a system changes what the system does.
These phenomena would be bizarre in an everyday setting, but they are routine in quantum mechanics.

The concept of measurement backaction is especially intriguing.
In our everyday experience, our measurements do not disturb the quantity we're measuring: No table has ever been widened by measuring its length.
Yet monitoring a quantum particle's trajectory irrevocably changes its momentum, injecting fundamental uncertainty about the particle's dynamics.

Introductory quantum mechanics posits that particles are described by a wavefunction composed of complex amplitudes.
Yet when we measure one of the particle's properties, say its position, we get a single, real number. 
This classical information neglects a great deal of the particle's state. 

What is so special about measurement that forces us to discard so much in bridging the quantum-classical divide?
The measurement apparatus must have been made up of particles described by their own quantum wavefunctions.
The measurement process must have involved an interaction between the apparatus and the system.
Both of these statements are well described by quantum mechanics. Where did the quantum information go?

The answer, in part, is that decoherence often rapidly distributes the quantum information throughout a wavefunction with many possible outcomes \cite{Zurek1981a}.
The system-apparatus interaction can cause the system's information to dissociate into distant components of the apparatus.
After the apparatus has completed the measurement, it continues to interact with its environment, and the state's information quickly erodes completely.
But decoherence is not an insurmountable challenge.

In systems with manageable decoherence, the quantum nature of measurement emerges.
Optics experiments were among the first to directly observe the continuous collapse of a system's wavefunction \cite{Guerlin2007}.
With modern circuit quantum electrodynamics (cQED) experiments, quantum trajectories under the influence of measurement have been directly observed \cite{Boissonneault2009, Hatridge2013, Murch2013, Weber2014}, even under the influence of multiple, non-commuting measurements \cite{Hacohen-Gourgy2016,Ficheux2018}.

In cQED, the microwave frequency range provides a timescale that is arguably as slow as possible to provide a view of quantum dynamics.
If the dynamics were slower (corresponding to lower frequencies), the thermal energy at 10 mK (the coldest temperature readily achieved with dilution refrigerators) would dominate the quantum energy scale.

With this in mind, the cQED platform has provided a remarkable view of quantum phenomena.
In particular, cQED experiments have directly observed quantum jumps, especially during spontaneous emission \cite{Vijay2011, Vool2016a, Campagne-Ibarcq2016b,Naghiloo2015}.
Moreover, in some situations, these jumps provide warning signs before they occur so that low-latency feedback controllers can reverse a jump while it is occurring \cite{Minev2018}.

Access to the continuous measurement process often comes with the ability to turn the measurement off.
In such a case, the quantum system may be partially measured. A system's free-evolution can be interrupted by a partial interrogation before continuing its free evolution.
Combined with unitary control of the system, partial measurements provide an excellent tool for control \cite{Balouchi2016,Hacohen-Gourgy2020}, measurement-based feedback both in real-time \cite{Campagne-Ibarcq2013,Bretheau2015}, and autonomously \cite{Arenz2016, Szombati2019}.

A particularly fascinating use case for partial measurements is when the partial measurement occurs between two fully projective measurements. 
The state begins in a pre-selected eigenstate (i.e.~protectively measured) and evolves under the presence of a weakly coupled apparatus. 
The system is then fully projected onto a different eigenstate. The pre- and post-selected expectation value called a weak value \cite{Aharonov1988}.
The weak value has excited debate across theory and experiment~\cite{Leggett1989,Aharonov1990,Hosten2008,Dressel2014, Jordan2014}.
Most remarkably, the magnitude of weak values can exceed the spectrum of the measurement operator \cite{Kocsis2011}. For example, a spin-1/2 operator (whose spectrum is $[-1,1]$) can have an anomalous weak value of greater than 1.


Thus, weak measurements provide unique views into quantum dynamics. 
In this thesis, we explore the statistics of weak measurement from two perspectives.
First, we look at how weak measurements affect uncertainty relations governing incompatible observables, similar to position and momentum.
We find that a weak measurement can reconcile the two observables, through the action of the weak value.
Second, we look at the reversibility (or lack thereof) of the wavefunction collapse process. In the process, we identify essential features of entropy generation in accordance with a generalization of the second law of thermodynamics.

\section{Quantum Mechanics with Circuits}
In this section, we describe the quantum nature of superconducting circuits.
We begin by first quantizing the electromagnetic fields of a microwave cavity, then we quantize the modes of a linear and a nonlinear LC oscillator.

	\subsection{Quantizing the cavity}
		\label{sec:cavityQuant}
	In this section, we describe the quantization of a single mode of the electromagnetic (EM) field.
	The quantum mechanical analysis of the classical equations of motion results in the quantum harmonic oscillator.
	
	Classical EM theory centers on Maxwell's equations, which can be written using the vector potential, $\vec{A}$:
	\begin{equation}\label{key}
	\left( \nabla^2 -\frac{1}{c^2} \partial_t^2\right) \vec{A} = 0.
	\end{equation}
	We have chosen the Coulomb gauge, $\nabla \cdot A=0$ so that the magnetic and electric fields may be written as:
	
	\begin{subequations}
	\label{eqn:EandB}
	\begin{align}
		\vec{E} &= \partial_t \vec{A}\\
		\vec{B} &= \nabla \times \vec{A}.
	\end{align}	
	\end{subequations}
	
	The EM mode we focus on is the fundamental mode of a three-dimensional microwave cavity \cite{Steck2007}.
	The linearity of this system and EM theory allows us to perform separation of variables on the vector potential: $A(\vec{r},t) = f(\vec{r})\, \alpha(t)$. 
	Considering a single mode of frequency $\omega$ allows us to calculate $\vec{E}  = -i\omega f(\vec{r}) \,\alpha(t) $.
	The spatial extent of the fields is set by $f(\vec{r})$, while the temporal variation is set by $\alpha(t)$. 
		
	Quantizing EM fields involves quantizing their Hamiltonian.
	Consider the pseudo%
	\footnote{The term pseudo refers to the fact that the effective mass which links $p$ and $q$ is the permittivity of free space, $\epsilon_0$, and that the spatial extent of the field (given by $f(\vec{r})$) is independent of $p$ and $q$.}
	 position and momentum coordinates,
	\begin{align}
		p(t)&= -\omega \epsilon_0 (\alpha(t) + \alpha^*(t))\\
		q(t)&= -i(\alpha(t) - \alpha^*(t)),
	\end{align}
	defined so that $p=\epsilon_0 \partial_t q$. We suppress the time-dependence for notational compactness.
	We can use these conjugate variables to transform the classical EM Hamiltonian:
	\begin{equation}
		\begin{split}
			H_{\rm EM,c}& = \frac{\epsilon_0}{2} \int dV\left(|\vec{E}|^2 + \omega^2 |\vec{A}|^2\right)\\
			& = 2 \epsilon_0 \omega^2 |\alpha|^2 \\
			&= \frac{p^2}{2\epsilon_0} + \frac{\epsilon_0}{2} \omega^2 q^2.
		\end{split}
	\end{equation}
	
	To more easily represent the above harmonic oscillator Hamiltonian, we use the creation and annihilation operators:
	\begin{equation}
	\begin{split}
		p &= i \sqrt{\frac{\epsilon_0 \hbar \omega}{2}}  (a^\dagger - a)\\
		q &= \sqrt{\frac{\hbar}{2 \epsilon_0 \omega}} (a + a^\dagger) .
	\end{split}
	\end{equation}
	Comparing to the quantum harmonic oscillator (QHO) Hamiltonian allows us to identify a quantization procedure to translate from classical to quantum descriptions:
	\begin{equation}\label{eqn:quantization}
	\alpha(t) \rightarrow i \sqrt{\frac{ \hbar }{2\epsilon_0 \omega}}  \,a,
	\end{equation}
	where $a$ is the typical creation operator whose explicit time dependence is understood in the Heisenberg picture \cite{sakurai}.
	
	The Hamiltonian of a single mode of the quantum mechanical EM fields thus becomes the QHO Hamiltonian:
	\begin{equation}
		\label{eqn:QHO}
	H_{\rm EM,q} = H_{\rm QHO} = \hbar \omega(a^\dagger a + \frac{1}{2})
	\end{equation}.
	
	We can also use Equation [\ref{eqn:quantization}] and Equations [\ref{eqn:EandB}] to identify the EM field operators:
	\begin{equation}
	\begin{split}
		E &= -\sqrt{\frac{\hbar \omega}{2\epsilon_0}} f(\vec{r}) \, \left(a +a^\dagger \right)\\
		B &= i \sqrt{\frac{ \hbar }{2\epsilon_0 \omega}} \left[\nabla \times f(\vec{r})\right] \, \left(a - a^\dagger\right)
	\end{split}
	\end{equation}
	The spatial extent of the fields retains its classical component unaffected. Thus, standard classical EM calculations lead to the spatial extent of quantum mechanical EM fields. In particular,
	quantum mechanical modes of complex EM structures are readily simulated with finite element modeling \cite{Minev2020}.
	Our analysis will primarily focus on the Hamiltonian form of the field from Equation [\ref{eqn:QHO}].

	\subsection{Quantizing the LC Circuit}
	\label{sec:LCQuant}
	Having quantized the EM field inside the cavity, we now describe how to attain a qubit out of a quantized circuit.
	Circuit quantization, like EM field quantization, has the structure of the quantum harmonic oscillator.

	The central circuit of interest is an LC oscillator.
	The state of the LC oscillator is defined in terms of the circuit's current, $I$, and voltage, $V$. 
	We transform these variables to get a canonically conjugate pair of variables.
	The voltage relates to the total charge, $q$, on the capacitor (with capacitance $C$):
	\begin{equation}
	V = \frac{q}{C},
	\end{equation}
	The current relates to the branch flux, $\phi$, passing through the inductor (with inductance $L$):
	\begin{equation}
	\label{eqn:branchFlux}
	\phi =  \int_{-\infty}^t V(t') dt' = I \, L.
	\end{equation}
	The capacitor contributes charging energy $q^2/2C$, while the inductor contributes inductive energy $\phi^2/2L$.
	Thus the Hamiltonian is given by:
	\begin{equation}
	H = 	\frac{q^2}{2C} + \frac{\phi^2}{2L}.
	\end{equation}
	
	The connection between the classical harmonic oscillator Hamiltonian above and the QHO Hamiltonian from Section \ref{sec:cavityQuant} suggests a similar variable transformation as Equation \ref{eqn:quantization}.
	We define raising and lowering operators for the quantized LC oscillator via:
	\begin{equation}\label{key}
	\begin{split}
		\phi &=  \phi_{\rm ZPF} \,(a+a^\dagger) \\
		q &= i \,q_{\rm ZPF} \,(a^\dagger -a),
	\end{split}
	\end{equation}
	where $\phi_{\rm ZPF}=\frac{\hbar Z}{2}$ and $q_{\rm ZPF}=\frac{\hbar }{2 Z}$ are the magnitudes of quantum zero-point fluctuations for the two operators with $Z=\sqrt\frac LC $.
	The operators obey the commutation relation $[\phi, q] = i\hbar$.
	With this representation, the LC oscillator also obeys the QHO Hamiltonian (Eqn.~[\ref{eqn:QHO}]).
	
	\subsection{Quantizing the Nonlinear LC}
	\label{sec:transmon}
	The LC oscillator described in the previous section has evenly spaced energy levels.
	To obtain a uniquely addressable qubit, the harmonic spacing must be broken.
	The non-linearity of a Josephson junction (JJ) provides a circuit with anharmonic energy levels.
	
	A Josephson junction is composed of two superconductors separated by an insulating barrier.
	As described in Section \ref{sec:JJFromBCS}, {JJ}s have a nonlinear inductance given by 
	\begin{equation}
	L_{\rm JJ} = \frac{I_0}{2 \pi\Phi_0 \cos\phi},
	\end{equation}
	where $\phi$ is the difference between the two superconductors' phases, $I_0$ is a fabrication-dependent critical current, and $\Phi_0= h/2e$ is the flux quantum. 
	
	The {JJ} contributes inductive energy
	\begin{equation}
	U_{\rm JJ} = \int V(t') I dt' = \frac{\Phi_0 I_0}{2 \pi} \left( 1 -\cos\phi\right) = E_J (1 -\cos\phi),
	\end{equation} 
	where we have defined the Josephson energy as $E_J =\frac{\Phi_0 I_0}{2 \pi}$. The {JJ} also contributes small capacitive energy which depends on the {JJ}'s geometry.
	So the Hamiltonian of the nonlinear LC circuit is:
	\begin{equation}
		\label{eqn:transmonHam}
		H = \frac{q^2}{2C} -\frac{\Phi_0 I_0}{2 \pi} \cos\phi = 4 E_C \,n^2 - E_J \cos\phi.
	\end{equation}
	We have represented the charge variable with the number of cooper pairs, $n$, (each having charge $2e$).
	We have defined $E_C= \frac{e^2}{2C_\Sigma}$ as the charging energy per electron with capacitance $C_\Sigma = C_{\rm shunt} + C_{\rm JJ}$ is the contribution from both the shunt capacitance as well as the {JJ}'s capacitance.
	
	Equation [\ref{eqn:transmonHam}] has an exact solution that depends on Mathieu’s characteristic value \cite{Koch2007}. 
	Because we will focus on the lowest energy levels where the $\phi$ variable is well-localized, we can Taylor expand the cosine potential:
	\begin{equation}\label{key}
	H = 4 E_C \,n^2 + \frac{E_J}{2} \phi^2 - \frac{E_J}{24} \phi^4.
	\end{equation}
	The first two terms recreate the LC Hamiltonian, and the last term provides the required anharmonicity.
	In the transmon limit, $E_J/E_C \gg 1$, the first transition frequency is $\omega_{01} = \sqrt{8 E_J E_C}-E_C$, the anharmonicity is $-E_C$, and charge noise is exponentially suppressed in $E_J/E_C$.
	The anharmonicity allows us to truncate the energy levels to the first two, which we use as our qubit. 
	
%
	
\section{Readout Physics}
	This section describes the measurement process for our quantum circuits.
	The cQED architecture utilizes cavity (or generic EM resonator) modes as an auxiliary quantum system to readout the qubit state. To measure the qubit, the qubit's state information first imprints onto the cavity's state, and then the cavity's state is measured.
	
	\subsection{Jaynes-Cummings Interaction}
		\label{sec:dispersive}
		The Jaynes-Cummings (JC) Hamiltonian describes a common interaction between an atom and a cavity.
		The cavity state is described by creation and annihilation operators, $a$ and $a^\dagger$. 
		We consider two levels of the atom and adopt the familiar pseudo-spin-1/2 notation where the operator $\sigma_z$ acts on the state in the energy basis. With this, the atom may be treated as a qubit.
		The total Hamiltonian is composed of three constituent Hamiltonians:
		\begin{align}
			H_{\rm cav}    &= \hbar \omega_c \left(a^\dagger a  + \frac{1}{2}\right)\\
			H_{\rm atom} &= \hbar \frac{\omega_q}{2} \sigma_z\\
			H_{\rm int} &= \vec{E}\cdot \vec{d}
		\end{align}
		In the JC model, the cavity and atom couple with a dipole interaction between the cavity's electric field, $\vec{E}$, and the atom's dipole operator, $\vec{d}$.
		
		The cavity's electric field operator was calculated in Section \ref{sec:cavityQuant}. The result is
		\begin{equation}
		\vec{E} = -\sqrt{\frac{\hbar \omega_c}{2 \epsilon_0}} \left[ f(\vec{r}) \,a + \mathrm{h.c.} \right]
		\end{equation}
		The spatial dependence, $f(\vec{r})$ is classical and presumed real, but the time dependence (through the Heisenberg picture of $a$) is quantum mechanical \cite{Steck2007}.
		
		The atom's dipole operator is 
		\begin{equation}
		\vec{d} = \vec{d}_{01} \big( \ket{0}\bra{1}+\ket{1}\bra{0} \big)  = \vec{d}_{01}(\sigma + \sigma^\dagger),
		\end{equation}
		where $\vec{d}_{01} = \braket{0|(-e) \vec{r}|1}$ is the dipole matrix element which depends on the dipole's position vector, $\vec{r}$, the fundamental charge $e$, and is presumed real.
		The coupling operates on the atom with $\sigma = \sigma_x + i \sigma_y$, which merely exchanges ground and excited states.
		Evaluating the dot product between $\vec{E}$ and $\vec{d}$ produces a coupling rate, 
		$\hbar g = -\sqrt{\frac{\hbar \omega_c}{2\epsilon_0}}  \, \vec{d}_{01}\cdot f(\vec{r})$.
				
		Combing the atom, field, and interaction Hamiltonians produces the JC Hamiltonian:
		\begin{equation}
			\label{eqn:JC}
			H_{\rm JC} /\hbar = \omega_c \left(a^\dagger a  + \frac{1}{2}\right) 
					+ \frac{\omega_q}{2} \sigma_z
			 + g(a+a^\dagger)(\sigma + \sigma^\dagger).
		\end{equation}
		
		The rotating wave approximation (RWA) replaces high-frequency terms with their average over the time scale set by the interaction rate.
		The approximation is not valid when the cavity photon number is high
		%
		The RWA approximation results in a new JC Hamiltonian:
		\begin{equation}\label{eqn:JCRWA}
			H^{({\rm RWA})}_{\rm JC} /\hbar = \omega_c \left(a^\dagger a  + \frac{1}{2}\right) 
			+ \frac{\omega_q}{2} \sigma_z
			+ g(a \sigma^\dagger+a^\dagger\sigma).
		\end{equation}
		From comparing Equations [\ref{eqn:JC}] and [\ref{eqn:JCRWA}], 
		the neglected high-frequency terms correspond to two-excitation processes, such as $a\,\sigma$ which annihilate one excitation in each of the cavity and atom.
		
		Because $a$ acts on an infinite-dimensional Hilbert space, finding energy eigenvalues of Equation \ref{eqn:JCRWA} can be challenging. 
		Fortunately, the Hamiltonian is block-diagonal and can be solved analytically. See References \cite{Naghiloo2019,Harrington2020} for insightful derivations.
		We elect to diagonalize the Hamiltonian numerically by truncating the Hilbert space, corresponding to only allowing for a fixed number of excitations. This is reasonable because the RWA doesn't apply for higher occupation numbers.
		The numerical treatment also provides a quick generalization when the multi-level transmon qubit replaces the two-level atom.
		
\begin{figure}[H]
			\centering
			\includegraphics[width=0.7\linewidth]{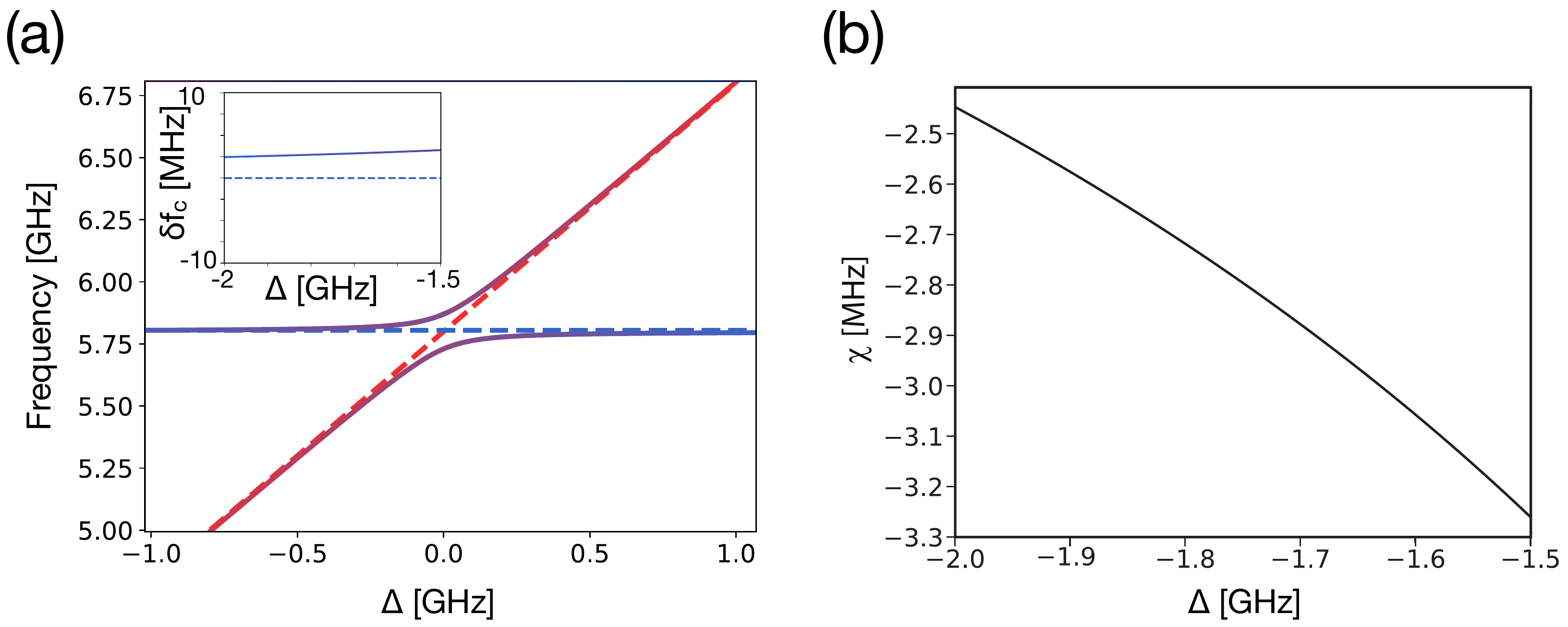}
			\caption[Jaynes-Cummings Hamiltonian Eigenenergies]{(a) The frequencies of the JC Hamiltonian [Eqn.~\ref{eqn:JCRWA}] as a function of the atom(qubit)-cavity detuning, $\Delta=\omega_q-\omega_c$ with $\omega/2\pi=5.8$ GHz and $g/2\pi=70$ MHz. The atom's frequency (red) increases while the cavity frequency (blue) remains fixed.  
			Compared to the uncoupled systems (dashed lines), atom-cavity coupling induces an avoided crossing at $\Delta=0$ of separation $2g$.
			Near resonance (purple) the atom-cavity constituents are indistinguishable.
			The inset shows the cavity behavior in the case of large detuning ($\Delta \gg g$).
			 (b) The dispersive interaction rate, $\chi$, as a function of the atom-cavity detuning.}
			\label{fig:JCDiag}
		\end{figure}

		We rely on the generative definitions for the matrix representation of $a$ and $a^\dagger$:
		\begin{align}\label{eqn:aDef}
			a\ket{n} &= \sqrt{n}\ket{n-1}\\
			a^\dagger\ket{n} &= \sqrt{n+1}\ket{n+1},
		\end{align}		
		In Figure \ref{fig:JCDiag}(a), we plot the lowest two eigenstates of the JC Hamiltonian after truncating the Hilbert space to five excitations.
		We see that when the atom and cavity are nearly resonant, their frequencies shift by $2g$.
		Near this avoided crossing, the atom and cavity are not well-distinguished: The energy eigenstates are mixtures of atom and cavity modes.
		Far from the avoided crossing, the cavity retains a small atomic component. 
				
		\subsubsection{Dispersive Jaynes Cummings}
		The Jaynes-Cummings Hamiltonian is further simplified when the atom and cavity are significantly detuned.
		In the dispersive limit, $|\Delta| \gg g$, a Schrieffer-Wolf transformation \cite{Schrieffer1966} with the unitary operator $U=e^{ \frac{g}{\Delta} \left(a\sigma^\dagger -a^\dagger \sigma\right)}$ results in the dispersive JC Hamiltonian:
		\begin{equation}\label{eqn:JCdisp}
		H^{({\rm disp})}_{\rm JC} /\hbar = \omega_c \left(a^\dagger a  + \frac{1}{2}\right) 
		+ \frac{\omega_q}{2} \sigma_z
		+ \chi a^\dagger a \sigma_z.
		\end{equation}
		The dispersive rate, $\chi = \frac{g^2}{\Delta}$, describes the small energy shift in the atom's and cavity's energy dispersion, shown in Figure \ref{fig:JCDiag}(b).
		
		The final term describes the atom-cavity interaction in this limit.
		Because the term is proportional to $a^\dagger a$, it can be incorporated into the cavity frequency:
		$\omega_c \rightarrow \omega_c + \chi \sigma_z$.
		The final term is also proportional to $\sigma_z$, so it can be incorporated into the atom's frequency:
		$\omega_q \rightarrow \omega_q + \chi a^\dagger a$.
		Each frequency shift occurs in tandem, affecting both atom%
		\footnote{The factor of $\frac{1}{2}\omega_c$ in the cavity Hamiltonian also shifts the atom's frequency by a small amount called the Lamb shift. But this is neither controllable nor observable, so we ignore it.}
		 and cavity frequencies.

		The cavity shift is of primary interest because it enables measurement of the atom's state.
		With the dispersive interaction, the cavity's frequency encodes the atom's state, being shifted by $+\chi$ if the atom's state is $\ket{0}$ and $-\chi$ if the atom's state is $\ket{1}$.
	
		This dispersive measurement technique is desirable for two reasons. First, dispersive measurements are non-destructive. 
		This contrasts with other platforms such as photon- or neutron-based qubits \cite{Hadfield2009,Chen2011, Demirel2019a}, which perform measurements by absorbing the system. 
		Such a measurement obviously prevents any further evolution or measurement.
		Dispersive measurements are non-destructive because the measurement interaction only shifts the frequencies of the atom and cavity, rather than transferring excitations between systems.
		
		However, merely non-destructive measurement can still significantly affect the system's dynamics.
		The second advantage of dispersive measurements is that they are quantum non-demolition {QND}. 
		QND measurements do not affect the system's dynamics during measurement. 
		This requires the system's Hamiltonian commutes with the interaction Hamiltonian \cite{Thorne1978, Poizat2007}.
		
		For the dispersive JC Hamiltonian, as described in Equation [\ref{eqn:JCdisp}], verifying QND is easy because the atom's Hamiltonian clearly commutes with the interaction Hamiltonian: $[\sigma_z,a^\dagger a \, \sigma_z]=0$.
		However, the calculation for transmon qubits coupled to microwave cavities is far more involved.
		Detailed calculations of the dispersive approximation \cite{Gambetta2006,Boissonneault2009} show that cQED measurement is approximately QND when the cavity contains less than the critical number of photons, $\bar{n}_c = \frac{ \Delta^2}{4 g^2}$.
		
		In this regime, cQED systems can be measured continuously. While the trajectory approach is outside the context of this dissertation, it provides a fascinating perspective on quantum dynamics \cite{Boissonneault2009, Hatridge2013, Murch2013, Weber2014,Vijay2011, Vool2016a,Foroozani2016, Campagne-Ibarcq2016b,Naghiloo2015, Campagne-Ibarcq2016c,Hacohen-Gourgy2016,Ficheux2018,Minev2018}.

	\subsection{Partial Measurements}
	\label{sec:partialMeasThy}
	\subsubsection*{Projective Measurement}
	In the simplest textbook approach, quantum measurement completely collapses the wavefunction and returns unambiguous state information.
	Consider a system in a general superposition state. The wavefunction may be written in the measurement basis as $\ket{\psi} = \sum_i \alpha_i \ket{i}$. The amplitudes, $\alpha_i$ are complex, and the states, $\ket{i}$, are eigenstates of the measurement operator. 
	This braket formulation describes pure states. However, partial measurements can result in incomplete information about the state. Therefore, we move to the density matrix formalism \cite{BJQuantum,qcqi}, where the state is $\rho = \ket{\psi}\bra{\psi}$.
	
	When a measurement occurs, the wavefunction collapses onto one of the eigenstates, chosen randomly with probability equal to the amplitude's square%
	\footnote{The square has been called the footnote that won a Nobel Prize. Max Born's 1926 paper \cite{Born1926} did not originally propose the squared magnitude. However, a footnote reads ``Addition in proof: More careful consideration shows that the probability is proportional to the square of the [amplitude].''}
	 magnitude.
	 The collapse is realized mathematically by applying a projector.
	 The projector representation decomposes the measurement operator, $A$, into a complete set of projectors:
	 \begin{equation}\label{eqn:projectorDecomp}
		 A = \sum \lambda_i \Pi_i \equiv \sum_i \lambda_i \ket{i}\bra{i},
	 \end{equation}
	 where $\lambda_i$ is an eigenvalue of the measurement operator with associated projector $\Pi_i$.
	If the random outcome is $\lambda_i$, then the state updates by applying the corresponding projector:
	\begin{equation}\label{eqn:projectorStateUpdate}
	\rho \rightarrow \rho' = \frac{\Pi_i \rho \Pi_i^\dagger}{\Tr[\Pi_i \rho \Pi_i^\dagger]} .
	\end{equation}
	The outcome occurs with probability $p_i = \Tr[\rho \Pi_i]$.

%

	\subsubsection*{Non-Projective Measurement}
	The projective model of measurement supposes that wavefunction collapse is immediate and irreversible.
	However, realistic measurements involve an interaction with a measurement apparatus over a finite amount of time.
	The interaction is unitary, and thus in principle reversible \cite{Jordan2010}.


		For partial measurements, we generalize the update operators from projectors to Kraus operators. 
		The Kraus operator formalism describes measurement as an interaction between the system and a measurement apparatus followed by the projection of the apparatus. The Kraus operator itself describes the effect of this process on the qubit state. See References \cite{Brun2001,Jacobs2005,qcqi} for detailed treatments.
		
%
		Upon obtaining an outcome $j$, the state updates via the Kraus operator $K_j$:
		\begin{equation}\label{eqn:krausUpdate}
			\rho \rightarrow \rho' = \frac{K_j \rho K_j^\dagger }{\Tr[K_j \rho K_j^\dagger]}.
		\end{equation}
		The probability of obtaining outcome $j$ is
		\begin{equation}\label{eqn:krausProb}
			P(j) = \Tr[ K_j \rho K_j^\dagger ].
		\end{equation}
		Furthermore, the probability of obtaining an outcome within a range, $[a,b]$, is given by 
		\begin{equation}\label{eqn:krausProbInterval}
			P([a,b]) = \sum_{j=a}^b \Tr[ K_j \rho K_j^\dagger ].
		\end{equation}
		The full set of Kraus operators, $\{K_j\}$, satisfies the relation $\sum_j K_j K_j^\dagger = 1$.
		This guarantees that $\{K_j\}$ is a valid (completely positive) density matrix transformation (ie that the $\rho'$ satisfies total probability $\Tr[\rho']=1$) and that the set of possible outcomes, $\{j\}$, is self-consistent.
		
		Using the relations above, we can construct a new operator $M$ = $\sum_{j=a}^b K_j K_j^\dagger$ which describes the effect of a coarse-grained measurement of all values on the interval%
		\footnote{Because this holds for any interval, the Kraus operator set forms a measure. Because each $K_j$ is positive, the new $M$ is positive. Hence, the Kraus operator map is a positive measure that is based on operators---a positive operator-valued measure.}
		$[a,b]$.
		If the interval is large enough, $M$ can describe the effect of a projector.
		Thus, Kraus operators can be thought of as building blocks for other measurement operators. Or conversely, that a single measurement, such as a projector, can be broken down into constituent Kraus operators.

		The Kraus operator formalism appears very similar to projective measurements (cf.~Eqn.~\ref{eqn:projectorStateUpdate} with Eqn.~\ref{eqn:krausUpdate}).
		The similarity arises because any valid measurement can be modeled as an interaction between the system and an auxiliary followed by a projective measurement of the auxiliary \cite{vonNeumann1932, Jacobs2006}.
		The distinction comes from the number of outcomes.
		Unlike the set of projectors, with cardinality equal to the Hilbert space dimension, a set of Kraus operators can have any number of elements.
		In particular, for the qubit-type systems we study in this dissertation, there are only two projectors for a given measurement operator (e.g.~$\sigma_z$), corresponding to the two eigenstates. 
		But our weak measurement has many more outcomes, thanks to the cavity's many pointer states.
		The multitude of outcomes allows for subtle state update (backaction) after a non-projective measurement.

		\subsubsection{Partial Measurement in cQED}	
		For a detailed treatment of obtaining Kraus operators in cQED platforms, see References \cite{Boissonneault2009,quantumMeasurement,Hatridge2013}.
		The key result is that because the cavity state is prepared as a coherent state, the probability of a partial measurement outcome is Gaussian-distributed.
		Therefore, the Kraus operator is given by:
		\begin{equation}\label{eqn:Kraus}
			K_j  
			=  \left(  \frac{1}{{2 \pi \sigma^2}}  \right)^{1/4} 
			\exp \left(  -\frac{[j I - \sigma_z]^2}{2 \sigma^2}   \right).
		\end{equation}
		The variance, $\sigma^2$ is given by:
		\begin{equation}\label{eqn:measurementRate}
			\sigma^2 = \frac{\tau}{\delta t},
		\end{equation}
		where $\delta t$ is the measurement duration, and the measurement rate $\frac{1}{\tau} = \frac{8 \chi^2 \bar{n}}{\kappa}$ \cite{Blais2004}, for a cavity of linewidth $\kappa$, populated with $\bar{n}$ photons and dispersively coupled to the qubit at a rate $\chi$.

\chapter{Direct-Write Qubits}

\section{Introduction}
This section details the development and results of a novel fabrication method for Josephson junctions (JJs) for use in superconducting qubits. 
The junctions are fabricated with an all-optical direct-write laser lithography system. 
%
Because the features written into a resist mask are significantly larger when using photolithography rather than e-beam lithography, photolithography-written {JJ}s often have large overlap areas.
Large lithographic areas lead to high numbers of two-level system (TLS) defects, as has been seen in a variety of experiments \cite{Martinis2005,Stoutimore2012}.

{JJ}s primarily function as a nonlinear inductance for LC circuits (see Sect.~\ref{sec:transmon}).
However, they also host part of the electric field which can dipole-couple to TLS defects.
To improve device quality, defect numbers and densities should decrease. 
Prior work has shown that decreasing {JJ} areas is an effective strategy for mitigating TLS loss.
However, as this chapter investigates, careful deposition and design  can still lead to low TLS and low loss even with large areas. 

TLSs have been a significant source of loss in state-of-the-art quantum devices, including both superconducting \cite{Muller2015, Klimov2018a,Burnett2019,McRae2020} and trapped ion \cite{Daniilidis2011} platforms.
Thus, devices have historically been made as small as possible while maintaining robust fabrication yield.
Though most of this shrinking has focused on the area of the junction, some work has also used extremely thin superconducting films (a few nm). Thinner films diminish the expected number of TLS at the cost of diminished superconductivity \cite{Shearrow2018}.

For our process, we seek to incorporate two lines of investigation which have proved beneficial in the superconducting circuit community: geometric variation and careful surface treatments.
For both coplanar-waveguide resonators and qubits, a device's geometry can be adjusted to minimize electric field storage in lossy materials.
While the primary materials (silicon, sapphire, aluminum, and niobium) have low loss \cite{Oliver2013}, other materials which find their way into devices do not have low loss. These primarily reside at metal-substrate, metal-air, and substrate-air interfaces, and they are often introduced during fabrication steps.

In resonators, modifying the form (either $\lambda/2$, $\lambda/4$, or lumped element) modifies how much of an electric field is stored at these interfaces \cite{Lindstrom2009, McRae2019}, as quantified by the filling factor \cite{Zmuidzinas2012,McRae2020}.
Detailed studies of resonator geometry have modified the filling factor by, for example, over-etching the substrate  to minimize the substrate-air interface \cite{Vissers2012,Bruno2015,Calusine2018, Woods2019,Nersisyan2019a}.
Other work has focused on the metal-substrate interface.
Larger center trace widths diminished TLS loss \cite{Barends2010,Geerlings2012,Jin2017,Niepce2020}. 
If TLS at interfaces can be suppressed, then large feature sizes can dilute electric field densities, decreasing TLS coupling and  improving coherence.

In qubits, altering the geometry has also led to improved understanding of the origin of TLS loss.
Besides micromachining substrates as in resonators \cite{Chu2016}, simple geometric modification can highlight sources of loss. 
In particular, a variety of alternative designs for transmon shunting capacitor pads led to an understanding of the loss tangents of these materials \cite{Wang2015,Dial2016}.
As is commonly used in the field, 3D transmons \cite{Paik2011} attain high coherence by moving electric fields away from the substrate.
Similar to the diluted electric fields in resonators, large-area {JJ}s have recently shown high coherence times in mergemon qubits \cite{Zhao2020, Mamin2021}.

The second line of investigation we follow is that of surface treatments.
Early experiments with resonators \cite{Gao2008} and recent direct measurements imply that around 60\% of TLS reside in interfacial surfaces \cite{Lisenfeld2019,Bilmes2020}.
This understanding has motivated surface-treatment experiments focusing on removing or avoiding oxides and other contaminants through the use of buffered-oxide etching (BOE) \cite{Zeng2015,Altoe2020}, other chemical cleanings \cite{Nersisyan2019a}, and inter-metal bandages \cite{Grunhaupt2017, Dunsworth2017,Osman2020}.
This focus on clean surfaces has improved resonator quality factors to well above $10^6$ \cite{Burnett2018, Calusine2018, Woods2019, Nersisyan2019a, Altoe2020}.

\section{Fabrication Theory}
\subsection{Josephson Junction Nonlinearity}
\label{sec:JJNonlinearity}
	Chapter 1 described how the nonlinear inductance provided by {JJ}s creates a platform for studying quantum information of pseudo-spin 1/2 systems.
	In this section, we describe the underlying physics that provides the nonlinear inductance.
	
Two superconductors separated by an electrically insulating tunnel barrier form a {JJ}. The insulator impedes electron transport, allowing conduction only via tunneling.
The superconducting wavefunction can coherently traverse the insulating barrier only when pairs of electrons tunnel (Cooper pair tunneling) \cite{BCS, Martinis2004}.

The key feature of a {JJ} is its nonlinear inductance. The inductive nature of the SIS junction comes from the Josephson relations, which will be derived below, in Section \ref{sec:JJFromBCS}.

The Josephson equations define the current-voltage relationship in terms of the superconductor phases at each end of the junction.
The difference between the phases, $\delta$, generates a tunneling current and a potential difference according to the Josephson equations:
\begin{align}
	V &= \Phi_0 \dot{\delta} \\
	I &= I_0 \sin \delta
	\label{eqn:josephsonCurr}
\end{align}
Fabrication fixes the critical current, $I_0$ (see Section \ref{sec:Simmons}), and $\Phi_0$ is the reduced flux quantum, $\Phi_0 = \hbar/2e \approx \mathrm{2.0\,10^{-15} ~\mathrm{Wb} \approx 3.0}$ GHz $h$/nA, which depends on (reduced) Planck's constant $h$ ($\hbar$) and the fundamental charge $e$.
Manipulating the above equations leads to  the inductance of the junction:
\begin{equation}
	\begin{split}
		L &= \frac{V}{\dot{I} } \\	
			&= \frac{\Phi_0 \dot{\delta}} {I_0 \cos \delta \dot{\delta} } \\
			& = \frac{\Phi_0 } {I_0 \cos \delta  }
	\end{split}
\end{equation}

Because the Josephson equations result from coherent tunneling, {JJ}s provide dissipationless nonlinear inductance.



\subsection{Josephson Relations from BCS Theory}
\label{sec:JJFromBCS}	
To derive the effect of tunneling Cooper pairs---the Josephson effect----we begin with the BCS model stated as a superconductor Hamiltonian \cite{SerniakThesis, Martinis2004}:
\begin{equation}
	\begin{split}
		H_{\rm BCS} = J \Sigma_k a^\dagger_k a_k + V a^\dagger_k a^\dagger_{-k} a_k a_{-k}.
	\end{split}
\end{equation}
The Hamiltonian describes single-electron excitations of momentum $k$ with creation (annihilation) operators $a^\dagger_k$ ($a_k$). The second term describes a potential that enforces net-zero-momentum interactions.

A Bogoliubov transform translates the second-quantized Hamiltonian into more familiar two-level-system operators.
The transform better represents the symmetry enforced by the zero momentum condition.
The new creation and annihilation operators, $c_k$ and $c_k^\dagger$ are
\begin{align}
	c_k &= u a_k + v a_{-k}^\dagger \\
	c_{-k} &= -v a_k + u a_{-k}^\dagger,
\end{align}
where $u=\cos\theta$ and $v=\sin\theta$ such that $2\theta = \arctan\left( \frac{J}{V}\right)$.
 
In the new representation, we can truncate the Hilbert space to get two-dimensional operators:
\begin{equation}
	\begin{split}
		H = \sum_k{ \xi_k \sigma^k_{z} }
		- \frac{V}{2}\sum_{k,l}{  \sigma^k_{x}\sigma^l_{x} + \sigma^k_{y}\sigma^l_{y} },
	\end{split}
\end{equation}
where we consider two momentum $k$ and $l$, $\xi_k$ is the kinetic energy relative to the Fermi energy, and $V$ is the potential difference across the junction leads.

The form of the Hamiltonian $H\sim \sigma_z + \sigma_x$ is a familiar qubit Hamiltonian under Rabi drive \cite{Naghiloo2019,Harrington2020}.
The energy eigenstates for this Hamiltonian can be written in the form 
\begin{equation}
\psi = \bigotimes_k{\left[ u_k \left(\begin{smallmatrix} 1 \\ 0 \end{smallmatrix}\right) 
	+ e^{i\phi_k} v_k \left(\begin{smallmatrix} 0 \\ 1 \end{smallmatrix}\right)	\right]}.
\end{equation}
The two states correspond to a filled ($v_k$) or unfilled ($u_k$) momentum-pair state. 


The mean-field solution \cite{Martinis2004} shows that the phase $\phi_k$ depends exclusively on the potential term in the Hamiltonian. Thus, with direct control of the external voltage across the junction, one can directly modify the phase. The modification transforms the phase:
$\phi_k \rightarrow \phi_k + 2 e/h \int V dt$. 
The   derivative of this voltage-altered phase is: 
\begin{equation}
	\dt{\phi_k} = \frac{2e}{\hbar} V,
\end{equation}
which is the AC Josephson effect, Equation [\ref{eqn:josephsonCurr}].



	




\subsection{Cabrera-Mott Oxidation Model}
		\label{sec:cabreraThy}

	A key development in this chapter is a novel oxidation schedule to achieve low critical currents in large-area {JJ}s. 
	Here we present a summary of the Cabrera-Mott model for metal-oxide growth dynamics on thin films \cite{Cabrera1949,Atkinson1985}.
	The resulting dependence of oxide thickness on time is logarithmic and self-limited. The model is commonly used to describe the growth of aluminum oxide \cite{Hayden1981,Jeurgens2002,Kang2014,Zeng2015,Gorobez2021}.
	
	Metals oxidize when exposed to molecular oxygen that has adsorbed onto the metal's surface.
 	The growth process can be summarized in three steps. First, electrons tunnel through the barrier and ionize adsorbed oxygen.
	Second, a Mott potential results from the displaced charges.
	Finally, metal ions hop across lattice sites, expanding the size of the oxide.
	We describe each step below.
	

	Cabrera-Mott theory depends on the free transfer of electrons from the metal to the adsorbed oxygen.
	In Section \ref{sec:Simmons}, we will describe the physical mechanism for tunneling and show that the tunneling probability decreases (exponentially) with the thickness of the film. This is the origin of the self-limiting nature of Cabrera-Mott growth and limits the discussion to thin films.
	
	Electrons donated by the metal ionize the adsorbed oxygen atoms or molecules on the oxide-gas surface.
	A positive charge at the metal-oxide interface results from the metal's electron donation.
	The charge difference creates the Mott potential, $\Delta \varPhi$, which drives metal ions across the oxide, increasing the oxide thickness.
	We can calculate the magnitude of the Mott potential by treating the charge separation as a capacitor:
	\begin{equation}
		\Delta \varPhi = Q/C = (2e n_0) \cdot X/ \epsilon_r,
	\end{equation}
	where $X$ is the oxide thickness, $n_0$ is the number of excess oxygen ions per unit area, $\epsilon_r$ is the relative permittivity of the oxide layer, and $e$ is the elementary charge. $n_0$ is calculated via 
	the equilibrium chemical reaction $\frac{1}{2} O_2 + 2e \rightarrow O^{2-}$. It determines the oxygen ion concentration as a function of the free energy difference of the reaction and the concentrations of each reactant.

	Cabrera-Mott theory then estimates the oxidation rate with the rate of the slowest process: vacancy formation due to metal atoms hopping across sites.
	Vacancies result from either metal atoms entering the oxide (combing with oxygen in the oxide) or from metal atoms leaving the oxide (combing with adsorbed oxygen on the oxide surface). 
	The energy cost of moving across interfaces is the activation energy $W$, but this cost is lowered due to the Mott potential.
	The hopping probability is thus given by $\exp\left[-\left( W - e a \Delta\varPhi/X \right)/kT\right]$.
	The growth rate depends on the interatomic spacing $a$ and the time scale of hopping, set by the vibrational frequency of the lattice, $\nu$.
	Thus, the Cabrera-Mott theory of oxidation predicts a growth rate:
	\begin{equation}
	\dot{X} =  \frac{D}{a} e^{\frac{X_{\rm max}}{X}},
	\label{eqn:CabreraMott}
	\end{equation}
	where $X_{\rm max} = e a \Delta \varPhi/ k T$ sets the maximum film thickness for which Cabrera-Mott theory applies and $D= a^2 \nu e^{-W/kT}$ sets a material-specific diffusion constant.
	
	The qualitative behavior of this type of film growth is exponential suppression of the growth rate as the film thickness increases.
	Equation [\ref{eqn:CabreraMott}] can be solved in the limit $X\ll X_{\rm max}$ to recover inverse-logarithmic growth of oxide thickness with time:
	\begin{equation}
	X(t) \sim \frac{1}{1 - \ln t}
	\label{eqn:CabreraTime}
	\end{equation}
	
\begin{figure}[h!]
		\centering
		\includegraphics[width=0.7\linewidth]{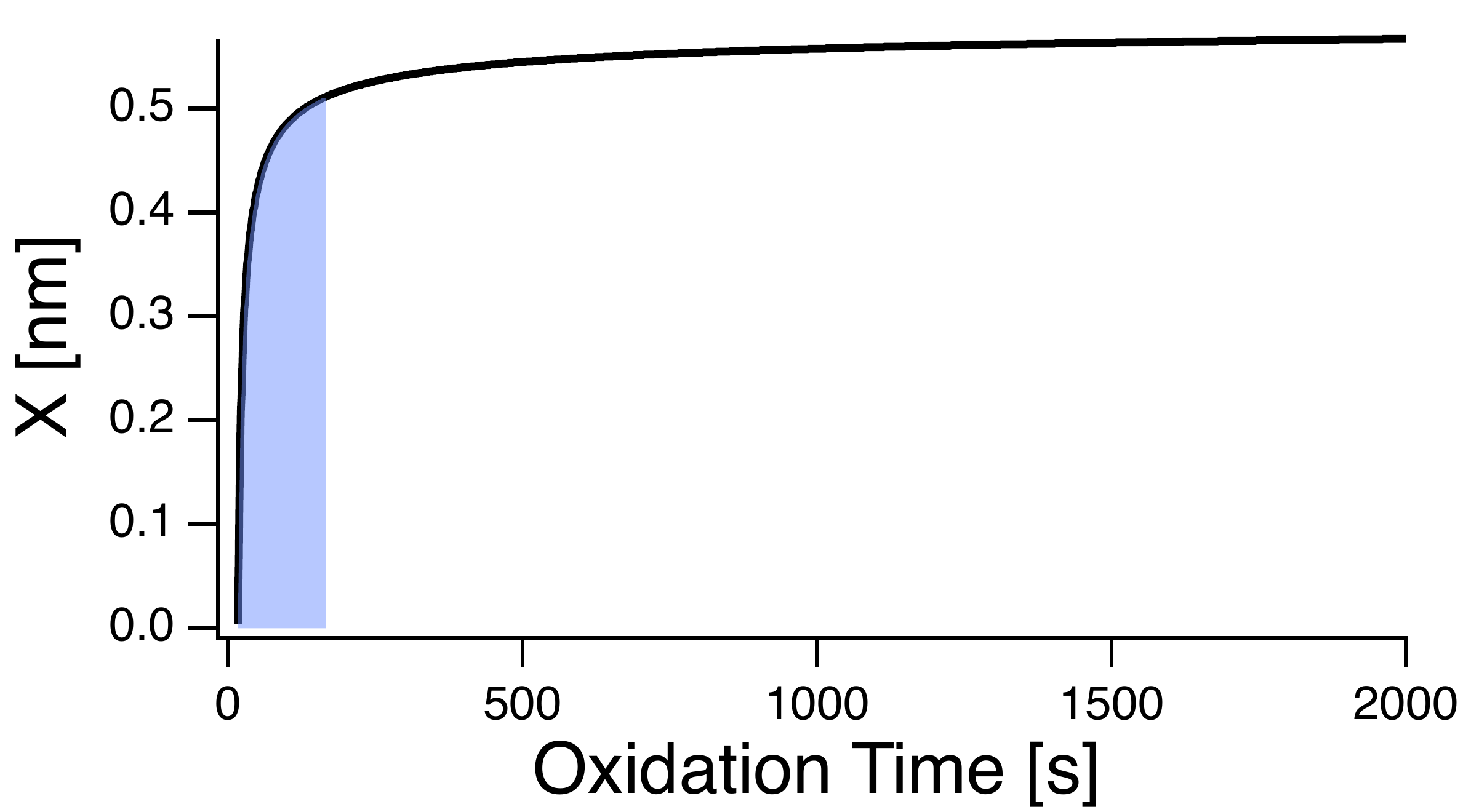}
		\caption[Cabrera Theory]{Oxide thickness, $X$, as a function of time under similar conditions to our oxidation environment. Two regimes are visible. A fast regime (blue) accounts for the first 90\% of the film grows, and the slow regime represents nearly saturated growth.
		Non-zero y-offset is due to numerical error.}
		\label{fig:cabreraTheory}
	\end{figure}

	This prediction has been recently measured directly with \textit{in situ} optical-transmission measurements during aluminum oxidation \cite{Gorobez2021}. Using constants fit to this data, the inverse-logarithmic behavior of Equation [\ref{eqn:CabreraTime}] is shown in Figure \ref{fig:cabreraTheory}. 
	In an initial fast-growth regime, the oxide thickness grows quickly. But the thickness quickly saturates to a maximal value within a few minutes, proceeding in a slow-growth regime.
	
	For the sake of tuning {JJ} critical currents, this strong logarithmic dependence is not insurmountable.
	The exponential sensitivity of critical current on oxide thickness [Eqn.~(\ref{eqn:Simmons})] cancels the logarithmic sensitivity of thickness on time, allowing reasonable scaling of critical current with oxidation time, up to a relatively thin limiting thickness.
	
	\subsubsection{Barrier Tunneling Resistance Theory}
\label{sec:Simmons}
We have discussed how the BCS Hamiltonian leads to eigenstates and dynamics of the current-voltage relationship for {JJ}s.
This section focuses on how junction geometry, especially the oxide barrier thickness, affects the critical current. The theory was originally developed by Stratton \cite{Stratton1962} and expanded by Simmons \cite{Simmons1963}.

Consider a {JJ} as a 1D tunneling barrier. With an applied potential difference, $V$, across the junction, electrons on one lead have a higher Fermi level than pairs on the other. However, the tunneling barrier  of height $\Phi$ inhibits conduction.
Provided the barrier is thin enough, electrons can tunnel through.
Using the WKB approximation \cite{BJQuantum}, the tunneling probability for an electron with energy $E$ is
\begin{equation}
p(E) = e^{-\frac{4\pi}{h}\sqrt{\Phi+\epsilon_F-E}}, 
\end{equation}
where $\epsilon_F$ is the Fermi energy of the metal.

To calculate the tunneling current, we need the number of electrons with energy $E$.
Modeling the electrons as a free Fermi gas model allows us to calculate the number of tunneling electrons.
The tunneling current density is given by 
\begin{equation}
	\begin{split}
J &= e \left(N_1 - N_2\right) \\
&= e \frac{4 \pi m^3}{h^3} \int \left[ f(E) - f(E-eV) \right] p(E) dE.
	\end{split}
\end{equation}
The integral can be solved for the specific case of a tunneling barrier and expressed as a resistance \cite{Simmons1963}:
\begin{equation}
R(X) = \frac{8\cdot2\pi R_0 \cdot X^2} {(1+2KX)\exp(2KX)},
\label{eqn:Simmons}
\end{equation}
where $K=\frac{1}{\hbar}\sqrt{2m \Phi}$ depends on the electron mass $m$, elementary charge $e$ and the tunneling barrier height, $\Phi$. $R_0=h/2e^2$ is the inverse of the conductance quantum.
For aluminum oxide, the estimates of the barrier range from $\Phi \sim 0.15-2$ eV \cite{Zeng2015,Paramonov2019,Kim2020a}.




The resistance derived above translates to a critical current via the Ambegaokar-Baratoff relation \cite{Ambegaokar1963}.
The relation comes from a detailed treatment of the superconducting wavefunction. However, the result has the familiar form of Ohm's law: 
\begin{equation}
I_c  = \frac{\frac{\pi}{2} \frac{\Delta}{e}}{ R_N}.
\label{eqn:ABRelation}
\end{equation}
The superconducting gap in volts is $\Delta/e$, and the normal resistance, $R_N$, can be probed at room temperature.
The full quantum mechanical treatment produces the factor of $\pi/2$ \cite{Ambegaokar1963}.

The joint requirements of frequency and transmon-regime set the scale for the required charging and Josephson energies, as discussed in Section \ref{sec:transmon}.
These specifications require $I_c \approx 10$ nA

\section{Recipe Development}
	\subsection{Design-of-Experiment Philosophy}
	\label{sec:DOE}
	When developing recipes and best practices, a common approach is to use ``A/B testing''. 
	A treatment (such as a cleaning step) can be applied (A) or not applied (B), and the final figure of merit (such as quality factor) can be directly evaluated based on the average outcome, while also considering standard deviations.
	This strategy can provide clear evidence for the efficacy of a single variable.
	However, when an outcome depends on multiple variables, the results can be misleading. For example, the A/B test results may differ when other variables changes. We will show an example in Section \ref{sec:sandwich}, where this was the case. 
	
	An alternative methodology is called ``Design of Experiment''.
	The most complete Design of Experiment test is a factorial test. 
	A factorial test sweeps each variable independently through its full range.
	Ideally, each sweep is repeated multiple times to quantify variability. 
	
	This test can be time- and labor-expensive, because for even just two treatments, the number of experiments grows like $2^n$ for $n$ number of variables.
	However, this methodology offers a complete statistical picture through analysis of variance (ANOVA).
	In particular, interactions between variables can be directly monitored in factorial design statistics.
	More advanced methods, such as nested factorial designs, offer more efficient approaches \cite{HicksDOE}.
	We will make use of a Design of Experiment test in Section \ref{sec:sandwich}.



\subsection{Resist Exposure Tuning}
\label{sec:ExposureRecipe}
\subsubsection{{Imaging Resist Only}}

In this section, we describe the lithography development process for our all-optical direct-write photolithography system. First, we focus only on one of the two resists in our dual-resist stack. Then, we describe the full resist stack.

We expose resist with a Heidelberg DWL 66+ system. To adjust the dose, the user may set the laser power, laser intensity, and filter.
Laser power and intensity combine to set the net dose. (Laser power is largely fixed for each write head.)
Filter percentage selects a neutral density filter to further attenuate the laser power, and thus further modifies the dose. 
The user may also adjust the focus of the optical write head.
The focus is a percentage value and sets the lens's relative Z-offset from its default position. This allows the laser to target different parts of the resist stack.
Each of these four parameters must be tuned to some precision to achieve target feature sizes. 

In addition to exposure calibration, the resist stack must be calibrated as well.
For example, we can also tune the resist behavior with softbake time and temperature along with development time, temperature, and mechanical agitation (such as by hand or with an ultrasonic bath).

Thus, the full calibration procedure can be quite involved. 
Degeneracy in the above nine variables confounds interpretation. For example, one can tune the exposure dose by using both the intensity and the filter percentage. 
While we have experimented with each of the nine parameters listed above, this section will only focus on changes in a few and leave the rest fixed to optimized values.


\begin{figure}[h!]
	\centering
	\includegraphics[width=0.75\linewidth]{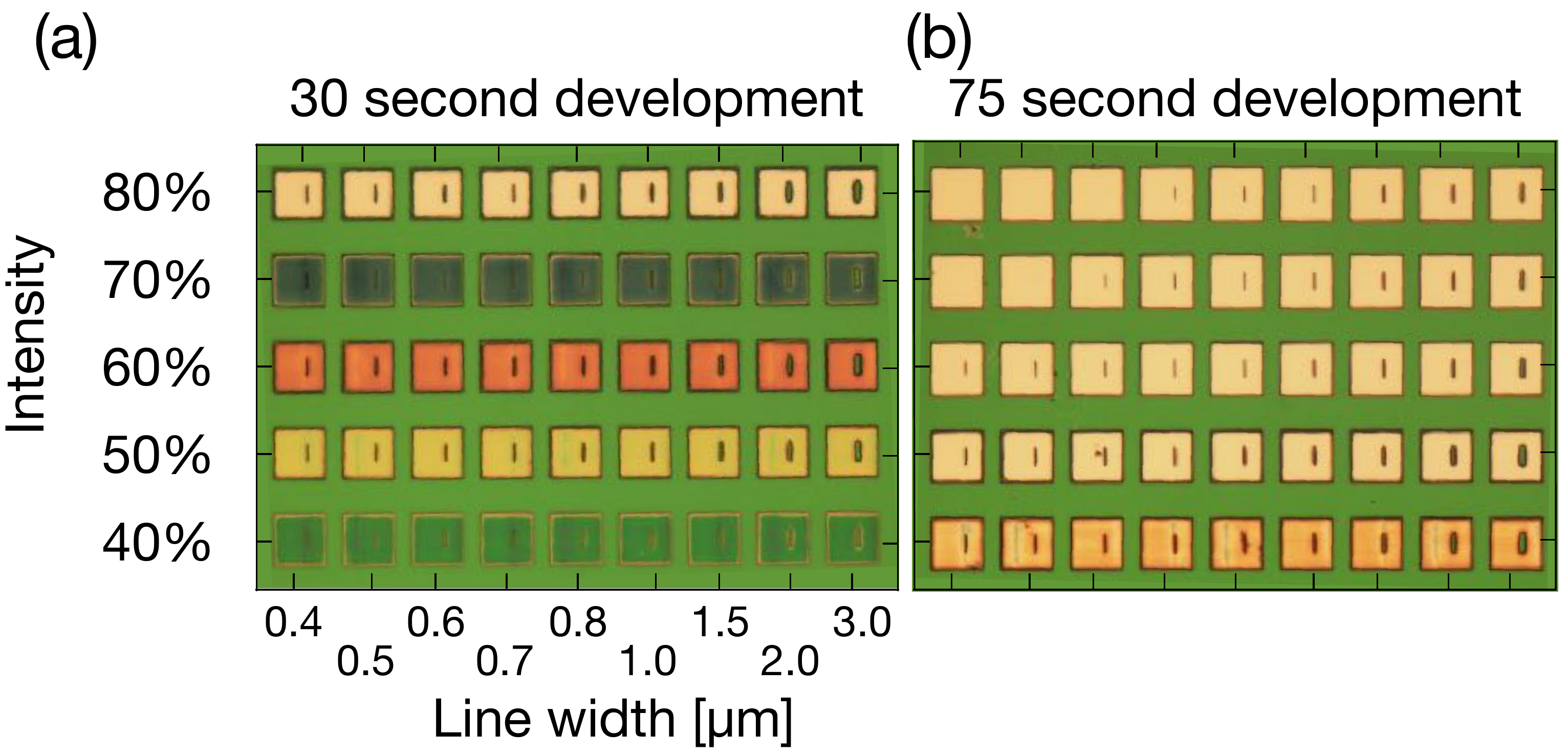}
	\caption[Exposure Test for S1805]{An optical micrograph of S1805 resist after an exposure dose test. The images display a three-dimensional slice of a five-dimensional sweep between the exposure intensity, the development time, minimal line width, filter percentage (fixed), and focus percentage (fixed).
	Colors (unaltered) indicate residual resist of varying thickness. Dark green corresponds to the full stack without exposure, and tan corresponds to the silicon substrate, where the resist is fully developed.
	}
	\label{fig:exposureTestShipley}
\end{figure}

First, consider a resist stack composed only of imaging resist---MicroChem Shipley S1805. The data presented here is from an early test that highlights the role of intensity and focus.
The pattern creates an array of tall resist stripes of varying widths between 0.3 and 3 \um.
This pattern is the inverse of what is necessary for a {JJ}, but it provides a basic test for the resist-exposure combination.
The pattern is written many times with a matrix sweep of laser intensity and focus percentage. The matrix repeats across the wafer, and matrix copies develop for various times.

Figure \ref{fig:exposureTestShipley} displays test results for three variables: intensity, development time, and minimum line width.
Degeneracy appears even in this relatively simple sweep. A high-intensity exposure developed for a short time (Fig.~\ref{fig:exposureTestShipley}a, 80\% intensity) produces similar results to a low-intensity exposure developed for longer (Fig.~\ref{fig:exposureTestShipley}b, 50\% intensity). 
This particular test showed no significant difference between adjusting the focus percentage from -50\% to -30\%, despite being quite different from the optimal (-17\% focus at the time).
Design principles such as working far from extreme values can help break the degeneracy and determine working parameters.

This test does not probe consistency, despite the importance to process reliability. Though many of the parameter vectors work here, this test does not identify if they will work consistently across a wafer or consistently throughout time. Such one-off tests should be repeated many times to provide yield information.

\subsubsection{Full Resist Stack}

The resist tests performed in the above section optimized the exposure and development procedures for a single layer of resist.
They set useful baselines for moving to the full double-resist stack.

	Our full resist stack includes two resist layers. The top imaging layer is S1805, and the bottom liftoff resist is MicroChem LOR 10B.
	Liftoff resists are more sensitive to exposure and thus overdevelop beyond the exposure region.
	This creates an undercut that can be used as a mask for angle-selective deposition of metals.
	We designed the resist stack to support Dolan bridge shadow mask evaporation \cite{Dolan1977}.
	Other mask techniques include the overlap and the Manhattan methods \cite{Potts2001,Costache2012,Zhang2017}, but the Dolan bridge technique allows us to utilize a multi-step oxidation scheme, as detailed in Section \ref{sec:oxyProcess}.
	
	 Our spin and softbake recipes target 1 \um~height in the liftoff resist and 0.6 \um~in the imaging resist (see Section \ref{sec:fullRecipe}).
	Our recipes are tailored for 1.5 \um~undercuts, but the liftoff photoresist supports undercuts up to 10 \um. 

\begin{figure}[H]
	\centering
	\includegraphics[width=0.7\linewidth]{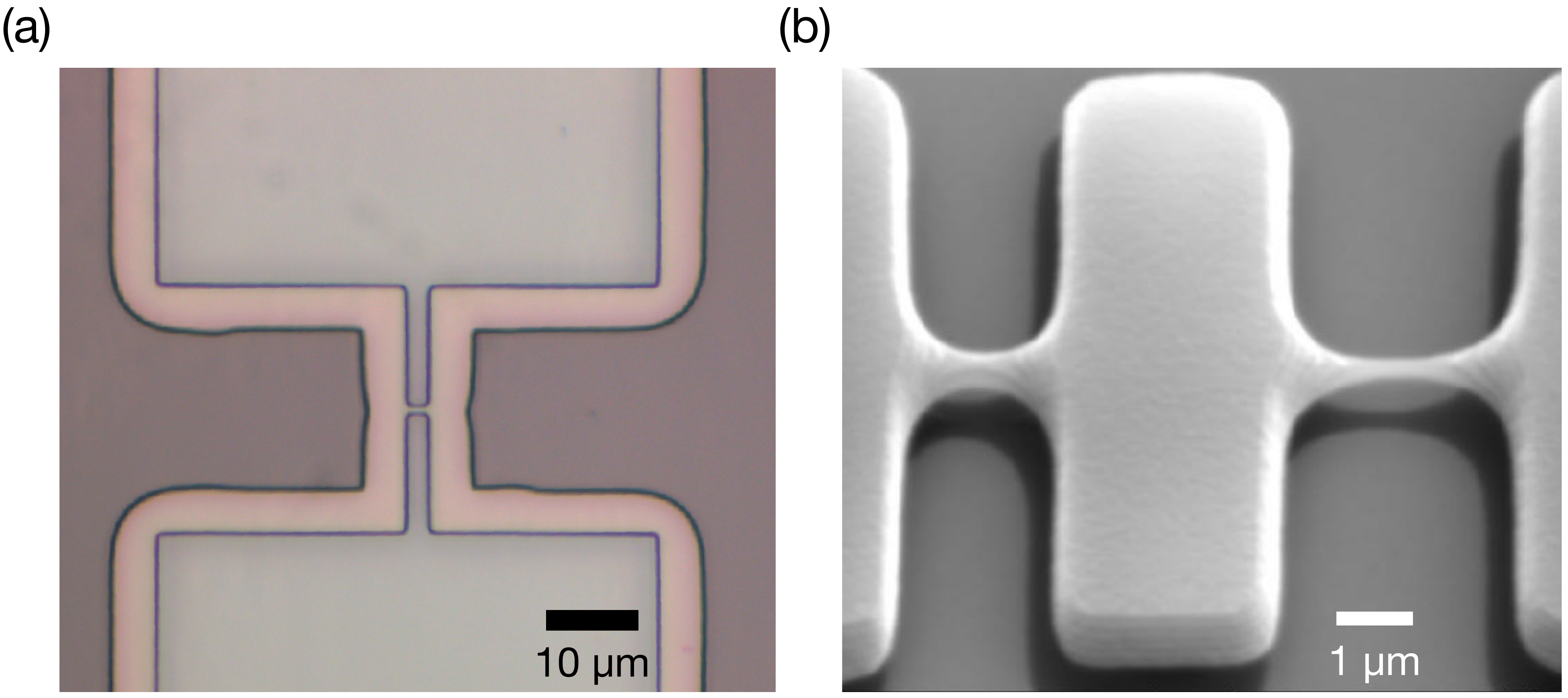}
	\caption[Resist Stack Imaging]{
		The developed resist stack after exposure and development.
		(a) An optical image of the resist after exposing a single {JJ}.
		The color indicates the number of resist layers. White corresponds to bare substrate, indicating both resist layers have cleared.
		Dark pink indicates both resist layers persist in unexposed regions. 
		Light pink indicates undercut regions where the bottom resist (LOR10B) has dissociated, but the top imaging resist (S1805) persists.
		Of central importance is the undercut region in the middle which creates the Dolan bridge.
		(b) A tilt-view SEM image of the resist stack (after depositing gold to prevent charge buildup and improve contrast).
		Shadows indicate a lack of gold due to resist masking.
	}
	\label{fig:resistImaging}
\end{figure}

	In Figure \ref{fig:resistImaging}, we show optical and scanning electron microscope (SEM) images of the resist stack after development.
	The feature sizes are large due to the wavelength limitation of photolithography.
	The SEM image  in Figure \ref{fig:resistImaging}(b) provides a perspective on the resist shadow-masking deposited gold.

%

%
%


%
%
	
	\subsubsection{Factorial Design of Process Variables}
	\label{sec:sandwich}
	With all the variables at work in the fabrication pipeline, understanding how each component behaves in isolation can tell a very different story when all variations can occur together.
	Such interacting processes require Design of Experiment principles, as introduced in Section \ref{sec:DOE}, we create a test that varies five critical variables in a factorial design: number of {JJ}s, exposure intensity, radial position, evaporation angle, and number oxidation layers.
	

	\begin{figure}[h!]
		\centering
		\includegraphics[width=0.7\linewidth]{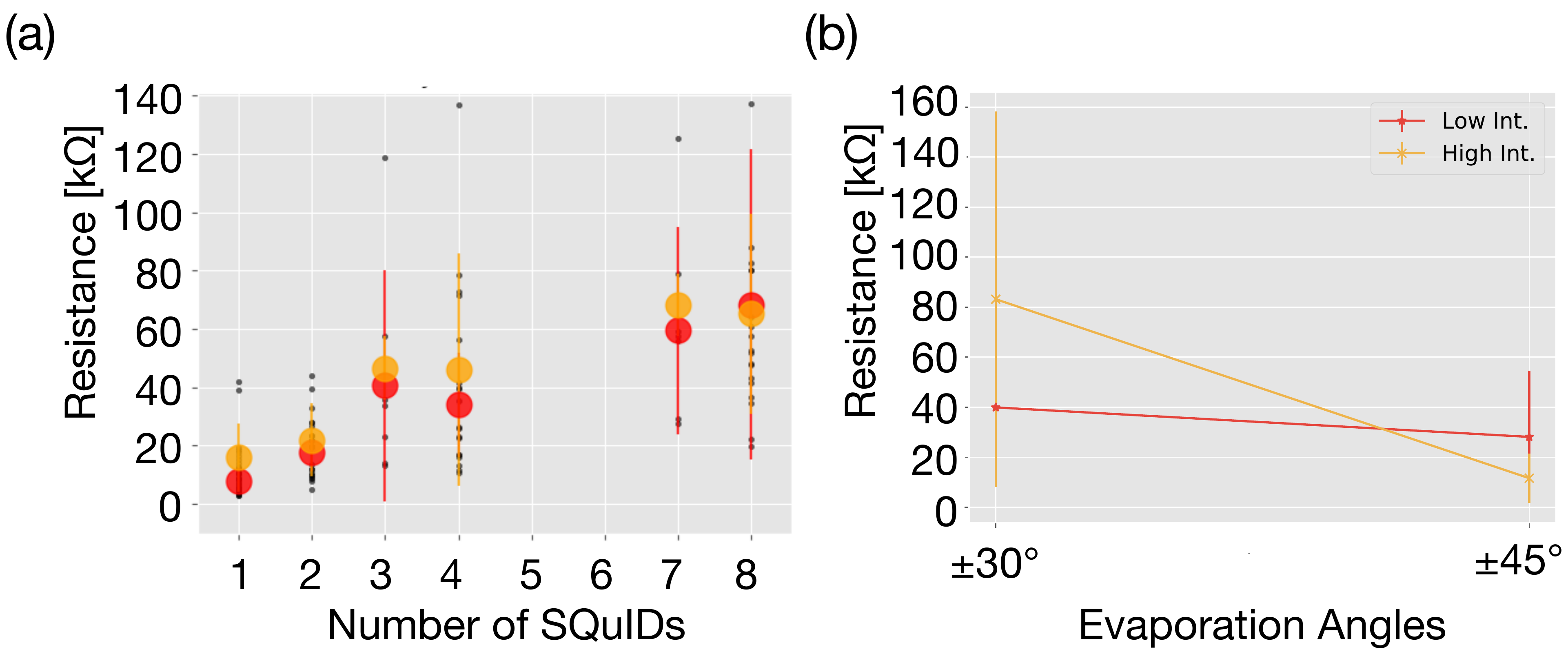}
		\caption[Codename Sandwich Factorial Design]{A selection of results for a factorial design of {JJ} fabrication parameters. 
		(a) The resistance scales with the number of {JJ} (SQuIDs) in series, as expected. For each number of SQuIDs, the oxidation and evaporation treatments are fixed. Black points indicate different positions. The average and standard deviation over the set of positions yields the colored circles and error bars.
		Colors indicate the development treatment: red corresponds to low-intensity exposure with long development time and orange corresponds to high-intensity exposure with short development time. Junction-number scaling is independent of development treatment.
		(b) The resistance for sets of single-SQuID devices averaged over the position and number of oxide layers. 
		The line is a guide to the eye showing the decrease with the area.
		The decrease is significantly stronger in the high-intensity development.
		Also, the high intensity development shows significantly larger variation.
	}
		\label{fig:codenameSandwich}
	\end{figure}
	
	The design includes a pattern that is repeated across the wafer and subjected to different treatments.
	A single pattern contains chains of up to 8 SQuIDs connected in series.
	This pattern repeats across the wafer. Quarters of the wafer receive different exposure treatments: either high-power exposure (54 mW effective power) with short development time (35 seconds) or low-power exposure (33 mW effective power) with long development time (60 seconds), similar to the treatment in Section \ref{sec:ExposureRecipe} (especially Fig.~\ref{fig:exposureTestShipley}).
	Horizontal dices across each quarter separate stripes of repeated patterns for different evaporation and oxidation treatments. The radial position varies along each stripe.
	The evaporation treatment entails evaporation at $30^\circ$ or $45^\circ$ relative to the wafer's normal vector.
	The oxidation treatment is either one or two layers of oxidation, detailed in Section \ref{sec:oxyProcess}.
	This test runs the gamut of tunable parameters within our standard process, although we have severely limited the range in order to retain tractability.
	
	The figure of merit for these devices is the room temperature resistance. Multiple repetitions of single-{JJ} chains provide statistics on resistance deviation. In the following results, we take a variety of slices through the parameter space and examine the impact on resistance and resistance deviation.
	
	First, consider how the resistance increases with the number of {JJ}s in Figure \ref{fig:codenameSandwich}(a).
	We focus discussion on the case of a single oxidation step with one evaporation angle while averaging over the radial position.
	The resistance trends nearly linearly with the number of SQuIDs. Deviation from the linear trend is largely due to the finite resistivity of the silicon substrate (15 ${\rm k\Omega\,cm}$).
	The trend is independent of the exposure treatment.
	
	This independence does not hold for the evaporation angle treatment, shown in Figure \ref{fig:codenameSandwich}(b). 
	For either exposure, increasing the evaporation angle increases the area of the junction, lowering the resistance.
	However, there is significant interaction between the evaporation angle and the treatment. 
	With low-intensity exposure and long development [red points in Fig.~\ref{fig:codenameSandwich}(b)], the decrease in resistance with angle is significantly less than in the case of high-intensity exposure.
	Longer development time allows the exposed resist to fully and uniformly dissolve.
	In contrast, short development time requires aggressive removal of resist and can lead to significant deviation within a single treatment.	 
	
\begin{figure}[H]
		\centering
		\includegraphics[width=0.7\linewidth]{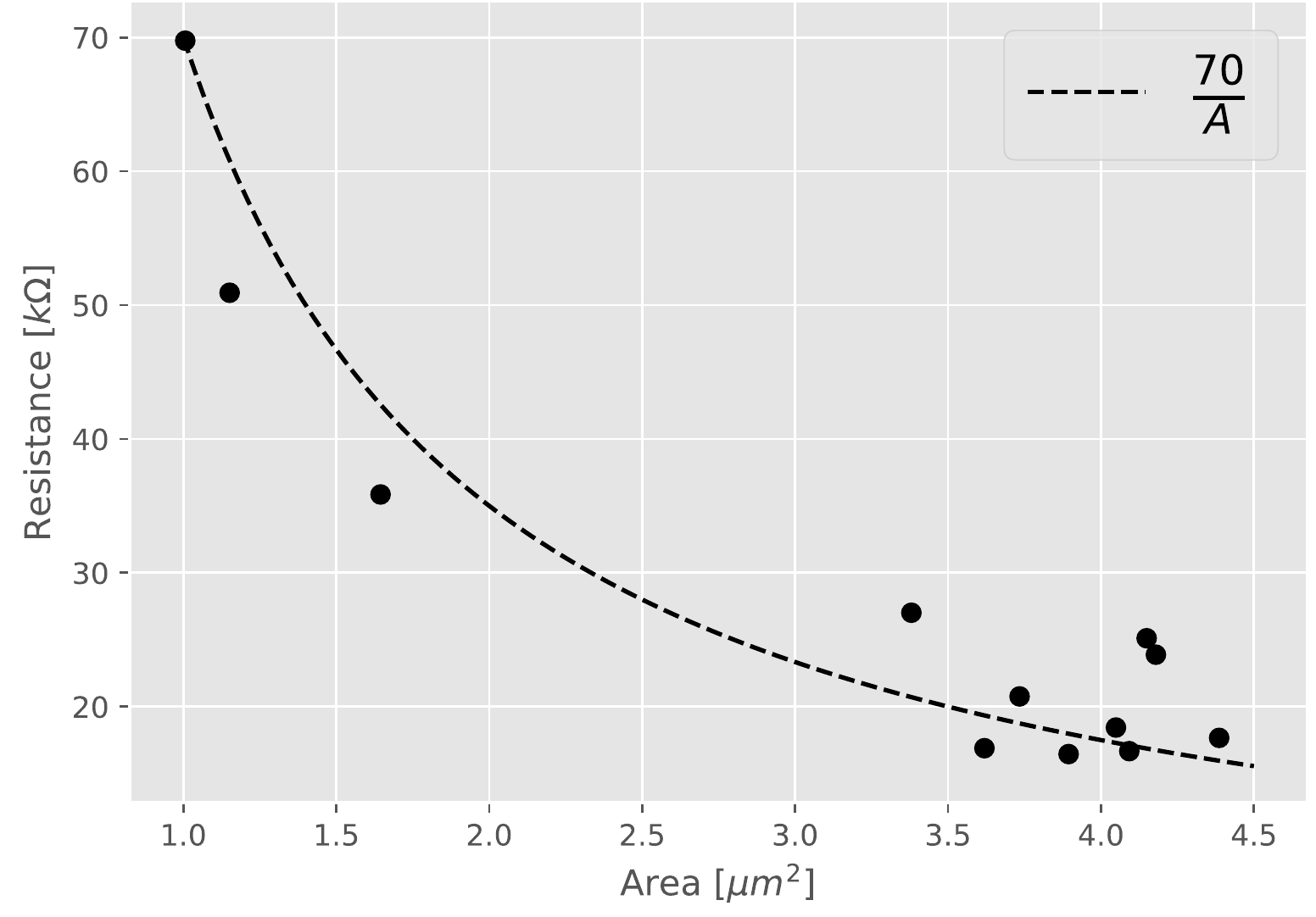}
		\caption[Resistance Variation from Area Variation]{Normal resistance measurements for a batch of junctions made under nominally identical conditions. The area, $A$, is based on SEM imaging [such as in Fig.~\ref{fig:SEMCompare}(a)]. The dashed line is a guide to the eye indicating the $1/A$ dependence of resistance.}
		\label{fig:sandwichArea}
	\end{figure}

	Throughout this design, we observed significant intra-treatment variation.
	Thus, several of the variables motivate specific studies as a single variable.
	For example, we found that a significant amount of variation resulted from position treatments.
	We corroborated the resistance measurements with SEM imaging [see Fig.~\ref{fig:SEMCompare}(a) for an example] for one wafer chip with fixed oxidation, evaporation angle, and development.
	Figure \ref{fig:sandwichArea} shows that this single treatment contained wide variation in the junction area, leading to predictable variation in the normal resistance.
	The variation in the area is discussed in detail in Section \ref{sec:JJConsistency}.
	In addition, Section \ref{sec:layerScaling} describes resistance scaling with the number of oxide layers.
	
	
\subsection{Junction Consistency Tests}
\label{sec:JJConsistency}
During recipe development, we identified significant variation among devices fabricated under identical processes.
The following tests illustrate a path towards treating the source of that variation.

\subsubsection{Manhattan Uniformity Tests}
\label{sec:Manhattan}
To quantify our process uniformity, we repeat a series of {JJ}s across a wafer.
The test includes variation on two distance scales. First, 16 junctions are located in a 50 \um~strip. Second, groups of junctions are separated throughout the wafer, each placed 6 mm apart.
The two scales provide different information about the process variability. Group-scale variation captures intrinsic variation, while wafer-scale variation captures extrinsic variation. These two types of variation suggest different treatment paths, as we describe in Sections \ref{sec:manhattanSummer} and \ref{sec:UniformityFuture}.

In this test we utilized Manhattan-style junctions \cite{Potts2001,Costache2012,Zhang2017} rather than Dolan-bridge-style junctions.
The Manhattan approach replaces the Dolan bridge with two orthogonal trenches in the resist. Deposition at more extreme angles (increased from $45^\circ$ to $70^\circ$) places metal on the electrode of the parallel trench but onto the sidewall of the orthogonal trench.
Manhattan-style junctions allow for better consistency because the junction area is not sensitive to the resist height \cite{Costache2012}.
However, the Manhattan process prohibits our two-junction oxidation method for large area {JJ}s (see Section \ref{sec:oxyProcess}). 

\begin{figure}[H]
	\centering
	\includegraphics[width=0.7\linewidth]{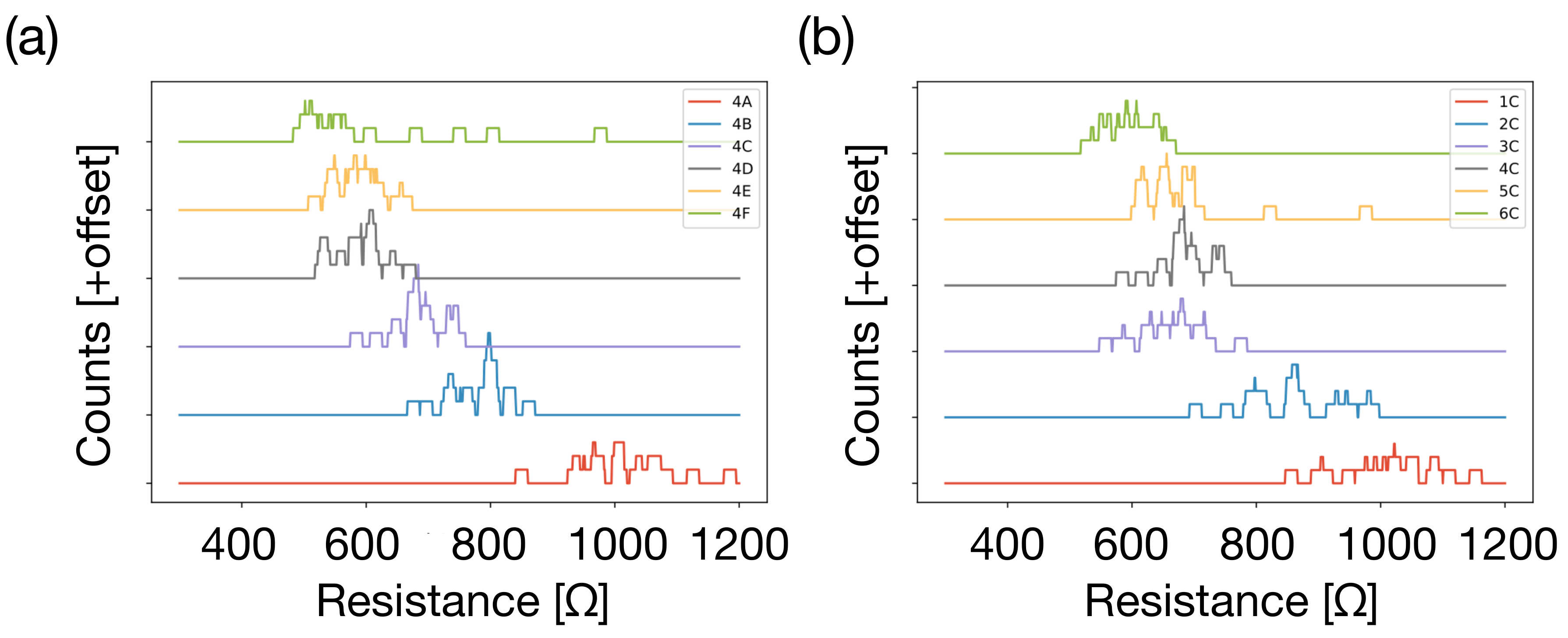}
	\caption[Variation in Manhattan Evaporation]{Histograms of resistance of identically fabricated {JJ}s using the Manhattan process. Each trace corresponds to a group of {JJ}s within a 50 \um~ strip. The traces are offset according to position either (a) vertically or (b) horizontally across a wafer. The variation across the wafer is larger than the variation within a single group.}
	\label{fig:manhattanVariation}
\end{figure}

The resistance results are shown in Figure \ref{fig:manhattanVariation}.
This test exhibits clear spatial dependence.
Moving across the wafer either horizontally [Fig.~\ref{fig:manhattanVariation}(a)] or vertically [Fig.~\ref{fig:manhattanVariation}(a)], increases the mean resistance by nearly a factor of two.
While the within-group variation is only 8\%, the between-groups variation is 20\%, after removing outliers such as open circuits. 

\subsubsection{Wafer Heating for Uniformity}
\label{sec:manhattanSummer}
Early in the recipe development, resist spinning showed clear deficiencies. Color changes of the resist on the wafer, due to interference of optical wavelengths, indicated variation in the resist thickness on the order of hundreds of nanometers.
Despite clean spins in the previous test, we continued to see large-scale variation.
Our testing led to suspicion of hotplate temperature uniformity.
An IR thermometer indicated regions $\pm 10^\circ$ C outside of the nominal temperature.
Although IR thermometer measurements are inaccurate due to the low emissivity of the ceramic platform, we take the measured range as an estimate of the variation's order of magnitude.
According to LOR10B's manufacturer specifications, this magnitude of temperature variation changes the dissolution rate by 45\%.
Thus, the local temperature variation can create significant differences in the resist patterns.

To test this variable's role, we added a copper sheet onto the hotplate while softbaking the resist. While the high thermal conductivity changes the heat capacity of the heating element, it improves the temperature uniformity.

\begin{figure}[H]
	\centering
	\includegraphics[width=0.7\linewidth]{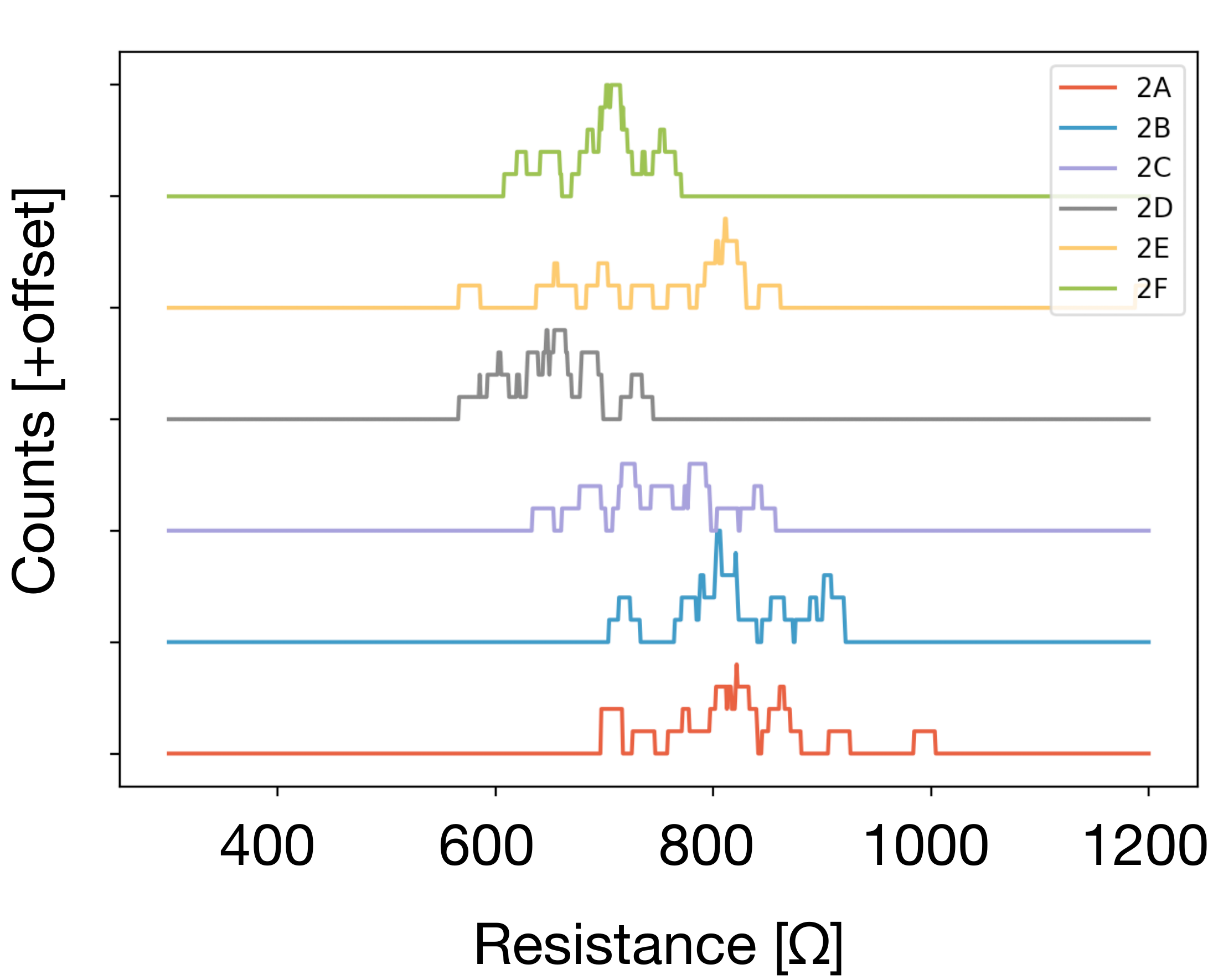}
	\caption[Wafer Variation with Improved Softbaking]{Similar to Figure \ref{fig:manhattanVariation}, each trace is a histogram of resistance within local groups of {JJ}s. Here, the wafer was baked on a copper sheet rather than directly on the hotplate's ceramic surface. The variability between groups is significantly improved.}
	\label{fig:manhattanSummer}
\end{figure}

The test results are shown in Figure \ref{fig:manhattanSummer}. We used the same pattern and process as in Section \ref{sec:Manhattan}. 
However, compared to the previous Manhattan-process test, the copper-plate softbake does not show a statistically significant shift across the wafer. The average within-group relative standard deviation (the standard deviation divided by the mean) was 7.5\%, similar to the bare-plate bake. However, the between-groups variation decreased from 20\% to 14\%.
Spatial variation is significantly mitigated using this softbaking approach.

\subsubsection{Further Directions for Uniformity}
\label{sec:UniformityFuture}
Monumental work towards the development of device consistency has identified several additional avenues \cite{Kreikebaum2019}.
Ultrasonic development, rather than by-hand mechanical agitation, can sufficiently displace dissolved resist while removing human error.
Other groups using a similar oxygen plasma asher device  have found non-radially-symmetric behavior during ashing \cite{Kreikebaum2019}.
Dynamic oxidation, in which oxygen is pumped in and out of the chamber also shows improved uniformity.
Cold development---as is common in e-beam lithography \cite{Naghiloo2019}---lowers the resist dissolution rate, allowing uniform pattern sizes.
Finally, our evaporation rate, 0.3 nm/s, is slow enough to allow large grain formation, improving uniformity and eliminating a site for lossy dielectrics \cite{Premkumar2020}.

\subsection{Full Recipe}
\label{sec:fullRecipe}
The recipe development described above culminates in the following detailed list of the full qubit fabrication process.

\begin{enumerate}
	\item {Silicon substrate}\\
		For all devices, we use single-side-polished 2-inch silicon substrates with 100 crystal orientation and high resistivity (${\rm >15 k\Omega\cdot cm}$). 
	\item {Substrate Cleaning}\\
		Clean the blank silicon substrate in a buffered oxide etch (BOE) consisting of ammonium fluoride, ammonium bifluoride, and hydrofluoric acid for 5 minutes at room temperature.
		A two-step DI water rinse removes and neutralizes residual HF.
	\item{Spin resist} \\
		Attach the wafer to the vacuum chuck of the Brewer Science CEE 200X spin coater.
		Dispense 3 mL of LOR10B, carefully maintaining a continuous stream to prevent bubbles.
		Resist dispensation can be a major source of device inhomogeneity.
		Spin at 3000 RPM for 50 seconds to achieve a target thickness of 1.0 \um.
		Soft bake the wafer at $195^\circ$C for 10 minutes. Lower temperatures can produce larger undercuts.
		Dispense 1 mL of Shipley S1805. 
		Spin at 2000 RPM for 30 seconds to achieve a target thickness of 0.6 \um.
		Soft bake the wafer at $115^\circ$C for 1 minute.
	\item{Expose transmon pattern}\\
		Load the wafer into the Heidelberg DWL 66+. Center the device, and focus the head with pneumatic mode\footnote{Pneumatic mode provides a height accuracy of $\pm100$ \um.}.
		Exposure parameters should be calibrated monthly for the particular write lens head.
		For the 10 mm head\footnote{10 mm refers to the focal length. This lens provides a 1.10 \um~ depth of field.}, 
		recent calibration required 67 mW power, 90\% intensity, 25\% filter and 0\% focus.
		The dose time is set with the beam's raster rate, 30 kHz.
	\item{Development} \\
		Mechanically agitate the wafer in Microposit Metal-ion Free (MF)-319 developer for 45 seconds at room temperature.
		Mechanically agitate the wafer in DI water for 30 seconds at room temperature.
		Blow dry with ${\rm N_2}$. 
		The sample may be visually inspected under an optical microscope to check nominal junction dimensions.
		Dolan bridges should be 1.5 \um~ long and 1 \um~ wide. Resist color (indicating thickness) should reflect cleared LOR-10B and intact S1805, as in Figure \ref{fig:resistImaging}(a).
	\item{Ashing} \\
		Use the Plasma Etch PE-50 to ash organics (residual resist) in 100 W oxygen plasma for 20 seconds.
	\item{Substrate Clean} \\
		Clean the wafer again in BOE for 30 seconds at room temperature.
		HF does not etch our photoresists \cite{Williams2003}.
	\item{Pump Down} \\
		Quickly mount the wafer and begin pumping down after the BOE clean.
		Apply two Ti getter steps, with about 5 minutes delay. The base pressure should be $<10^{-9}$ Torr.
		Allow at least 18 hours in this ultra-high vacuum (UHV) environment. We believe this removes any residual surface contaminants.
		Our e-beam evaporator is an AJA ATC-Orion-8E with an 8.5 kV electron source.
	\item{Evaporation of Bottom Electrode}\\
		Align the wafer's in-plane angle to align the wafer's flat vertically, orthogonal to the normal-angle rotation axis. 
		Position the normal angle to $+45^\circ$.
		With the shutter blocking the sample, turn on the automated ramp to increase the e-beam source filament from 0 mA to 180 mA over two minutes.
		When the targeted current is reached, open the shutter to begin deposition.
		A crystal oscillator monitor tracks the deposited film thickness, and a PID controller tunes the filament current to match the targeted rate, 3 \AA/s.
		When the target thickness for the bottom electrode, 30 nm, is reached, the shutter blocks the sample, and the current ramps down in a few seconds.
		\label{Evapstep}
	\item{Oxidation} \\
		Remove the sample from the UHV chamber into the load lock. Flood the load lock with 99.99\% pure ${\rm O_2}$ at 4.3 Torr. Use a dose time calibrated to the targeted critical current. Large-area JJs in our transmons require 1800 seconds (30 minutes). Josephson parametric amplifiers (${\rm I_c\sim1 \mu A}$) require $\sim 600$ seconds.
		Our transmon junctions also utilize a multi-step oxidation process. See Section \ref{sec:oxyProcess}.
	\item{Evaporation of Top Electrode}\\		
		Repeat Step \ref{Evapstep}, but with the normal angle set to $-45^\circ$. The target thickness is 60 nm.
	\item{Protective Oxide} \\
		After removing the sample to the load lock, another 4.3 Torr oxidation for 60 seconds creates a clean protective oxide. We believe oxidation in a vacuum environment improves aging compared to atmospheric oxidation.
	\item{Liftoff}\\
		Place samples in N-Methyl-2-pyrrolidone (NMP), heated to $60 ^\circ$C for at least one hour.
		Use a pipette filled with NMP to squirt sacrificial aluminum away.
		Other groups sonicate during liftoff.
		Spray-clean the sample with isopropyl alcohol.
		Blow dry with $\mathrm{N_2}$.
	\item{Dice}\\
		Use a diamond scribe to score the wafer parallel with the crystallographic axis.
		Cleave the wafer at the score.
	\item{Mount Device} \\
		For transmons, adhere the silicon to the cavity shelf with GE varnish.
		For other planar devices, adhere the sample to the enclosure and wire bond the connections.

\end{enumerate}

\subsubsection{Multi-step Oxidation for Large-area JJs}
\label{sec:oxyProcess}
To attain the required critical current for transmon qubits ($\sim10$ nA), a new oxidation scheme is required.
Cabrera-Mott Theory (Section \ref{sec:cabreraThy}) predicts a saturation thickness which results in $\sim$1.5 nm thick oxides for our temperature and pressure conditions \cite{Kang2014,Zeng2015}. With a $1.5~ \um^2$ area, this creates a {JJ} with $\sim\mathrm{1\mu A}$ of critical current \cite{Simmons1963}.
While suitable for parametric amplifiers \cite{Vijay2009, Slichter2011,Macklin2015}, the inductance of such large junctions is unsuited for transmon qubits.
Instead, the alternative oxidation scheme laid out below creates two junctions. One of which has the requisite 10 nA critical current.

Our {JJ}s consist of a 30-nm-thick aluminum bottom electrode, an aluminum oxide insulating barrier, and a 60-nm-thick aluminum top electrode.
After depositing the bottom electrode, we create the oxide tunnel layer by exposing the electrode to oxygen at 4.3 Torr for 300 seconds. 
Standard evaporation procedures would proceed with the top electrode \cite{Naghiloo2019,Harrington2020}.
Instead, we add a 0.5 nm filler layer of aluminum. We then oxidized the filler under the same conditions. 
300 seconds ensures the full layer is oxidized \cite{Kang2014,Zeng2015,Gorobez2021}.
We add as many fully oxidized filler layers to achieve the desired critical current, typically three.

Two junctions form through this process. 
When the bottom electrode oxidizes, both its top face and side face develop oxide layers with the same thickness. 
The filler aluminum layer only covers the top electrode due the sharp resist [see Figure \ref{fig:resistImaging}(b)].
Thus, the top-face oxide layer grows by an additional 0.5 nm, owing to the fast initial growth phase of Equation [\ref{eqn:CabreraTime}]. However, the side-face oxide grows only a small amount, owing to the slow growth phase (see Figure \ref{fig:cabreraTheory}).

The top electrode encapsulates both top- and side-face junctions.
The top-face junction's thickness implies that its resistance and inductance are exponentially larger than the side-face junction, according to Equation [\ref{eqn:Simmons}].

\section{Devices}
\label{sec:devices}
In this section, we describe results for large-area transmons \cite{Monroe2021b}.
The results provide insights into the loss mechanisms for state-of-the-art  quantum computing devices.
In particular, we show JJ areas need not be small to have good coherence. Instead, surface cleaning is essential.

The results are broken down into four categories.
First, in Section \ref{sec:JJstructure}, we describe the physical structure of the junctions, including a cross-sectional view of junctions and tests of the oxidation method.
Then, Section \ref{sec:spectra} presents spectroscopy results, which verify the transmon nature of the circuit. In the process, we observed several of strongly-coupled TLS defects resulting from uncleaned surfaces. 
Next, Section \ref{sec:TimeDomain} presents coherence times, including for uncleaned devices.
Finally, Section \ref{sec:otherTests} provides a few checks against additional loss mechanisms and summarizes the results.

\subsection{Structure of Large-Area Junctions}
\label{sec:JJstructure}

\begin{figure}[H]
	\centering
	\includegraphics[width=0.7\linewidth]{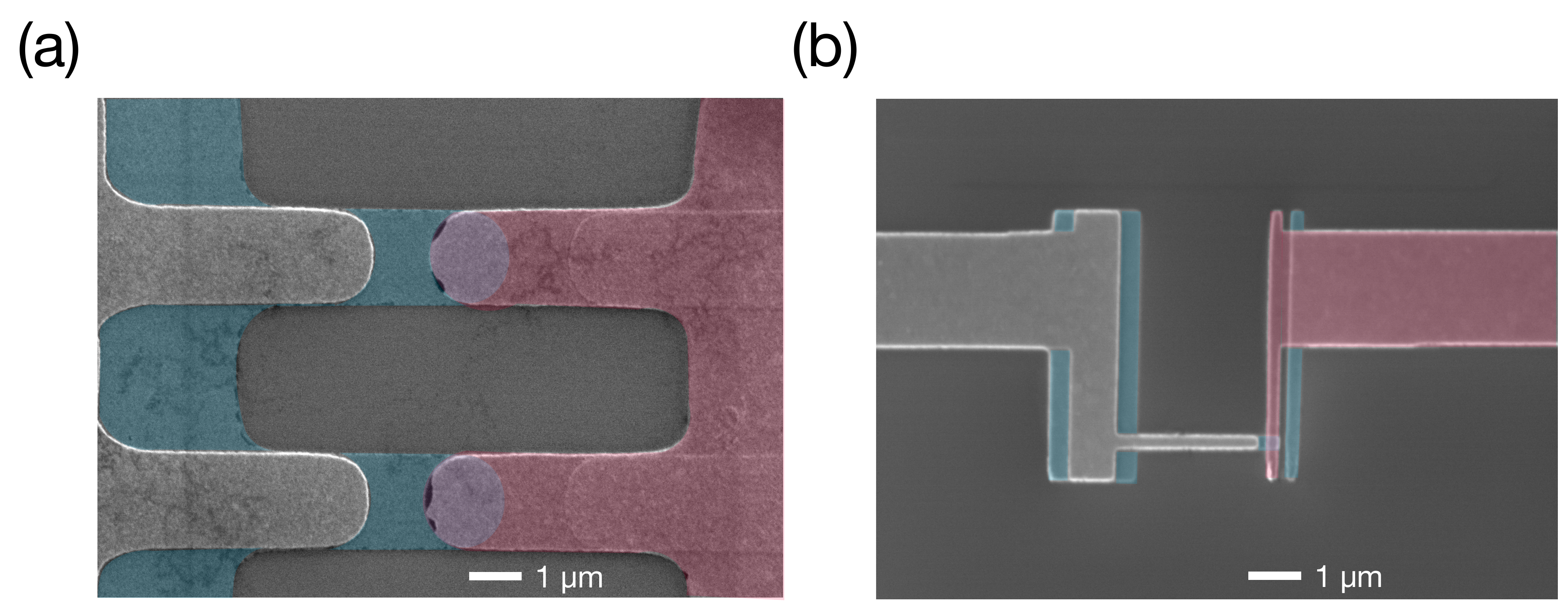}
	\caption[SEM image of a Large-area Junction]{SEM images of junctions made with (a) photolithography and (b) eBL lithography. Images are at the same magnification.
	In each image, the bottom electrode is false-colored blue, and the top electrode is false-colored red. The overlap between electrodes defines the junction.
	The area of each junction in (a) is $2.88~ \um^2$ whereas the junction's area in (b) is 0.08 $\um^2$. Our oxidation process achieves similar critical currents despite the significant disparity in area.}
	\label{fig:SEMCompare}
\end{figure}

In this section, we discuss the physical structure of large-area {JJ}s. Because their patterns are defined with photolithography, the feature sizes are large compared to electron-beam lithography (eBL), see Figure \ref{fig:SEMCompare}. 
The large junction area significantly shifts the oxide barrier requirements, namely thickness and cleanliness.

\subsubsection{TEM sample}
\label{sec:TEM}
\begin{figure}[H]
		\centering
		\includegraphics[width=0.7\linewidth]{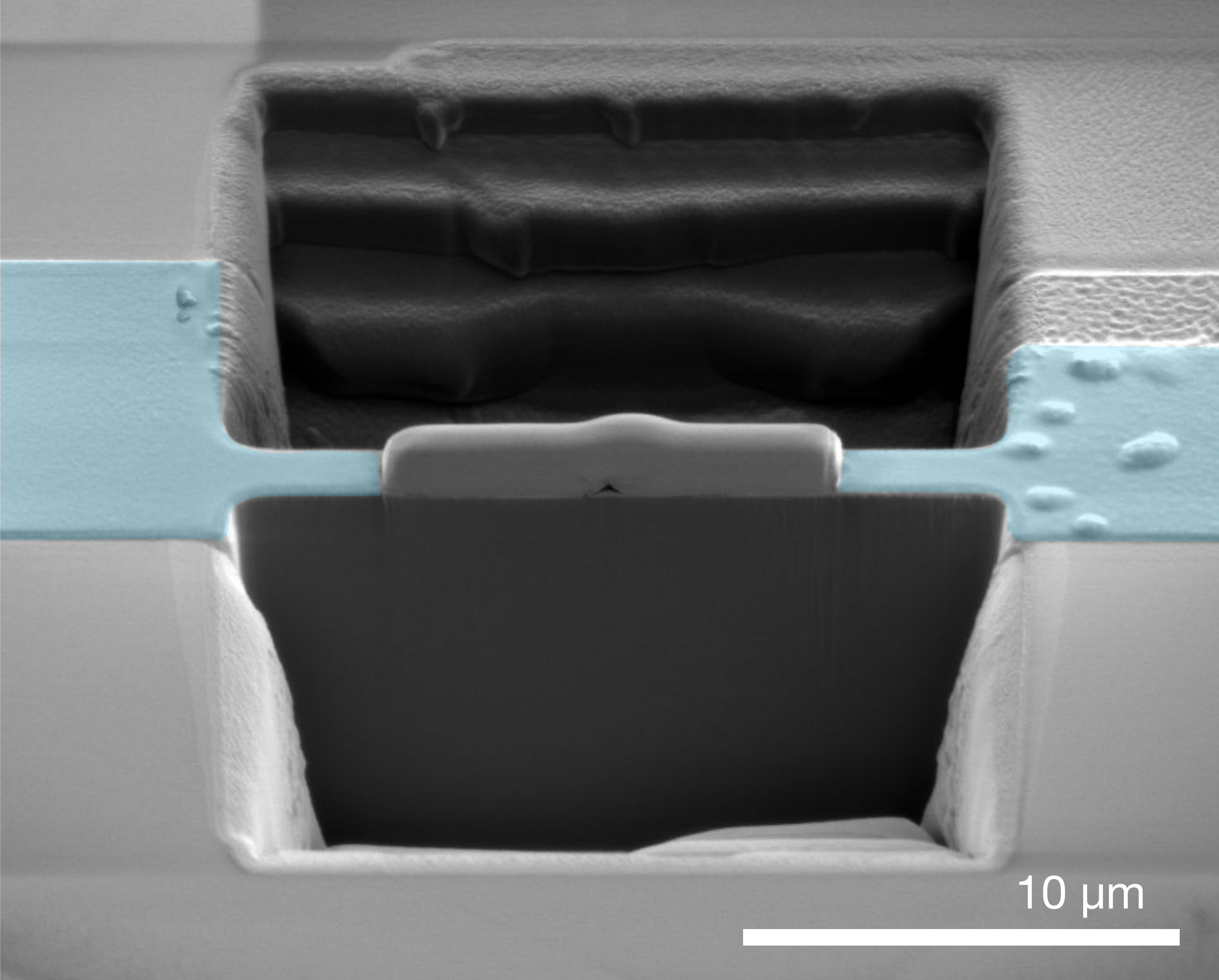}
		\caption[Preparing TEM Sample]{
			An SEM image of preparing a {JJ} for cross-sectional viewing. At this intermediate state, the FIB has cut away the majority of material and is preparing to separate the sample before thinning.
			The aluminum electrodes are highlighted in cyan. 
			The junction (covered in a protective platinum layer) is visible as the black chevron in the image center . (This junction appears to have formed as a cusp on top of resist, see Figure \ref{fig:resistImaging}b.) 
		}
		\label{fig:TEMprep}
	\end{figure}

	To confirm the structure of the large-area JJ, we image a {JJ} cross-section with a transmission electron microscope (TEM). A focused ion beam (FIB) creates the cross-section using gallium ions to selectively mill regions of the JJ, as seen in Figure \ref{fig:TEMprep}.
	The FIB first cuts away material surrounding the {JJ}. The FIB then thins the cut-away to $\sim 100$ nm to make the sample electron-transparent.
	
	With the sample prepared, we image the atomic configuration with a TEM.
	Electrons' small de Broglie wavelength allows TEM imaging to image with exquisite spatial resolution.
	Imaging capabilities can be combined with energy-dispersive x-ray spectroscopy (EDXS) to attain atomic-precision chemical abundance.
	
\begin{figure}[H]
		\centering
		\includegraphics[width=0.7\linewidth]{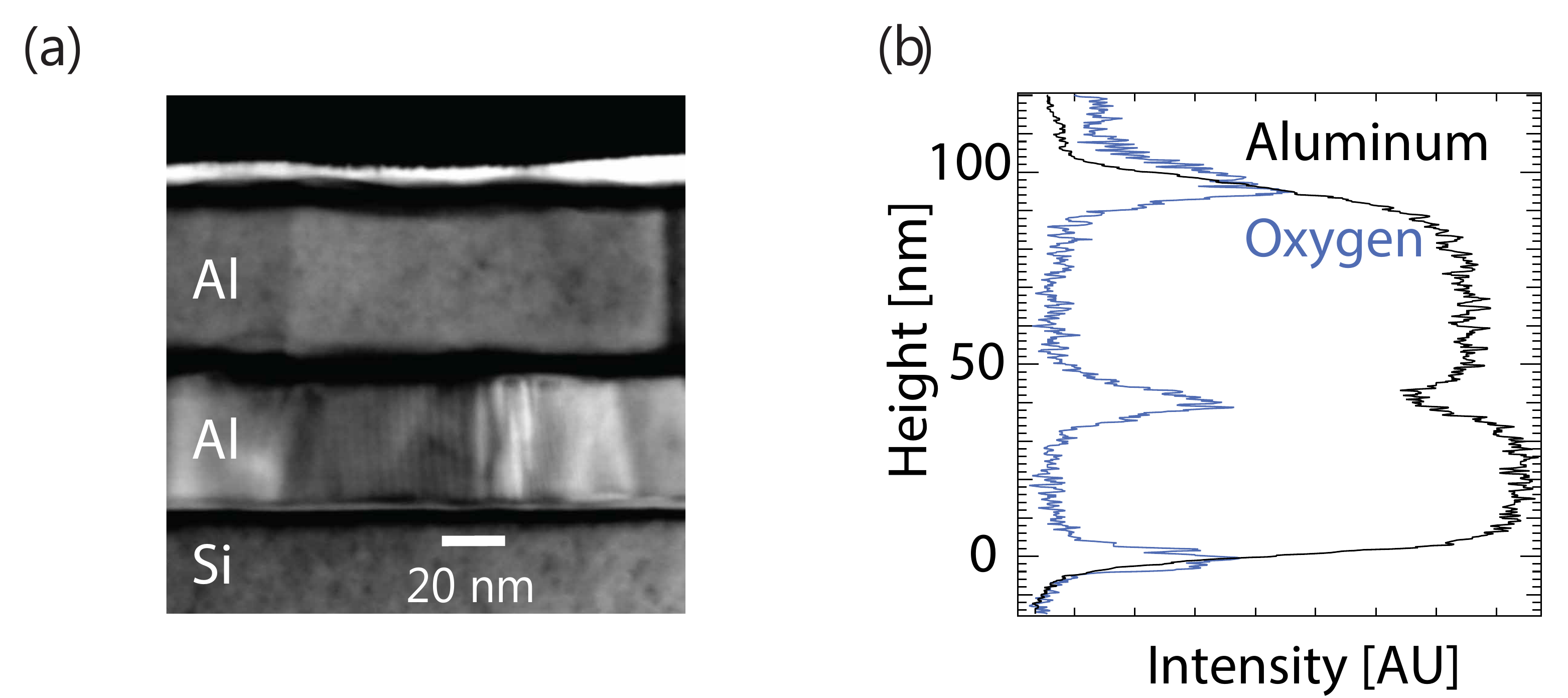}
		\caption[EDXS Data for Multi-layer Junction]{(a) Bright-field image from the EDXS measurement highlights the structural properties of the junction. Each aluminum layer and the silicon substrate are labeled, with respective oxides appearing as dark. Note the grain boundaries in the bottom aluminum layer.
		(b) EDXS chemical abundance averaged over the columns of panel (a). Blue indicates oxygen abundance, and black indicates aluminum abundance. The oxygen abundance in between the aluminum layer fits a Gaussian with a standard deviation of 4.5 nm.}
		\label{fig:TEMdata}
	\end{figure}

	In Figure \ref{fig:TEMdata}, we show EDXS data for a {JJ} made with two layers of oxidation, as described in Section \ref{sec:oxyProcess}. 
	The magnification of this image is sub-optimal, but we nonetheless examine the {JJ} structure.
	We estimate the thickness of the aluminum oxide barrier with a Gaussian fit to the averaged oxygen abundance [see Figure \ref{fig:TEMdata}(b)]. The fit returns an estimate of 4.5 nm, which is much larger than expected.
	However, we note that TEM imaging significantly overestimates the size of the effective barrier \cite{Kim2020a}.
	TEM images reflect the structural properties (atom locations) of the barrier, but they do not capture the electronic properties.
	 Quantitative X-ray photoelectron spectroscopy (XPS) would provide a better estimate of the true barrier thickness.
	This TEM imaging thus serves as an upper bound for the junction thickness.
		
	\subsubsection{Resistance Scaling: Evidence of Multiple Junctions}
	\label{sec:layerScaling}
	Section \ref{sec:Simmons} described the dependence of normal resistance on the thickness of the junction barrier.
	This section describes a series of measurements investigating the resistance scaling with the number of layers applied with our multi-step oxidation scheme (see Section \ref{sec:oxyProcess}).
	
\begin{figure}[H]
		\centering
		\includegraphics[width=0.7\linewidth]{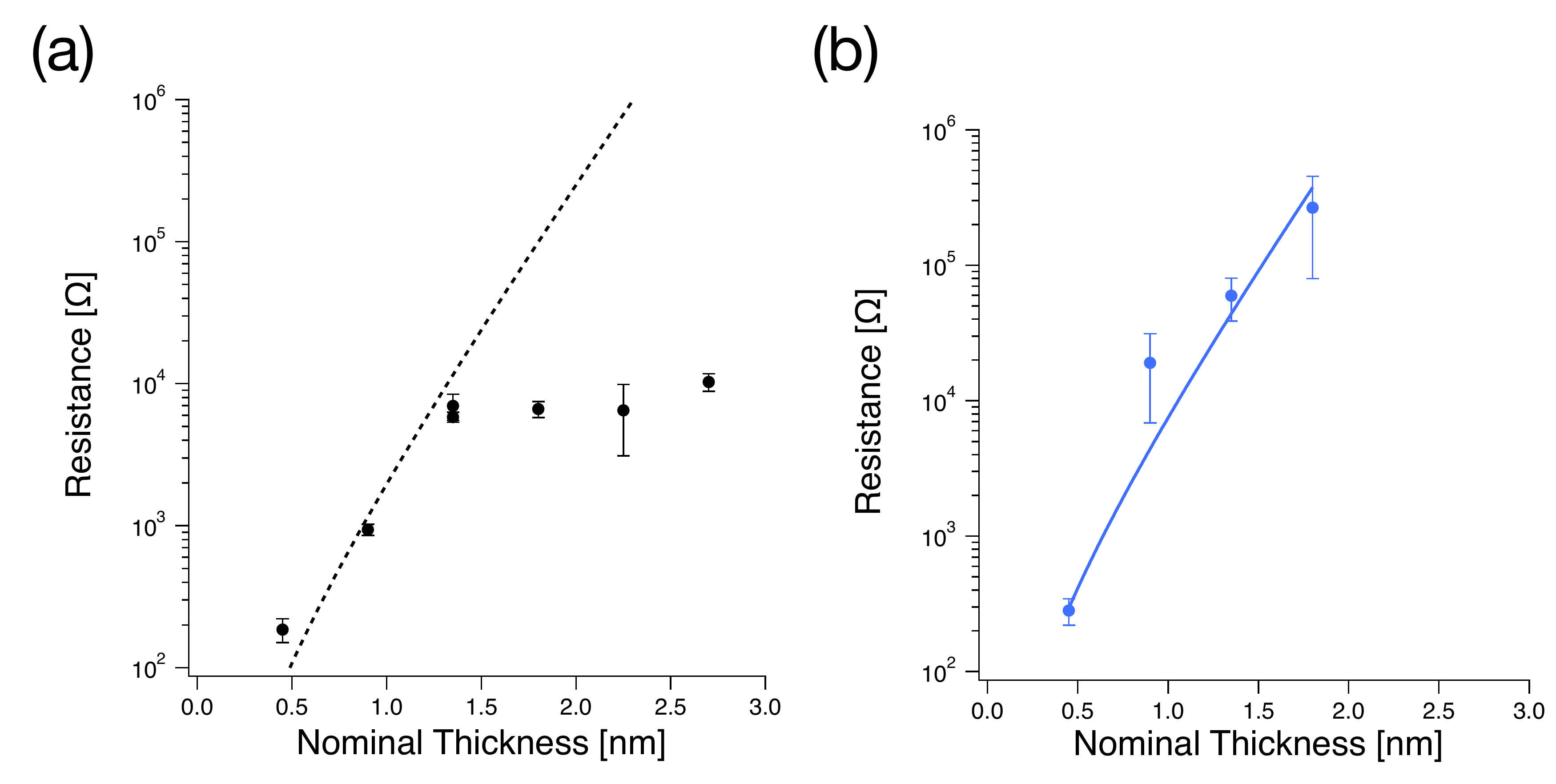}
		\caption[Resistance Scaling with Oxide Thickness]{The room-temperature normal resistance for junctions with two evaporation schemes. (a) Electrodes are evaporated at $\pm45^\circ$ with a number of oxide filler layers. As a function of the nominal oxide thickness (number of filler layers), the resistance does not scale exponentially, indicating the parallel resistance of a second junction. The dashed line is a guide to the eye based on the Simmons equation (Eqn.~[\ref{eqn:Simmons}]). (b) Electrodes are evaporated at $10^\circ$ and $60^\circ$, respectively, with a variable number of intermediate filler layers. The results can be fit to the Simmons equation, indicated by the solid blue line.
		Error bars in both figures correspond to multiple devices evaporated together.}
			\label{fig:layerScaling}
		\end{figure}

	Figure \ref{fig:layerScaling}(a) shows the resistance of {JJ}s with a variable number ($N_{\rm layer}$) of total oxide layers.
	Each layer after the first ($N_{\rm layer}\geq 2$) is created with a 0.5 nm filler layer aluminum, so we expect the film thickness to be $\left(1.5 + 0.5 N_{\rm layer}\right) \,\mathrm{[nm]}$, based on an initial oxide layer's thickness \cite{Kang2014,Zeng2015,Gorobez2021}.
	Despite the predicted exponential scaling, we see significantly sub-exponential scaling.
	Resistance in parallel with the junction can lower the measured resistance. We attribute the parallel resistance to the presence of two junctions.
	
	Consider the geometry of the evaporation.
	After the first deposition at the standard angle of $45^\circ$ relative to normal, a 30 nm electrode lies underneath the junction. The oxygen exposure oxidizes every face of the electrode.
	The next layer of filler aluminum only covers the top face of the electrode.
	The electrode's side face does not receive additional aluminum.
	The next oxidation contributes significantly to the top-face oxide's thickness because the filler aluminum oxidizes in the fast regime of Cabrera-Mott oxidation theory (see inset of Figure \ref{fig:cabreraTheory}).
	However, the side-face oxide does not grow significantly because the oxide has entered the slow regime (see Figure \ref{fig:cabreraTheory}).
	The process repeats for additional filler layers.
	When the second electrode evaporates at $-45^\circ$, two junctions are created. The oxide layer on the top face is thick, leading to high resistance. The resistance of this top-face junction' scales exponentially in $N_{\rm layer}$.
	The oxide layer on the side face is thin, leading to lower resistance which is nearly independent of $N_{\rm layer}$.
	The parallel is resistance is dominated by the lesser resistance of the side junction.
	
	To confirm this model, we modify the oxidation procedure.
	Rather than evaporating layers at $\pm 45^\circ$, we evaporate layers at $+10^\circ$ and the final layer at $+60^\circ$. Because the evaporations approach from the same side of the bridge (ie the angles have the same sign), the filler layers build the full junction.
	The scaling of resistance with $N_{\rm layer}$ is shown in Figure \ref{fig:layerScaling}(b). The exponential scaling indicates a single, thick junction has been created.
	We note that the finite resistivity of the silicon substrate slightly decreases the measured resistance.
	

\subsection{Circuit Spectroscopy}
\label{sec:spectra}
These devices are intended for use in cQED experiments. Thus, we embed large-area {JJ}s in a standard 3D cQED circuit. The circuit contains a SQuID (two JJs in a loop) and a shunt capacitor. The shunt capacitor should be large compared to the junction's capacitance so that the charging energy, $E_C$, is low enough to
suppress charge noise (via low charge dispersion) \cite{Koch2007}.
The criteria for $E_C$ and $E_J$ is the transmon approximation: $E_J/E_C \gg 1$ \cite{Koch2007}.

$E_J$ depends on the {JJ} critical current.
Were we to follow the conventional transmon fabrication recipe for large-area {JJ}s, the critical current would be too high, according to Equation [\ref{eqn:ABRelation}]. 
But JJs fabricated with the procedure outlined in Section \ref{sec:oxyProcess} have a transmon-suitable critical current ($\sim 10$ nA).

\begin{figure}[H]
	\centering
	\includegraphics[width=0.7\linewidth]{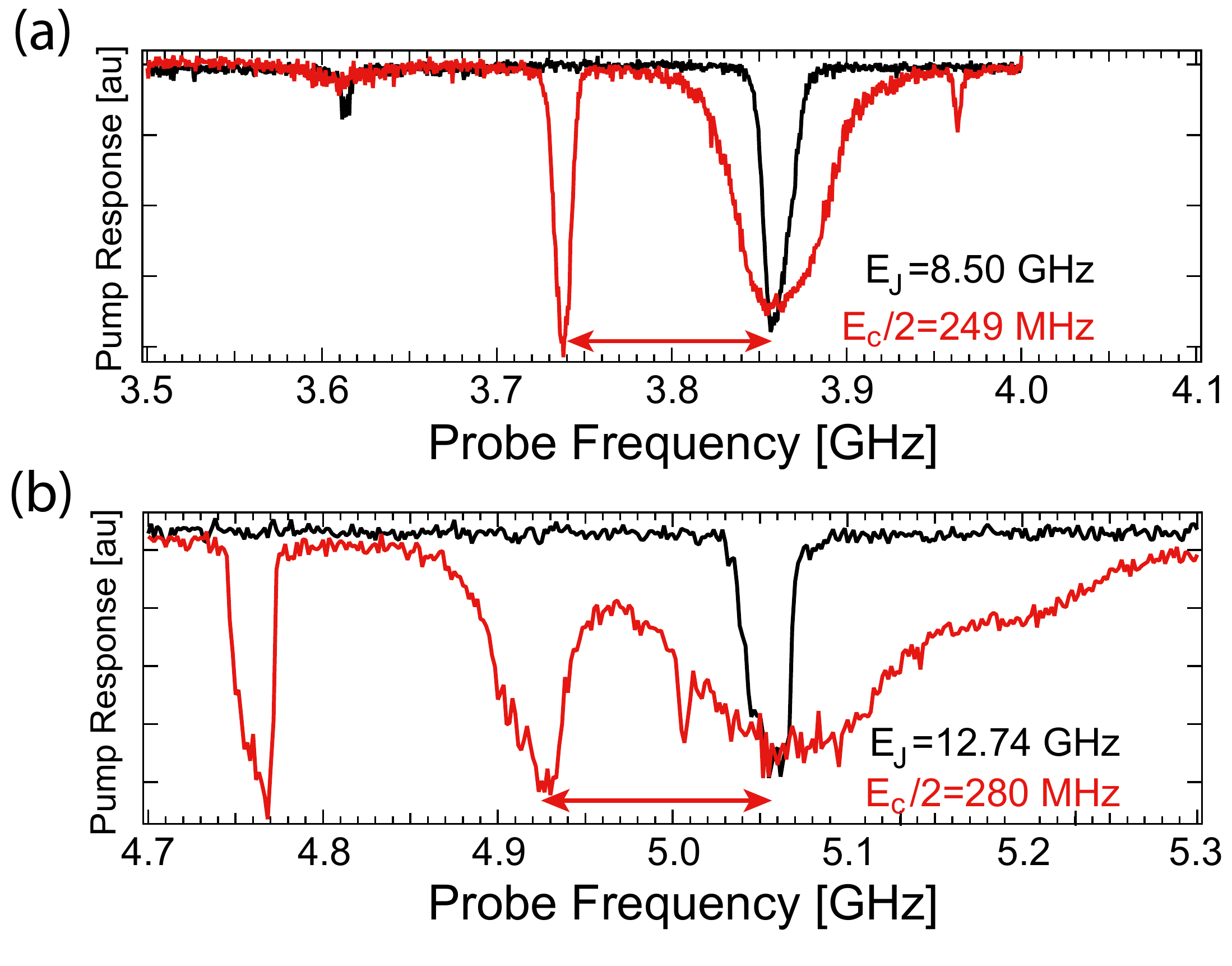}
	\caption[High Power Spectroscopy]{In both panels, two-tone spectroscopy probes the energy structure of the circuit. Lower probe power (black) only excites the first transition, $f_{01}=\sqrt{8E_J E_C}-E_c$. Higher probe powers (red) excite higher transitions separated by the circuit anharmonicity, $\alpha=-E_C$. (High power also induces significant power broadening). Arrows indicate $\alpha/2$, the frequency spacing of the two highest-frequency peaks. The extracted circuit energies, $E_J$ and $E_C$ are indicated for each device. Panel (a) corresponds to a device made with multiple oxide steps, and panel (b) corresponds to a device with a single oxidation step. Note the similar charging energies.}
	\label{fig:highPowerSpec}
\end{figure}

To verify the transmon approximation, we measure $E_J$ and $E_C$. $E_C$ can be found by measuring the circuit's  anharmonicity. Moderate-power two-tone spectroscopy excites multi-photon transitions in the circuit. 
The highest frequency corresponds to the $\ket{0} \leftrightarrow \ket{1}$ transition. High probe power excites the next two transitions: $\ket{1} \leftrightarrow \ket{2}$ and $\ket{0} \leftrightarrow \ket{2}$, which are partially populated due to thermal excitation.
Spectra like the one shown in Figure \ref{fig:highPowerSpec} exhibit dips at frequencies corresponding to each energy-level difference\footnote{The factor of 2 is the number of photons required for the transition.}: $f_{01}$, $f_{12}$, and $f_{02}/2$.
The difference $\alpha \equiv f_{12}-f_{01}$ is the anharmonicity, and $E_C \approx -\alpha$ \cite{Koch2007}.
The transmon energy levels can be solved for $E_J$ as a function of $E_C$ and $f_{01}$.

Surprisingly, large-area JJs do not exhibit charging energies significantly different from their small-area counterparts, despite a 100 times enlargement of the area. 
The reason for similar charging energies is the presence of the two junctions, detailed in Section \ref{sec:oxyProcess}.

\subsubsection{TLS Evidence}
\label{sec:TLSSpec}
Early experiments in superconducting qubits with large areas showed evidence of many TLS defects \cite{Martinis2005,Stoutimore2012}.
We show that insufficient surface cleaning may have significantly contributed to these results. We do so by excluding the BOE surface treatments and searching for spectroscopy signatures of TLSs.

At minimal probe power, two-tone spectroscopy does not saturate TLS, allowing excitations to swap between the qubit and the TLS.
Strong TLS-qubit coupling significantly shifts the qubit frequency \cite{Mueller2013,Klimov2018a}, resulting in avoided crossings in spectroscopy.
Spectroscopy of four devices with sufficiently low power exhibited four total resolved (strongly coupled) avoided crossings.
Two devices exhibited no avoided crossing, and two devices exhibited two avoided crossings each.
This finding is consistent with the colloquial wisdom that TLS occurrences are stochastic so that some devices are ``unlucky''.

\begin{figure}[H]
	\centering
	\includegraphics[width=0.7\linewidth]{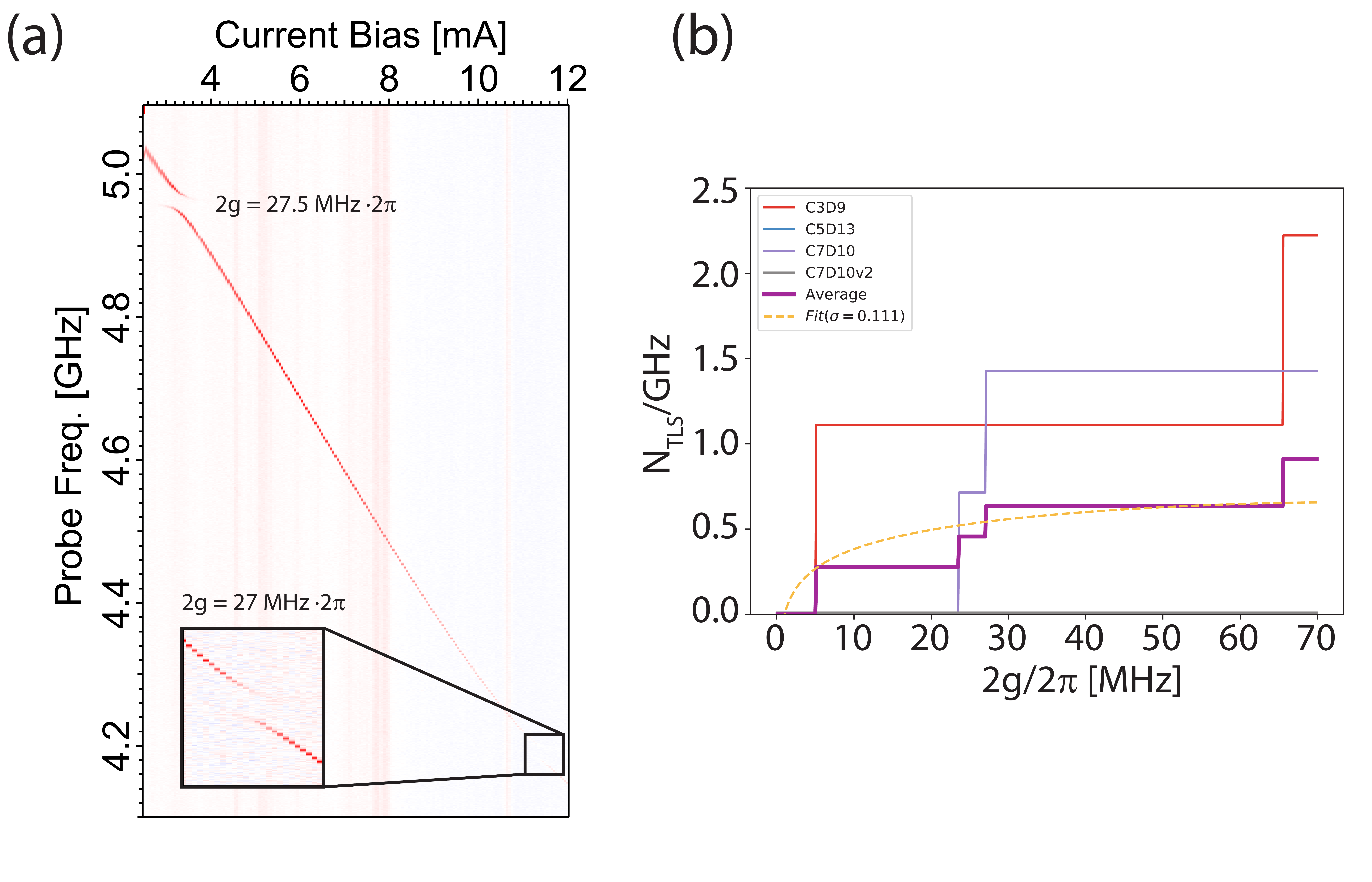}
	\caption[TLS in Spectroscopy]{(a) An example of avoided crossings as a qubit transition frequency comes into resonance with a strongly coupled TLS. The coupling strength is given by the maximal separation, $2g$, as indicated. (b) The cumulative density of splitting for multiple qubits (see legend). The average density (purple) and a fit to the standard TLS model (yellow dashed) estimates the total defect density, $\sigma =0.11 ~\mathrm{N_{TLS}/GHz/\um^2}$.}
	\label{fig:TLSCDF}
	\end{figure}

An estimate for the total TLS density results from combining the TLS count over the swept qubit-frequency range with the known area of the junctions \cite{Martinis2005}.
Figure \ref{fig:TLSCDF} contains the cumulative distribution function of TLS coupling strengths, along with a fit to the standard TLS model \cite{Phillips1987,Martinis2005,Muller2017}:
\begin{equation}
	N_{\rm TLS} = A\, \sigma \, \sqrt{\frac{1}{g^2}  - \frac{1}{g_{\rm max}^2}}
\end{equation}

The TLS density in these non-BOE cleaned devices is $\sigma = 1.1/{\rm GHz~\mu m^2}$, consistent with other measurements \cite{Martinis2005,Stoutimore2012}. 
Because the JJs in these devices have 100-times larger insulating barriers, we expect these devices should host significantly more defects.


For devices that received surface treatments, we do not observe strongly-coupled TLS with sufficient statistics to fit. 
However, the absence of spectroscopically-resolved (strongly coupled) defects does not imply the absence of defects. 
Recent experiments using DC electric fields to directly tune TLSs have shown that only 3\% of TLS are strongly coupled to the qubit\cite{Lisenfeld2019}.

\subsection{Time Domain Measurements}
\label{sec:TimeDomain}
Having examined the energy properties of the circuits with large-area {JJ}s, we have confirmed their transmon nature.
Now, we turn to the time domain to measure their coherence properties, especially $T_1$.


\subsubsection{$\mathbf{T_1}$ Without Surface Cleaning}
\label{sec:dirtyT1}
\begin{figure}[H]
	\centering
	\includegraphics[width=0.7\linewidth]{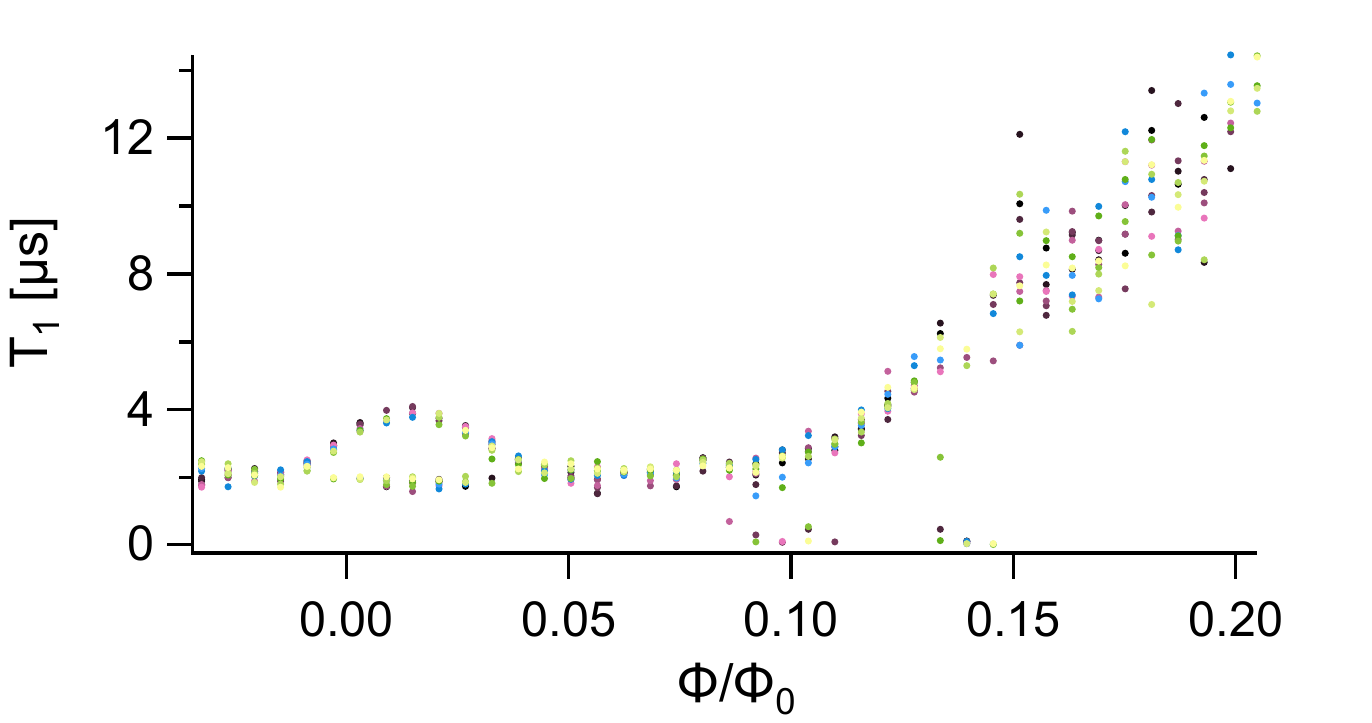}
	\caption[$T_1$ vs Flux and Time]{The $T_1$ time of a device that did not receive surface cleaning. We repeatedly sweep through flux biases for several hours and measure $T_1$. Near 0 bias, $T_1$ oscillates between two steady states--- a clear signature of TLS state-switching. $T_1$ is further suppressed by the Purcell effect, given the small detuning, $\Delta_{\rm min}$ = 0.235 MHz.}
	\label{fig:T1vsTimevsFlux}
\end{figure}

When a qubit is near resonance with a TLS, the decay rate is highly sensitive to the TLS state, due to an increase in the environmental density of states.
With fluctuations of the TLS state, $T_1$ fluctuates in the presence of TLS \cite{Muller2015, Klimov2018a,Burnett2019}.
Having identified regions where TLS strongly couple to the qubit through spectroscopy, we can compare the $T_1$ behavior to these regions.
For the same device pictured in Figure \ref{fig:TLSCDF}(a), we measure $T_1$ across a range of qubit frequencies multiple times over the course of 40 hours. Figure \ref{fig:T1vsTimevsFlux} shows the effects of TLS state-switching.
 Our sweeping rate for flux is relatively slow, so the sampling rate for $T_1$ at a particular qubit bias is only once per five minutes. Therefore, the measurement is not sensitive to diffusive or telegraphic statistics of the TLS \cite{Klimov2018a}.
We note that the $T_1$ values for this qubit are also significantly influenced by Purcell decay. 

\subsubsection{$\mathbf{T_1}$ With Surface Cleaning}
\label{sec:cleanT1}
TLSs reside at interfaces \cite{Bilmes2020}.
Thus, surface cleaning processes \cite{Zeng2015a, Keller2017, Nersisyan2019a,Altoe2020} are essential.
The next generation of large-area transmons was  cleaned with BOE both before spinning resist and before evaporation pump down.
BOE-cleaned devices exhibit significantly fewer strongly-coupled defects in spectroscopy.
The absence of strongly-coupled defects does not imply the absence of a weakly-coupled bath of defects \cite{Muller2017,McRae2020}.
However, the difficulty of probing these kinds of TLS forces us to turn to less direct probes of TLS activity.

We perform similar repeated measurements of $T_1$ over hour-long timescales.
The statistics of these measurements hint at the abundance of TLS.
Because the qubits have significantly different frequencies, $f_{01}$, $T_1$ should not be directly compared between the devices.
Instead, we utilize the quality factor:
\begin{equation}
	\frac{1}{Q} =  \frac{1}{T^{\rm tot.}_1 \,2\pi f_{01}}  = \sum_i p_i \tan\delta_i +\frac{1}{T_1^{\rm other}\,2\pi f_{01}}.
\end{equation}
$Q$ provides a time-scale invariant metric for qubit loss by directly connecting to dielectric losses, $\tan\delta_i$, weighted by participation ratios, $p_i$  \cite{Wang2015, Muller2017, Nersisyan2019a, McRae2020,Minev2020}.
$T_1^{\rm other}$ indicates the coherence time as limited by other factors such as Purcell decay and  quasi-particles.

\begin{figure}
	\centering
	\includegraphics[width=0.7\linewidth]{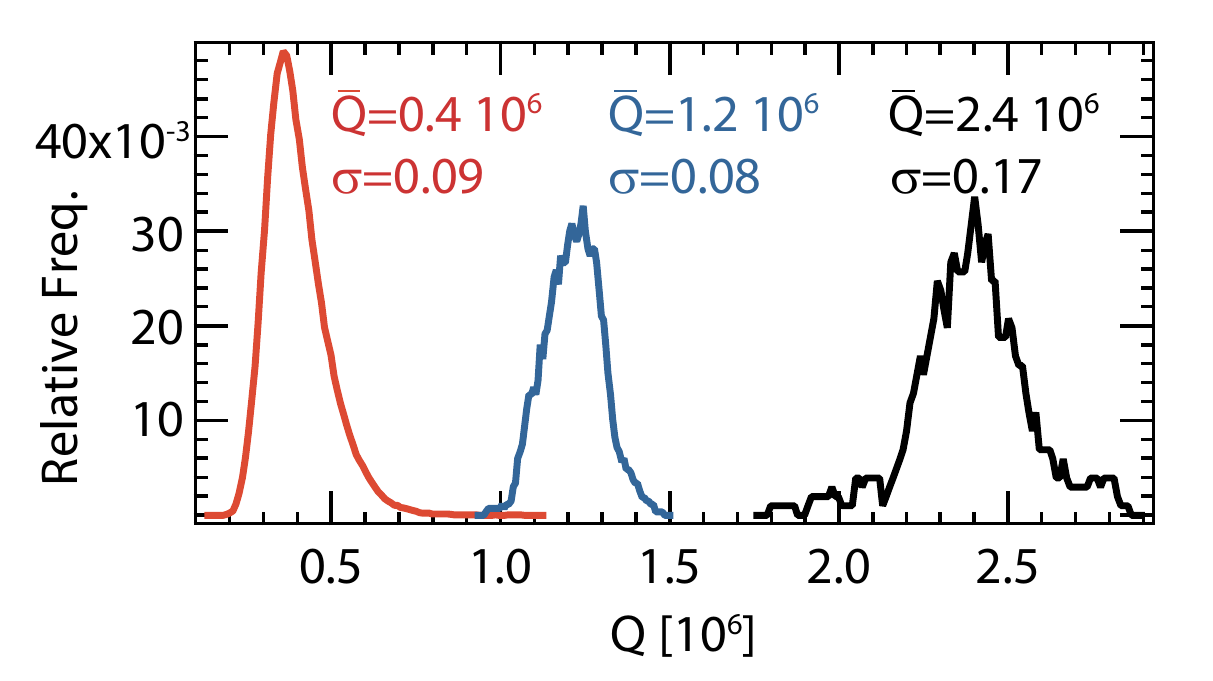}
	\caption[Quality Factor Histograms]{Distributions of quality factors measured for multiple hours for three devices under different cleaning conditions. Red curve: (same as in Fig. \ref{fig:T1vsTimevsFlux}) no surface cleaning; blue curve: BOE dip for 5 minutes before spin coating, a 30-second dip before evaporation; black curve: Piranha cleaned for 10 minutes, BOE clean before spin coating. The average Q and standard deviation of Q are indicated for each device.}
	\label{fig:Qhist}
\end{figure}

In Figure \ref{fig:Qhist}, we plot histograms of $Q$ measurements for each of three devices receiving different cleaning treatments.
The substrate of the first device received no surface cleaning upon receipt from the manufacturer.
The next substrate was cleaned with a 5-minute BOE dip before spin coating. After pattern development and ashing, the wafer received another BOE dip for 30 seconds.
The third substrate was cleaned with a Piranha solution (3:1 mixture of $H_2SO_4$ and $H_2O_2$ at $120^\circ$) for 10 minutes along with a 5-minute BOE dip before spin coating. This device also received a 30-second BOE dip after development and ashing.

Additional cleaning steps lead to a marked improvement in $Q$.
The uncleaned device which exhibited strongly coupled TLS (Fig.~\ref{fig:T1vsTimevsFlux}) exhibits a long, non-Gaussian tail in its $Q$ distribution. We associate this with the TLS state switching away from the qubit, allowing higher-than-average $Q$ ($T_1$).
However, we note that the device which received multiple cleaning treatments also exhibits extended tails.
With an overall improvement in the loss rate, the qubit appears more sensitive to fluctuations of weakly coupled defects.
Using Fermi's golden rule, the density of (TLS) states has decreased, but the transition matrix element is unaffected.
The low decay rate can thus significantly shift with TLS fluctuations.
We also note that $T_1$ for the doubly-cleaned was not monitored for as long as for the other devices, so the statistical uncertainty is higher.


\subsubsection{Aging $T_1$}
\label{sec:agingT1}
\begin{figure}[H]
	\centering
	\includegraphics[width=0.7\linewidth]{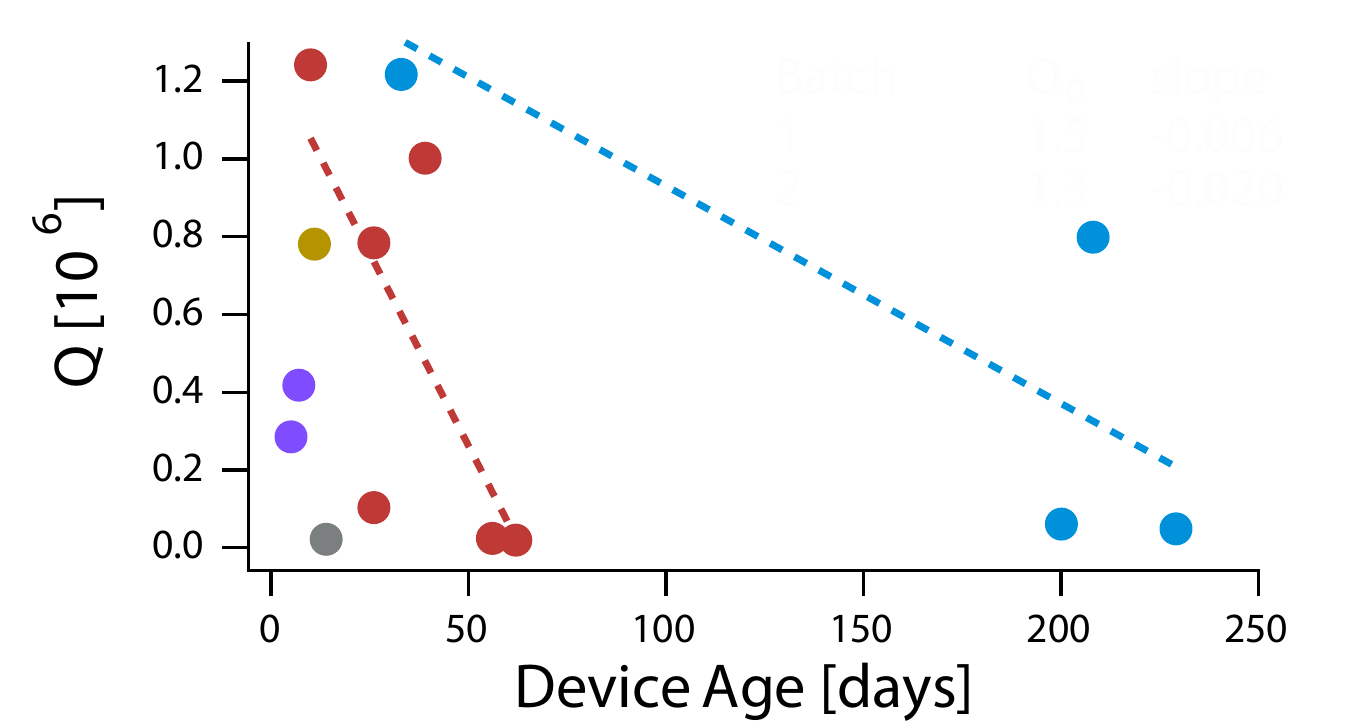}
	\caption[Quality Factor Aging]{Quality factor measurements for qubits from the same batches (color) as a function of age (time between fabrication and cooldown). Dashed lines are guides to the eye, signifying mostly negative trends.}
	\label{fig:QAging}
\end{figure}

While $T_1$ can vary significantly over a batch of devices, $T_1$ of a single device can also vary significantly from run to run.
In Figure \ref{fig:QAging}, we show $Q$ data as a function of a qubit's age: the time between fabrication and cooldown. Most qubits are cooled within a few days of fabrication, but we have occasionally returned to old batches.
While not cold, qubits are stored in ambient conditions.
We find that in these few tests, $T_1$ can significantly degrade. 
We suspect that large areas {JJ}s are more susceptible to aging due to the increased surface area on which oxygen and other contaminants can adsorb \cite{Moffat1989,Geng1999}.
Similar effects have been observed in resonators \cite{Vogt2013} and qubits \cite{Axline2018} after multiple cooldowns.

\subsection{Additional Noise Sources}
\label{sec:otherTests}
We have primarily focused on $T_1$ effects due to TLS defects. 
While other mechanisms also affect $T_1$ such as Purcell decay, other material losses, packaging modes, their contributions are expected to be small (see Refs.~\cite{Martinis2014a, SerniakThesis, Axline2018} and references therein for detailed reviews).
However, at the current level of $T_1$, quasiparticle (QP) tunneling can significantly contribute to $T_1$ loss.

QP tunneling occurs when a single electron is excited out of the superconducting ground state and tunnels across the {JJ} barrier.
The tunneling event drives transitions between the qubit's energy states (favoring excited state to ground transitions), leading to $T_1$ decay \cite{Lenander2011, Catelani2012}.
At thermal equilibrium with a 10 mK environment, the number of QP tunneling events is effectively 0, due to exponential scaling in the temperature. 
However, incident photons with frequencies above the superconducting gap of aluminum ($\Delta$ = 50 GHz) can induce nonequilibrium QP transitions \cite{Lenander2011,Catelani2012, Levenson-Falk2014,Serniak2019a}.

To mitigate QP, we extensive shielding and cable filtering.
At the dilution refrigerator's base plate, we surround qubit-cavity enclosures with a black-coated copper shield to absorb tray infrared radiation.
The coating is a mixture of carbon-lack powder, stycast epoxy, and SiC granules \cite{Barends2011,Corcoles2011,Kreikebaum2016}.

Shielding limits radiation from nearly the full $4\pi$ steradians, but RF lines remain as an entry path.
We include absorptive low-pass filters to further attenuate infrared radiation in the lines.
Our homemade filters are either stripline or coax transmission lines with a lossy dielectric. The dielectric contains magnetic particles which absorb high-frequency radiation \cite{Wollack2008}. Typical attenuation is 3 dB at 6 GHz and 70 dB at 50 GHz.
Evidence suggests that filters are most effective against QP when placed inside the can \cite{SerniakThesis}. 

\subsection{Limits on Loss Tangents}
\label{sec:dielectricLosses}
Devices with high coherence times can be used as upper bounds for the intrinsic loss tangents of materials \cite{Kim2011a, Rigetti2012, Murch2012a}.
High coherence devices have focused on loss tangents for substrate materials such as silicon \cite{Krupka2006,Kopas2020} and sapphire \cite{Braginsky1987,Kamal2016,Minev2013}.
But our devices emphasis the loss tangent of the aluminum oxide barrier.

Our highest $Q=2.8~10^6$ device allows us to contribute to the body of estimates for the loss tangent of aluminum oxide \cite{Kim2011a, Deng2014}.
Assuming published ($p$, $\tan\delta$) values for metal-substrate interface ($10^{-4},10^{-4}$), metal-air interface ($10^{-5},10^{-2}$), surface-air interface ($10^{-4},10^{-4}$), and silicon ($10^{-1},10^{-7}$) \cite{Martinis2005,OConnell2008,Wang2015,Woods2019,Axline2018} (contributing a net $1/Q=1.1\cdot 10^{-7}$), we estimate an upper bound on the loss tangent of $\mathrm{AlO_x}$ of $2.3~10^{-6}$.	
While this is lower than the current limit, $4~10^{-8}$ \cite{Kim2011a}, further coherence improvements may set a lower bound.

\chapter{Entropic Uncertainty Relations}

\section{Introduction}
Uncertainty relations provide limits to precision metrology \cite{Aasi2013,Degen2017}, amplification \cite{Caves1982, Clerk2010}, and measurement-based feedback \cite{Hacohen-Gourgy2020}.
The minimal amount of noise achievable is lower-bounded in uncertainty relations.
They highlight how quantum noise arises from 
disagreement between, or incompatibility of, quantum operations.
They predict lower limits on the measurement uncertainty of incompatible measurement operators, such as the position and momentum of an electron, as in Heisenberg's original thought experiment \cite{Heisenberg1927}.
The uncertainty extends to measurement operators which are not conjugate pairs.
For any Hermitian quantum operators, $A$ and $B$, Robertson proved \cite{Robertson1929} the textbook uncertainty relation:
\begin{equation}
\Delta A \Delta B \geq   \frac{1}{2} \big| \bra{\psi }  \left[A,B\right] \ket{\psi} \big|
\label{eqn:robertson}
\end{equation}
where $\Delta A$ ($\Delta B$) denotes the standard deviation of repeated $A$ ($B$) measurements for identically prepared systems with state $\ket{\psi}$ \cite{Robertson1929}.

%




\subsection{Problems with Variance-based Uncertainty Relations}
Variance is often used to quantify uncertainty.
The variance of a probability distribution for a random discrete variable $x$ is
\begin{equation}
\text{var}\left( \{p_x\}\right) =  \sum_i p_i \cdot(x_i-\bar{x})^2.
\end{equation}
Variance measures a probability distribution's expected deviation from its mean and thus quantifies the spread of a distribution. However, this deviation 
 is an inappropriate metric for uncertainty when applied to finite-dimensional operators \cite{Deutsch1983, Bialynicki-Birula2011}.

Consider a spin-1 particle in a maximally mixed state between its three possible values: $s_j \in \{-1,0,+1\}$.
The variance is $\frac{1}{3} (-1)^2 + \frac{1}{3} (0)^2 + \frac{1}{3} (1)^2 = \frac{2}{3}$.
Suppose we perform a measurement that excludes the possibility that $s_j=0$. 
We have gained information by decreasing the number of possibilities. Yet the variance has \textit{increased} to 
$\frac{1}{2} (-1)^2 +  \frac{1}{2} (1)^2 = 1$.

Variance inherits the units of its underlying random variable.
This makes variance undesirable for uncertainty relations because eigenvalue labels can confound the underlying goal of uncertainty relations.
For example, in the above spin-1 example, the story changes if the measurement excludes $s_j=+1$ instead of $s_j=0$.
The same number of outcomes have been excluded, yet the change in variance is different.

As a unitful quantity, variance is also incompatible with nonnumerical quantities. 
Neutrino mass-eigenstates oscillate between flavors: ``electron'', ``muon'', and ``tau'' \cite{SuperKamiokande1998}.
Treated as a three-outcome observable, a flavor measurement's uncertainty cannot be quantified with variance \cite{Coles2014}.
Entropy, on the other hand, only applies to the probability of an outcome, not to the value of the outcome.

The bound of Inequality [\ref{eqn:robertson}] has a few shortcomings as well \cite{Deutsch1983}.
From the relation, we would like to gather that quantum measurements are noisy by virtue of measurement operators themselves, namely their failure to commute.
However, the bound depends also on the prepared state. Thus, the Robertson-Heisenberg uncertainty relation, unfortunately, doesn't relate quantum measurement uncertainty with quantum measurements alone.
Besides lacking aesthetic purity, state-dependence makes the bound inapplicable to cryptography applications, where the state is often unknown and/or prepared by an adversary \cite{Tomamichel2012}.

Of particular interest is the uncertainty relation for $\sigma_x$ and $\sigma_z$. Inequality [\ref{eqn:robertson}] implies that the bound is $\frac{1}{2}\left|\bra{\psi} \sigma_y \ket{\psi}\right|$. For any state in the $X$--$Z$ plane of the Bloch sphere, the bound is zero---equivalent to the trivial bound.
For an eigenstate of $\sigma_x$ or $\sigma_z$, the trivial bound is tight (the LHS of the inequality matches the RHS). However, in this case, nothing is conveyed about the second measurement's uncertainty, which is multiplied by zero---the first measurement's variance.
The bound is tight only because the product of a minimal uncertainty state with a maximal uncertainty state is zero. So the tightest bound conveys no information about the outcomes.

In our measure of uncertainty, we desire a quantity which appropriately represents the information content (or uncertainty due to lack thereof) of measurements.
We then want to make meaningful statements of the bounds for such a measure.



\subsection{Entropic Uncertainty Relations}
\label{sec:entropy}
Modern uncertainty relations replace the product of variances with the sum of entropies \cite{Coles2017}.
Entropy, unlike variance, directly pertains to the information content of a probability distribution.
The Shannon entropy of a probability distribution is
\begin{equation}
\text{H}\left( \{p_x\}\right) =  \sum_i p_i \cdot \log_2\left(\frac{1}{p_i}\right).
\end{equation}
As the expected value of ``surprisal''($\log\frac{1}{p}$), entropy ideally quantifies the \textit{information content} of a probability distribution.

Besides improving the deficiency for variance-based uncertainty relations, entropic uncertainty relations make connections between fields like thermodynamics and quantum cryptography \cite{Tomamichel2012}.
In addition to answering the above concerns, entropic relations also provide bounds for connected sectors of quantum information science.
For example, quantum cryptography findings often base the capacity of quantum channels on entropic uncertainty relations \cite{Coles2017}.

A seminal result from Maassen and Uffink \cite{Maassen1988} displays the advantages of entropic uncertainty relations. For an $A$ measurement of a system described by density matrix $\rho$, suppose outcome $a$ occurs with probability $p_a$. The operator $A$ has a Shannon entropy via the $a$ probability distribution:
\begin{equation}
H(A)_\rho = -\sum_a p_a \log_2 p_a.
\label{eqn:entropy}
\end{equation}
With an analogous $H(B)_\rho$, Maassen and Uffink proved
\begin{equation}
H(A)_\rho + H(B)_\rho > -\log c.
\label{eqn:muEUR}
\end{equation}
$c$ denotes the \emph{maximum overlap} between any eigenstates, $\ket{a}$ and $\ket{b}$, of the observables' eigenstates: 
$c  =  {\rm max}_{a, b}  \left\{|\braket{b| a}|^2\right\}$.

Inequality [\ref{eqn:muEUR}] improves the weakness of Inequality [\ref{eqn:robertson}].
First, the new bound is independent of the state $\rho$. 
Second, entropy depends only on probabilities, in contrast with variance which depends on both probabilities and outcome. 
Finally, the smallest minimum in the bound of Inequality [\ref{eqn:muEUR}] is zero: minimum uncertainty in all cases occurs when observable eigenvectors coincide, leading to a maximum overlap of $c=1$ and thus a bound of zero.

The work of this chapter centers on including a weak measurement in an entropic uncertainty relation.
Weak measurements illuminate quantum dynamics, as in 
the tracking of the progress of spontaneous emission \cite{Naghiloo2015, Campagne-Ibarcq2016b}, the catching and reversing of quantum jumps \cite{Minev2018}, and observations of noncommuting observables' dynamics \cite{Hacohen-Gourgy2016}.
Here, we ask what role weak measurements play in operator incompatibility. We find that weak measurements can decrease the uncertainty between otherwise incompatible observables.

In the process of including a weak measurement in the uncertainty relation, we find that a weak value naturally appears. 	Weak values’ significance and utility have been debated across theory and experiment~\cite{Leggett1989,Aharonov1990,Hosten2008, Dressel2014, Jordan2014}.
Our work demonstrates how weak values have physical meaning in uncertainty relations.
As we will see, weak values decrease the incompatibility between incompatible observables.

Other experimental work has explored entropic uncertainty relations with various platforms, 
including neutrons, optics, and nitrogen-vacancy centers~\cite{Li2011,Prevedel2011,Xing2017,Demirel2019a}.
The measurements in~\cite{Demirel2019a}, though nonprojective,
are probabilistic projections.
In contrast, our measurements are weak and experimentally demonstrate 
the weak value's role in reconciling incompatible operations.

Uncertainty relations occupy two categories~\cite{Coles2017}, 
one centered on measurement outcomes' unpredictability~\cite{Xing2017, Demirel2019a} 
and one centered on measurements' disturbance of quantum states~\cite{Li2011, Prevedel2011}.
Our uncertainty relation occupies both categories, similar to some optical photon experiments~\cite{Rozema2012}:
on the one hand, we prepare an eigenstate of one measurement operator and perform a measurement of an incompatible measurement operator.
On the other hand, we take advantage of the weak measurement's disturbance of the initial eigenstate.
This work identifies weak measurements as a means to unify the classes of uncertainty relations.

Let us now turn to the underpinnings of our entropic uncertainty relations for weak measurements. 

\section{Theory}
The progress of uncertainty relations is the gradual tightening of bounds for ever-more-complicated operators.
In this section, we describe the derivation of our central entropic uncertainty relation governing weak measurements.
The path will begin with several bounds for familiar projective operators (called projector-valued measures, PVMs). Then, we will generalize the operator to positive operator-valued measures, POVMs.

In every example, the uncertainty relation centers on the disagreement between two operators. 
The LHS of the inequality is constructed as the sum of two entropies---one for each of the operators.
The calculation of the entropy depends on the details of the operator, but the process boils down to calculating the Shannon entropy for probabilities of outcomes.

The RHS of each inequality is an attempt to leverage various facts about entropies to derive as tight of a bound as possible. 
The relevant facts are often geometric. The arguments rely on the correspondence between entropies and distances:
Consider a probability distribution over $N$ outcomes has an $N$-dimensional vector.
The (R\'enyi) entropy of the probability distribution is proportional to the logarithm of the vector's norm (of degree $\alpha$):
\begin{equation}
H_\alpha(\{p\}) = \frac{\alpha}{1-\alpha} \log ||\vec{p}||_\alpha
\end{equation} 
For reference, the 3-norm of vector $\vec{x}$ is $||\vec{x}||_3 = \left( \sum_j x_i^3 \right)^\frac{1}{3}$, and the 2-norm of a vector is the well-known Euclidean distance.
The Shannon entropy is equal to the R\'enyi entropy of degree 1.
The above equation implies that the uncertainty of a probability distribution is related to how extended the distribution is over its support: If the probability amasses on one outcome, the extent is small and the uncertainty is low.

The profound mathematical fact of uncertainty relations is that not all vector spaces (quantum operators) that host the probability vector (quantum state) are equally efficient.
In terms of entropy, some operators require more bits of information to represent the same intrinsic probability distribution (quantum state). 

Let us work up to this fact through simpler examples.

\subsection{Trivial Bound}
The simplest bound for entropic uncertainty relations uses elementary properties of probabilities \cite{Hirschman1957}. Probabilities\footnote{Certain representations of quantum states such as the Wigner function use quasi-probabilities which can be negative. }
 are quantities between 0 and 1.
The logarithm of a quantity between 0 and 1 is always negative, and the probability is always positive. 
The entropy, $-p\log p$, is a negated product of the logarithm with the  probability---a positive quantity. This relation gives the trivial bound for the sum of two entropies:
\begin{equation}
H_\rho(A) + H_\rho(B) \geq 0.
\end{equation}
Equality depends on the limit $-p\log_2 p \rightarrow 0$ as $p\rightarrow 0$.

\subsection{Deutsch Bound}
David Deutsch used simple algebraic features of quantum states to improve the bound \cite{Deutsch1983}.
Consider the expanded expression for the sum of entropies:
\begin{equation}
H_\rho(A) +H_\rho(B)= \sum_a- \left| \braket{a|\psi}\right|^2 \log_2\left( \left| \braket{a|\psi}\right|^2\right)
+ \sum_b- \left| \braket{b|\psi}\right|^2 \log_2\left( \left| \braket{b|\psi}\right|^2\right)
\label{eqn:deutschExpand}
\end{equation}
where the presumed pure state $\rho=\ket{\psi}\bra{\psi}$ has been replaced with it's ket representation.
With the identity, $\sum_x \left| \braket{x|\psi}\right|^2 =1$, Equality [\ref{eqn:deutschExpand}] rearranges into an expression that conjoins probabilities of the two operators.
\begin{equation}
\begin{split}
H_\rho(A) +H_\rho(B) &= \sum_{a,b}- \left| \braket{a|\psi}\right|^2\left| \braket{b|\psi}\right|^2
 \left[		\log_2\left( \left| \braket{a|\psi}\right|^2\right) +   \log_2\left( \left| \braket{b|\psi}\right|^2\right) 	\right] \\
& =  \sum_{a,b}- \left| \braket{a|\psi}\right|^2\left| \braket{b|\psi}\right|^2
\left[		\log_2\Big(  \braket{\psi|a}\braket{a|\psi} \braket{\psi|b}\braket{b|\psi} \Big) \right].
\label{eqn:deutschNew}
\end{split}
\end{equation}
Now we may leverage some properties of quantum states. The maximum value for the square-bracketed factor corresponds to the state which bisects the eigenstates. 
The state corresponds to $\psi_{\rm bisector} = \frac{1}{\sqrt{2(1+|\braket{a|b}|)}} \left( \ket{a} + e^{-i\theta} \ket{b}  \right)$, where we've used $\theta = \arg(\braket{a|b})$.
This state bounds the logarithm factor in Equality [\ref{eqn:deutschNew}]: 
\begin{equation}
\log_2\Big(  \braket{\psi|a}\braket{a|\psi} \braket{\psi|b}\braket{b|\psi} \Big) 
\leq 2 \log_2\left[ \frac{1}{2} (1+|\braket{a|b}|) \right].
\end{equation}

Replacing each term in the sum with its maximal%
 \footnote{The maximum becomes a minimum with the leading negative sign.}
 value leads to the Deutsch entropic uncertainty relation:
\begin{equation}
H_\rho(A) +H_\rho(B) \geq  \min_{a,b} \left\{ -2 
\log_2\left[ \frac{1}{2} (1+|\braket{a|b}|) \right]  \right\}
\end{equation}
Note how state removal transformed the equality into an inequality.

\subsection{Maassen-Uffink Bound}
Hans Maassen and Jos Uffink slightly improved this bound while extending the relation to include an entire class of (R\'enyi) entropies \cite{Maassen1988}.
The class of entropies corresponds to the different norms of the probability distribution (treated as a vector).
Our targeted (Shannon) entropy is a special case of the generalization (using the 1-norm).

Maassen and Uffink apply a result from complex analysis known as the Riesz-Thorin interpolation theorem. The theorem applies to transformations of complex-valued vectors which preserve length.
Here, the complex-valued vectors are inner products $x=\braket{a|\psi}$, and the transformation is a basis transform: $T: \braket{a|\psi} \rightarrow \braket{b|\psi}$.
Thus formulated, the Riesz theorem states:
\begin{equation}
c^\frac{1}{n} \left(\sum_j |Tx_j|^n \right)^\frac{1}{n}  \leq c^\frac{1}{m} \left(\sum_j |{x}|^m \right)^\frac{1}{m} ,
\end{equation}
where $c$ is the maximum transformation matrix element: $c = \max_{a,b} |\braket{b|a}|$.
For choices of $m$ and $n$ which select the Shannon entropies, the Maassen-Uffink relation states:
\begin{equation}
H_\rho(A) + H_\rho(B) \geq 
\min_{a,b} \left\{ -2 \log_2 |\braket{b|a}|  \right\}.
\end{equation}

This uncertainty relation provides a direct comparison to our uncertainty relation in the absence of the weak measurement. We will refer to it elsewhere.

\subsection{Tomamichel Bound}
The primary bound for this thesis is an application of a bound derived by Marco Tomamichel \cite{Tomamichel2012} and adapted by Nicole Yunger Halpern, Anthony Bartolotta, and Jason Pollack \cite{YungerHalpern2017}.
The bound generalizes the previous results from projector-valued measures (PVM) such as projectors of Pauli operators to positive operator-valued measures (POVM). 
The generalization is essential in describing weak measurements which do not project the system of interest.

The general uncertainty relation follows a similar path to the Maassen-Uffink relation. The bound depends on the ``distance'' (inner product) between the quantities of interest. The Maassen-Uffink bound depends on the distance between eigenvectors of the operators. The Tomamichel bound depends on the distance between POVMs.
The operator-norm quantifies the distance between two POVMs, $\mathcal{A}_a$ and $\mathcal{B}_b$:
\begin{equation}
\Vert  \mathcal{A}_a \mathcal{B}_b \Vert
= \lim\limits_{\alpha\rightarrow \infty}
 \Tr \left\{ \left[\sqrt{(\mathcal{A}_a\mathcal{B}_b)^* (\mathcal{A}_a\mathcal{B}_b) }\right]^\alpha\right\}^\frac{1}{\alpha}.
\end{equation}
The derivation of the bound follows again from the Riesz-Thorin interpolation theorem \cite{Krishna2001}, now applied to this operator norm.
The bound states 
\begin{equation}
H_\rho(\mathcal{A}) + H_\rho(\mathcal{B}) \geq 
\min_{a,b} \left\{ - \log_2 \Vert  \mathcal{A}_a \mathcal{B}_b \Vert ^2   \right\}.
\end{equation}

The next section describes the application of this bound to the particular POVMs of interest.

\subsection{Weak Measurement Bound}
The first operator in our weak measurement uncertainty relation is a projective Pauli operator. We choose to fix this operator as $\sigma_z$, amounting to a choice of the coordinate system.
Measurements of this initial operator result in an outcome $i\in \{-1,1\}$. 
The second operator is a composition POVM. It combines a weak-measurement Kraus operator with a final projective Pauli operator: ${A}F$. 
As such, measurements of this POVM result in a tuple of outcomes, $(j,f)$, for the weakly-measured component and the strongly-measured component.

Let us consider the entropies of each operator in detail.
The entropy of the initial $\sigma_z$ measurement follows from Equation [\ref{eqn:entropy}].
The entropy of the joint-measurement POVM is calculated for the joint probability of $(j,f)$ outcomes.
$j$ is a continuous variable, and so the Shannon entropy%
\footnote{The differential entropy quantifies information in the case of continuous-variable probability distributions.}
would not be well-defined.
However, weak measurements are experimentally discrete quantities due to finite detector resolution.
Thus, the Shannon entropy is calculated over a finite sum: 
\begin{equation}
H({A}F)_{\rho} = \sum_{j,f} p_{j,f} \log_2 p_{j,f}
\end{equation}
The range of $j$ depends on the detector settings (see Sect. \ref{sec:EURexp}).
The sum of the two entropies yields the LHS of the EUR.

Now consider the bound.
The  {Yunger Halpern}-Bartolotta-Pollack bound \cite{YungerHalpern2017} is an application of the Tomamichel bound to the POVM describing weak measurements.
We aim to describe the uncertainty between two POVMs.
The POVM for the first strong (projective) measurement of $\mathcal{I}$ is $\{\Pi_i\}$. 
The set contains elements corresponding to each outcome $i=\pm1$.
The second POVM for the joint weak-strong measurement of ${A}F$ is $ \{\Pi_f K_j\}$, where the Kraus operator 
\begin{equation}
\label{eq_Kraus1}
K_j  
=  \left(  \frac{\delta t}{{2 \pi \tau}}  \right)^{1/4} 
\exp \left(  -\frac{\delta t}{4\tau} [j I - A]^2  \right) 
\end{equation}
describes the weak measurement of operator $A$ with outcome $j$ \cite{Jacobs2006}.

In the Tomamichel form, the calculation of the operator norm results in an expression that depends on a maximal eigenvalue and is not directly accessible to experiments \cite{YungerHalpern2019}. Instead, experiments access expectation values of eigenvalues.
We use the monotonicity%
\footnote{Monotonicity states that norms decrease with increasing order: $||x||_i > ||x||_j$ when $i<j$.}
of vector norms to translate the expression into an experimentally accessible form.
In particular, the operator norm (which uses infinite degree) is less than the 1-norm:
\begin{equation}
\Vert \mathcal{O}\Vert = \lim\limits_{\alpha\rightarrow \infty}
\sqrt{\Tr \left[ \mathcal{O}^\frac{\alpha}{2}\right]^\frac{2}{\alpha}} < \sqrt{\Tr \left[ \mathcal{O}\right]}.
\end{equation}
By measuring expectation values (the RHS above), we get a quantity which upper-bounds the infinite norm.
This inequality admits experimental tests at the cost of a loosened bound.

With this expression, we can formulate the bound of our uncertainty relation. 
We use the norm between the POVMS $\Vert \mathcal{I\cdot W}F \Vert$.
The entropy of the two POVMs is bounded as:
\begin{equation}
H_\rho(\mathcal{I}) + H_\rho(\mathcal{WF}) \geq 
	\min_{i,j,f} \left\{ - \log_2 \left(  \Tr\left[ \Pi_i K_j^\dagger \Pi_f K_j \right]\right)   \right\}.
\label{eqn:EURnotSimplified}
\end{equation}

Some simplification is in order. 
Because the measurement is weak, we can Taylor expand the Kraus operator using the strength of the measurement as a smallness parameter:
\begin{equation}
K_j = \sqrt{p_j}\left( I + g_j A\right) + O(g_j^2).
\label{eqn:KrausTaylor}
\end{equation}
$\sqrt{p_j} = 	 \left( \frac{ \delta t }{ 2 \pi \tau } \right)^{1/4}  
\exp \left( - \frac{ \delta t }{ 4 \tau }   \left[ j^2 + 1 \right]  \right)$ is the prefactor to $I$ and describes the probability of obtaining outcome $j$ if the detector is not coupled to the system.
$g_j  =  \frac{ \delta t }{ 2 \tau } j $ is the prefactor to $A$ (combined with $\sqrt{p_j}$) and is thus related to the amount of backaction from the weak measurement.
Section~\ref{sec:RHSBound} discusses the measurement of these quantities.

The detailed simplification below shows how the weak value directly appears in the bound of our uncertainty relation.
We plug the Taylor-approximated form of the Kraus operator into the bound of Inequality [\ref{eqn:EURnotSimplified}].
The result is:
\begin{equation}
\begin{split}
&- \log_2 \left\{  \Tr\left[ \Pi_i K_j^\dagger \Pi_f K_j \right]\right\}   \\
	=& - \log_2 \left\{  \Tr\left[ \Pi_i \sqrt{p_j}(I+g_j^*A^\dagger) \Pi_f \sqrt{p_j}(I+g_jA) \right]\right\}   \\	
	=& - \log_2 \left\{ p_j\Tr\left[ \Pi_i \Pi_f\right] + p_j\Tr\left[ \Pi_i g_j^*A^\dagger\Pi_f + \Pi_i\Pi_f g_jA \right]\right\}   \\
	=& - \log_2 \left\{ p_j\Tr\left[ \Pi_i \Pi_f\right] 
	+ p_j \Tr\left[ \Pi_i \Pi_f\right]
			\left( 
				  \frac{\Tr\left[ \left(\Pi_f g_jA\Pi_i\right)^*\right]}{Tr\left[ \Pi_f \Pi_i\right]} 
				 +\frac{\Tr\left[\Pi_f g_jA\Pi_i \right]  }  {Tr\left[ \Pi_f \Pi_i\right]} 
			\right)	  \right\}   \\
 	=& - \log_2 \left\{ p_j \, p_{f|i} \right\} 
	-\log_2  \left\{1+ 2 \Re\left(g_j A_{\rm wv}\right)\right\} \\ 
  	\approx& - \log_2 \left\{ p_j \,p_{f|i}\right\} 
			-\frac{2}{\ln{2}} \Re(g_j A_{\rm wv}) 
\end{split}
\label{eqn:deriveRHS}
\end{equation}

With this, we have the final form of our entropic uncertainty relation:
\begin{equation}
\boxed{
H_\rho(\mathcal{I}) + H_\rho(\mathcal{WF}) \geq 
\min_{i,j,f} \left\{
 - \log_2 \left\{ p_j\, p_{f|i} \right\} 
-\frac{2}{\ln{2}} \Re(g_j A_{\rm wv}) 
\right\}
}
\label{eqn:EUR}
\end{equation}

\section{Experiment}
\label{sec:EURexp}
Here we describe the experimental results for measuring each side of the uncertainty relation derived in the last section \cite{Monroe2021}.
We first focus on entropies before moving onto bounds.
The entropy results highlight two distinct components of uncertainty relations.
First, \textit{preparation uncertainty} results from incompatibility between a measurement and the underlying state.
For example, a position eigenstate is incompatible with momentum measurements.
Second, \textit{disturbance uncertainty} relates to the measurement-induced backaction imparted on the state.
For example, a linear combination of position states collapses into a single position upon measurement.

We measure the entropic uncertainty relation with a transmon superconducting qubit.
 The qubit couples to one mode of the electromagnetic field in a three-dimensional microwave cavity. 
 The qubit frequency, $\omega_q/(2\pi) =3.889$ GHz, is far detuned from the cavity frequency, $\omega_c/ (2\pi) =5.635$ GHz, enabling a dispersive interaction. 
Dispersive interactions do not exchange energy, allowing for quantum-nondemolition measurements (see Sect.~\ref{sec:dispersive}).

\subsection{Entropies of the Entropic Uncertainty Relation}
\begin{figure}
	\centering
	\includegraphics[width=0.7\linewidth]{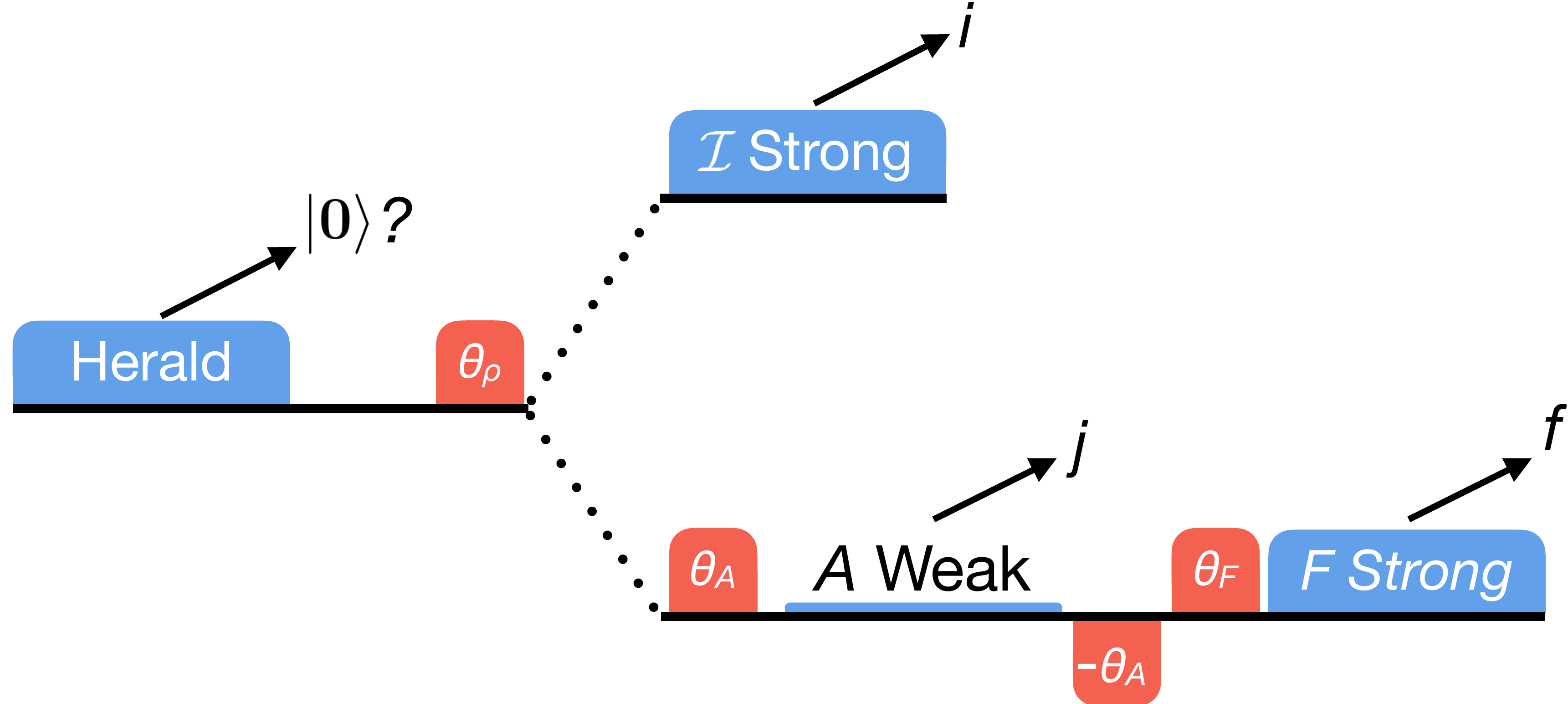}
	\caption[Pulse sequence for measuring entropies]{Pulse sequences measuring both $H(\mathcal{I})_\rho$ (top branch) and $H({A}F)_\rho$ (bottom branch). Blue pulses gate the cavity frequency, and red pulses gate the qubit frequency (single-sideband modulation not shown). The outcomes of each measurement $i$, $j$, and $f$ are shown with their respective cavity pulses.}
	\label{fig:EURseq}
\end{figure}

Entropy measurements of the two observables are performed with the following sequence (see Figure \ref{fig:EURseq}).
We herald the ground state by strongly measure $\sigma_z$ at the beginning of each repetition. Having projected the qubit into either $\ket{0}$ or $\ket{1}$, we discard experimental runs in which the qubit was not in the ground state.
A resonant qubit drive rotates the qubit state by $\theta_\rho$ to prepare the initial state, $\rho$.
We then perform one of two measurements, $\mathcal{I}$ or ${A}F$.
The first measurement, $\mathcal{I}=\sigma_z$, results in outcomes $i=\pm 1$. 
The relative frequencies sample the probability distribution of $\mathcal{I}$, $p_i$.

\begin{figure}[h]
	\centering
	\includegraphics[width=0.7\linewidth]{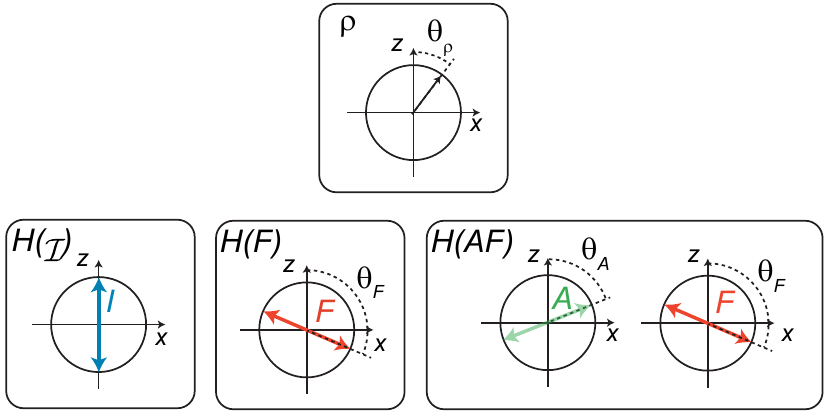}
	\caption[Measurement Angles]{Each preparation and measurement is characterized by an angle in the $X$--$Z$ plane of the Bloch sphere. Measurements at a specified orientation are proceeded with  to align the measurement axis with the $Z$-axis. With this scheme, we effectively measure the desired operator by measuring $\sigma_z$.
	Another state-rotation with negated amplitude (see Fig.~\ref{fig:EURseq}) after the measurement undoes the first rotation.
	}
	\label{fig:eurangles}
\end{figure}

The measurement ${A}F$ is a joint weak strong measurement.
The weak measurement axis is $\theta_A$ radians away from the $+Z$-axis, and the strong measurement axis is $\theta_F$ radians away from the $+Z$-axis (see Fig.~\ref{fig:eurangles}).
Our experiment only measures the $\sigma_z$ component of the qubit (see Sect. \ref{sec:dispersive}).
We can measure a different operator by first rotating the state so that the operator's measurement axis aligns with the $Z$-axis. Our $\sigma_z$ measurement then effectively measures the desired operator.
For the weak measurement, after which other operations will be performed, we rotate the state by the negative angle, to reset the orientation (see Fig.~\ref{fig:EURseq}).
The joint $AF$ measurement procedure results in a tuple of outcomes, $(j,f)$. 
The relative frequency of each tuple samples the two-dimensional joint probability distribution of ${A}F$, $p_{j,f}$.

The outcome $f$, being the measurement result of a Pauli operator, can equal $\pm1$.
The outcome $j$, on the other hand, is the result of a cavity state measurement. Because the cavity's quadrature values are continuous variables (see Sect.~\ref{sec:cavityQuant}), $j$ is a continuous variable.
However, in practice, we discretize the continuous outcome into 52 bins. Of these, approximately half typically receive counts. 
This choice contributes a baseline amount of $\log_2(26)\simeq4$ bits of entropy and also raises the bound.


\subsubsection*{Entropies of Strong Measurements}
\begin{figure}[h]
	\centering
	\includegraphics[width=0.7\linewidth]{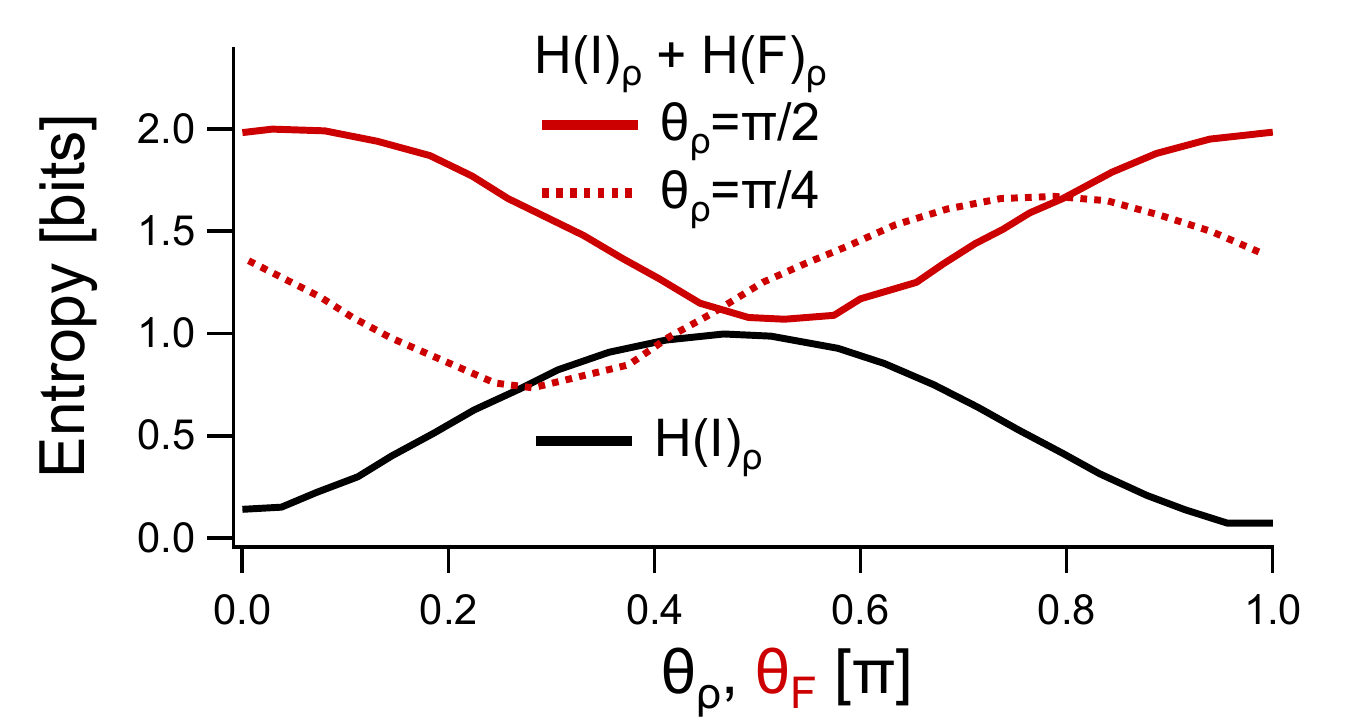}
	\caption[Projective measurement entropies]
	{In black, the entropy of projective $\mathcal{I}$ measurements as a function of $\theta_\rho$ displays preparation uncertainty.
		This entropy (for fixed $\rho$) adds to the entropy of the other projective measurement $F$, in red, for two different initial states. 
		The near-equality between the solid curves at $\pi/2$ demonstrates the uncertainty trade-off between the two measurement operators.
		Finite readout fidelity (98\%) is reflected in 0.14 lower bounds in each entropy.}
	\label{fig:EURlhsRho}
\end{figure}

Consider the entropy of the first measurement, $H(\mathcal{I})_\rho$, shown as the black trace in Figure \ref{fig:EURlhsRho}.
Different initial states result in different amounts of uncertainty for the $\sigma_z$ measurement.
When $\theta_\rho=0$, the initial state is $\ket{0}$---an eigenstate of the measurement operator.
Measurement outcomes thus result in one outcome with near-unity probability. (In practice, measurement infidelity reduces the probability from 100\%.)
The entropy of this probability distribution is nearly zero. The same argument applies when $\theta_\rho=\pi$. 
At the other extreme, when $\theta_\rho=\pi/2$, the initial state $\rho = \left(\ket{0}\bra{0} +\ket{1}\bra{1}\right)/\sqrt{2}$ is an eigenstate of $\sigma_x$. 
Here, $i$ outcomes of the $\mathcal{I}=\sigma_z$ are nominally equiprobable. The uncertainty of this probability distribution, quantified as entropy, is maximal, with a value of 1 bit due to the number of possible outcomes (2).

The red traces of Figure \ref{fig:EURlhsRho} examine the sum of entropies as a function of $\theta_F$ for a fixed initial state. Because there are now two binary outcomes---each of which can misalign with the state---the maximal entropy is now two bits. The state can introduce additional uncertainty with respect to both of these measurements.

The changes in entropy for different state preparations exemplify preparation uncertainty. 
A given state can be represented in any (complete) basis. However, there is one particular basis in which the state's description is most efficiently represented---the state's eigenbasis.
Efficiently represented information has minimal entropy.
An operator which shares the state's eigenbasis will have measurements with minimal representation uncertainty.

Because the uncertainty relations we consider have two measurement operators, there are two sources of preparation uncertainty---one for each measurement.
Consider an initial state prepared with $\theta_\rho=\pi/2$ and measured with $\mathcal{I}=\sigma_z$ or with $F=\sigma_z$ (solid red line in Fig.~\ref{fig:EURlhsRho} at $\theta_F=0$). This choice of measurements leads to two bits of uncertainty because the state's eigenbasis is mutually unbiased with respect to both measurements.
The uncertainty improves when $\theta_F$ becomes $\pi/2$, because now the $F$ measurement shares the state's eigenbasis.

Compare to the case of a $\theta_\rho =\pi/4$ preparation (dashed red line in Fig.~\ref{fig:EURlhsRho}). As a function of $\theta_F$, the sum of entropies has a minimum at $\theta_F=\pi/4$, unsurprisingly because $F$ and $\rho$ share an eigenbasis.
However, for this preparation, the new maximum is only 1.67 bits, compared to 2.0 bits. 
The maximum decreases because the state's eigenbasis better aligns with both of the operator eigenbases.

The total entropy decreases when the state's eigenbasis aligns with either measurement operator.
Because we chose to fix the configuration for $\mathcal{I}$, we can minimize the entropy overall by always initializing the state in an eigenstate of $\mathcal{I}$, e.g.~with $\theta_\rho=0$.
Initializing the state to an eigenstate of one of the operators is the most interesting case because it allows the uncertainty relation to quantify uncertainty between the operators, rather than the uncertainty between each operator and the state.
With the state initialized to an eigenstate of an operator, we focus only on preparation uncertainty resulting from operator-operator disagreement (rather than state-operator disagreement).

Let us turn to this operator incompatibility, quantified by the Maassen-Uffink relation (Ineq. [\ref{eqn:muEUR}]).
For our chosen initial state, this uncertainty relation quantifies the degree of operator-operator disagreement.
The relation includes two entropies and provides a lower bound for their sum. 
Consider the entropies.
For the Maassen-Uffink case, the operators are Pauli operators, $\mathcal{I}=\sigma_z$ and $F=\sigma_\theta$. $F$ is a linear combination of $\sigma_z$ and $\sigma_x$, as in Figure \ref{fig:eurangles}.

\begin{figure}
	\centering
	\includegraphics[width=0.7\linewidth]{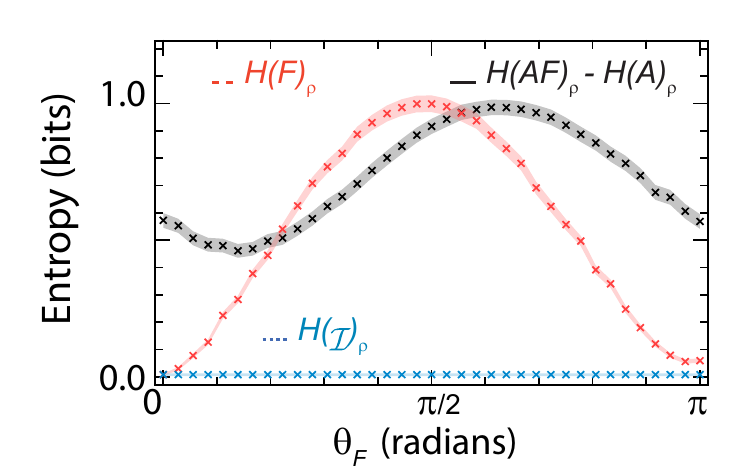}
	\caption[Entropy in the Presence of Weak Measurement]{Three entropies as a function of the $F$ measurement orientation. The blue and red curves display the entropy of $\mathcal{I}$ or $F$ measurements, respectively, without any weak measurement. The black curve represents the entropy of the joint ${A}F$ measurement with the baseline entropy due to the many-outcome ${A}$ subtracted off. Here, the weak measurement axis is set to $\theta_{A} = \pi/4$. Error bands indicate the propagated uncertainty due to finite sampling.
	}
	\label{fig:eurLHSthetaf}
\end{figure}

Each of the entropies is shown in Figure \ref{fig:eurLHSthetaf}.
Because of the initial state, the first entropy term is nominally zero.
However, state preparation and measurement errors (especially finite-temperature qubit populations) cause $\mathcal{I}$ measurements to result in both outcomes, with 98\% and 2\% probability. This probability distribution has 0.14 bits of entropy, as the blue line in Figure \ref{fig:eurLHSthetaf} shows.

The second entropy, shown in red in Figure \ref{fig:eurLHSthetaf} shows how entropy depends on the measurement axis, $\theta_F$. 
When $\theta_F$ increases from 0 to $\pi/2$, the two operators become less compatible, and entropy increases.

The dependence illustrates the second type of uncertainty relations owing to operator-operator disagreement called preparation uncertainty.
That is, quantum mechanics disallows preparation of a state which minimizes uncertainty of both $\sigma_x$ and $\sigma_z$. Preparing a definite eigenstate of one operator results in a completely mixed state of the other.


\subsubsection*{Entropies of Weak Measurements}
To consider the entropy of the weak measurement, we first describe the weak measurement outcome.
The outcome results from probing the state of the cavity.
We correlate the cavity signal by scaling it so that when the qubit initialized to the ground or excited states, the average outcome is $+1$ or $-1$, respectively.
 We scale the weak measurement outcomes such that an initial ground or excited state results in an average outcome of $+1$ or $-1$, respectively.
 With the measurement strength chosen to be weak enough to satisfy the Taylor approximation in the bound (see Eqn.~[\ref{eqn:KrausTaylor}]), the distribution of weak measurements has a standard deviation of 5.5.
We discretize the distribution of $j$ outcomes into 52 bins.
%
%

First, let us examine the case of a weak measurement fixed to an axis $\theta_{A}=\pi/4$, while varying $\theta_F$ as above. 
To facilitate comparison to the Maassen-Uffink relation, we subtract off the entropy at $\theta_F=0$. This point amounts to a normalization.
This measurement entails preparing the ground state, then immediately measuring $\sigma_z$ weakly before strongly measuring $\sigma_z$ again.
The outcomes are as certain as possible with a weak measurement. However, there is uncertainty because weak measurements only provide partial state information.
The entropy of the joint $WF$ measurement thus has a minimal but nonzero value of 4.53 bits at $\theta_F=\theta_W=0$.
We subtract this normalization value off of the relation in Figure \ref{fig:eurLHSthetaf}.

Thus normalized, the value $H_\rho({A}F) - H_\rho({A}_{\sigma_z} \sigma_z)$ compares directly to $H_\rho(F)$ and $H_\rho(\mathcal{I})$.
With $\theta_W=\pi/4$, Figure \ref{fig:eurLHSthetaf} characterizes $H_\rho({A}F)$ as a function of $\theta_F$.
Here, the minimal value is nonzero because the state is misaligned with the weak measurement axis.
The maximum value is one bit, corresponding to the entropy added by the strong measurement---just as was the case without the weak measurement.
Despite a $\sigma_z$-eigenstate preparation, the maximum does not occur at $\theta_F=\pi/2$, when $F=\sigma_x$.
The weak measurement has shifted the eigenbasis which is mutually unbiased to $\sigma_z$.

\begin{figure}
	\centering
	\includegraphics[width=0.65\linewidth]{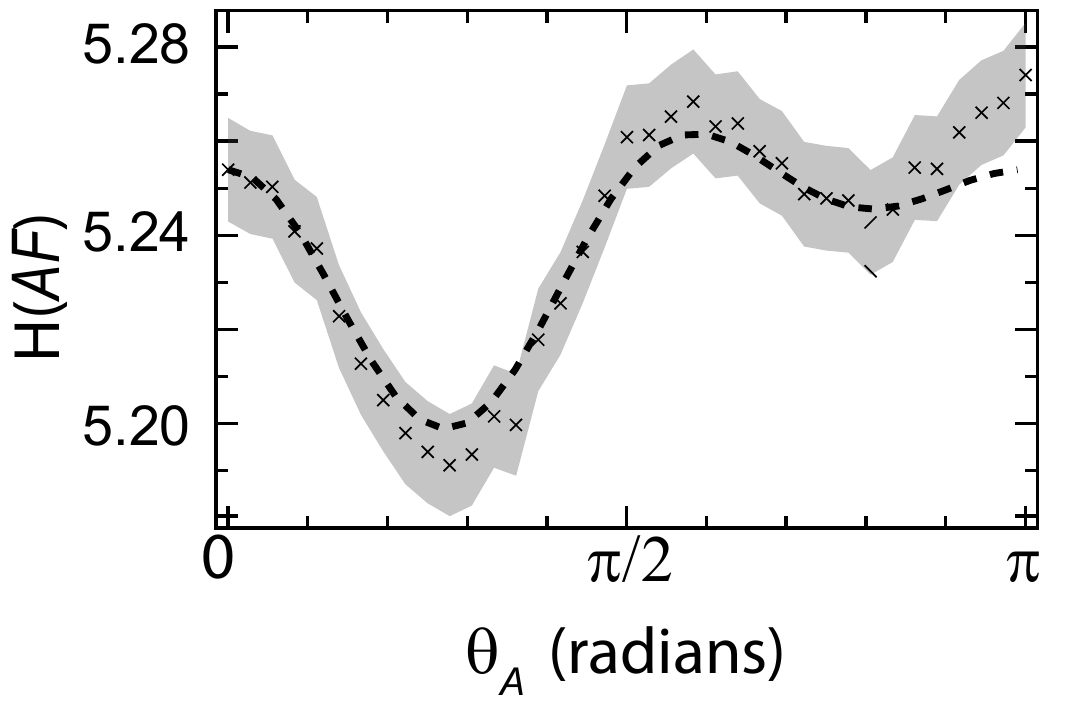}
	\caption[Entropy Under the Influence of Weak Measurement]{Entropy of the joint ${A}F$ measurement as a function of $\theta_{A}$ with the strong-measurement component fixed at $\sigma_x$. The prepared state is the $\ket{0}$ eigenstate of $\sigma_z$. 
		The dashed line indicates the prediction of a modified theory which accounts for measurement induced dephasing as well as $T_1$ decay. Dephasing enhances the dip at $\pi/4$ and $T_1$ decay raises the entropy with higher $\theta_{A}$.
		}
		\label{fig:EURLHSthetaW}
\end{figure}

From another perspective, with a preceding weak measurement, $\sigma_x$ no longer maximally disagrees with $\sigma_z$, despite the two operators' failure to commute. 
We now explore the amelioration of $\sigma_z$ and $\sigma_x$ by considering $H_\rho({A}F)$ as a function of $\theta_{A}$ with fixed $\theta_F=\pi/2$ in Figure \ref{fig:EURLHSthetaW}.

As before, the entropy at $\theta_{A}=0$ serves as a base case.
The value results from preparing $+z$, weakly measuring $\sigma_z$ then measuring $F=\sigma_x$.
Weakly measuring an eigenstate does not change the state, so the $F$ outcomes are unaffected.
The entropy is thus the combination of uncertainty from unbiased $F$ outcomes and the multiplicity of ${A}$ outcomes.

Remarkably, the entropy resulting from a weak measurement halfway between the two strong measurements ($\theta_{A}=\pi/4$) is lower than the unbiased value.
The decrease is due to the measurement-disturbance component of uncertainty relations.
Weak measurements cause slight backaction on the state. The state partially collapses towards one of the eigenstates of the measurement operator. 
The state dephases, and off-diagonal components (in the measurement basis) decay.

When $\theta_{A}=\pi/2$, the dephasing is maximal because the state is fully orthogonal to the measurement axis. 
However, as Figure \ref{fig:EURLHSthetaW} shows, the resulting entropy is not minimal.
This is because the dephasing does not significantly change the $\sigma_x$ component of the state.
There is no entropic advantage to ``pre-measuring'' the state. 

However, when $\theta_{A}=\pi/4$, the entropy is minimal.
Although this orientation does not maximize dephasing, the orientation does optimally shift the state towards $\sigma_x$ eigenstates.
%
This appears to contradict the results of a direct calculation of $H_\rho({A}F)$ which predicts that $\theta_{A}=\pi/4$ should be a \textit{maximum} in the entropy.
The direct calculation results in a subtle shift in the state.
However, when we estimate the probability by collecting an ensemble of measurements, inefficient detection plagues the distribution.
The ensemble-level dephasing overwhelms the subtle changes at the single trajectory level.
Thus, the dephased ensemble-level dynamics dominate the measurement bias.
Note also how Figure \ref{fig:EURLHSthetaW} exhibits skew with increasing $\theta_{A}$.
This asymmetry about $\theta_{A}=\pi/2$ is due to small but pronounced $T_1$ decay effects.


\begin{figure}
	\centering
	\includegraphics[width=0.7\linewidth]{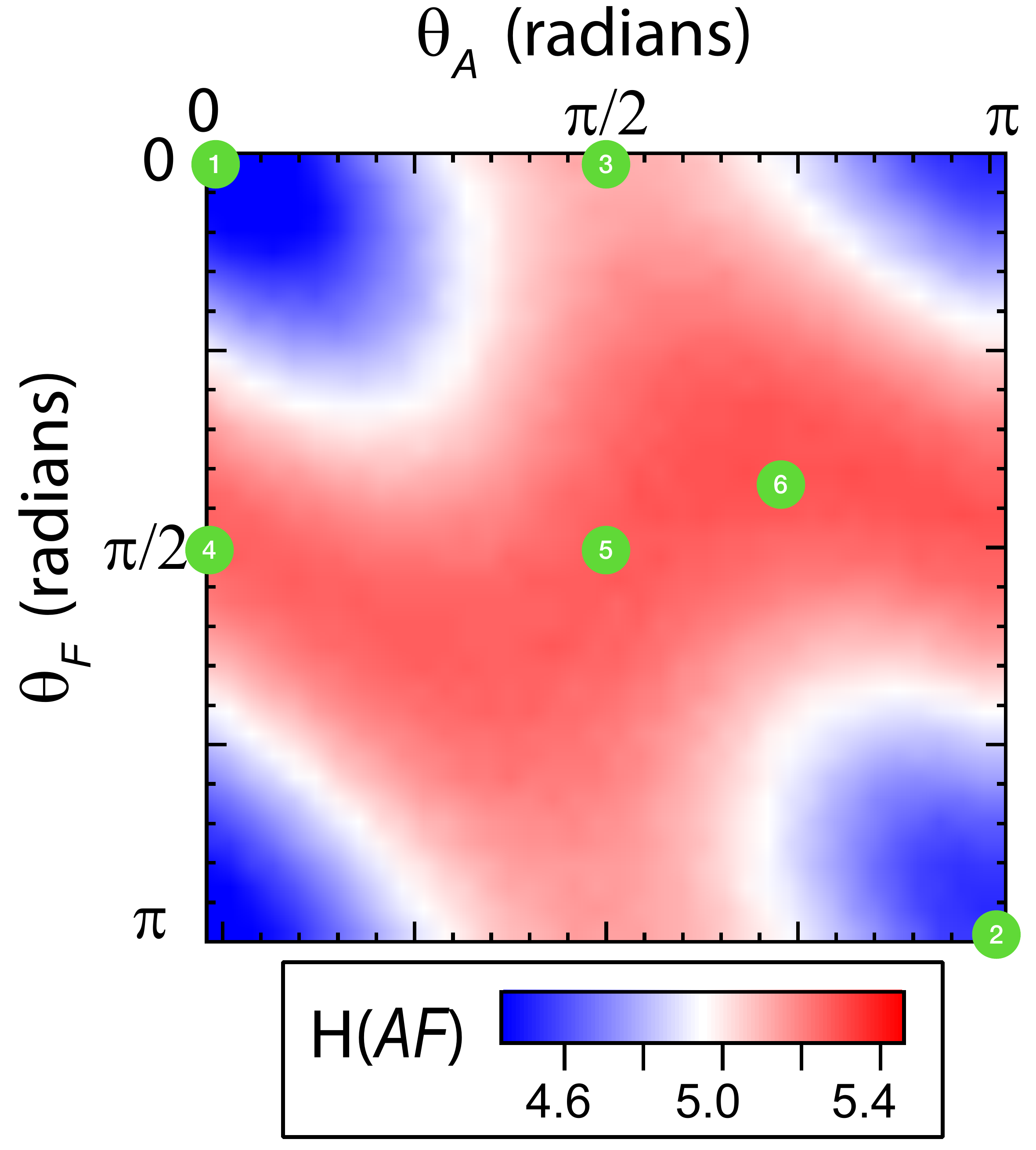}
	\caption[Entropies of the Entropic Uncertainty Relation]{The entropy of the ${A}F$ measurement outcomes as a function of the configuration angles for each measurement component. See Figure \ref{fig:eurangles} for definitions of the measurement angles. The minima in entropy, colored blue, appear when both measurement axes align with the state, initialized to a $\sigma_z$ eigenstate ($\theta_\rho$=0). See Table \ref{tab:HWFVals} for explicit values at marked points. Maxima in the entropy, colored red, occur when either the weak measurement ${A}$ or the strong measurement $F$ are oriented $\pi/2$ radians away from the initial state, due to these measurements being unbiased bases for the initial state.
	Across the parameter range, $H(\mathcal{I})_\rho$ contributes 0.01 bits.}
	\label{fig:eurLHSall}
\end{figure}

The above results for $H({A}F)_\rho$ as a function of the two joint-measurement configuration angles are a part of a larger parameter space.
Figure \ref{fig:eurLHSall} displays measured entropies for each angle \tw\ and $\theta_F \in \left[0,\pi\right]$. 
See Table \ref{tab:HWFVals} for specific values.
Broad features match expectations;
Minima in the entropy occur when both \tw\ and $\theta_F$ are either $0$ or $\pi$. This configuration involves $\sigma_z$ measurements (weak and strong) of a $\sigma_z$ eigenstate. 
The outcomes are minimally uncertain because each form of entropy is minimized: (i) the representation is efficient (ii) the preparation is not incompatible and (iii) no backaction occurs in either measurement.
Measurements with either component near $\theta=\pi$ tend to have slightly higher entropy due to the $T_1$-induced skew explained above.

Measurement configurations with maximum entropy occur near $\theta_F=\pi/2$ when the strong-measurement component disagrees maximally with the incompatible prepared state. Many measurement orientations result in statistically similar values of entropy. 
Surprisingly, this region of maximum entropy does not follow the line $\theta_F=\pi/2$ for all \tw. 
Instead, for each value of \tw, the maximum entropy occurs for a slightly different $\theta_F$.
At $\tw=0$, the maximum entropy occurs at $\theta_F=0$, but at $\tw=\pi/4$, the maximum entropy occurs at $\tw=0.66\pi$. 
The entropy-maximizing $\theta_F$ sweeps back and forth like a sinusoid as a function of \tw with an amplitude of 0.11$\pi$.
This result is an alternative way to describe Figure \ref{fig:EURLHSthetaW}: Varying \tw\ along $\theta_F=\pi/2$, the entropy decreases because the maximally disagreeing $F$ operator is no longer $\sigma_x$.
Although the absolute maximum occurs at $(\tw, \theta_F) = (0.69\pi, 0.42\pi)$, this appears to be a statistical fluctuation on top of $T_1$-induced skew.

\begin{table}
	\centering
\begin{tabular}{|c|c|c|c|c|}
	\hline 
	Index & $\theta_{A}$ & $\theta_F$ & $H({A}F)_\rho$[bits]  & Notes \\ 
	\hline 
	1& 0 & 0   &  4.30  & Absolute minimum \\
	\hline 
	2& $\pi$ & $\pi$   &  4.56  &Theoretical minimum, biased by T1 noise \\
	\hline 
	3& $\pi/2$ & 0   &  5.11 &  Disagreement with ${A}$ alone \\
	\hline 
	4& 0 & $\pi/2$   &  5.25 & Disagreement with $F$ alone\\
	\hline 
	5& $\pi/2$ & $\pi/2$   &  5.26 & Disagreement with both ${A}F$ \\ 
	\hline 
	6& 0.69$\pi$ & 0.42$\pi$   &  5.29  &	Absolute maximum \\
	\hline 
\end{tabular} 
\caption[Select values of $H({A}F)_\rho$]{Explicit values for the entropy $H({A}F)_\rho$ (Fig.~\ref{fig:eurLHSall}) for various points of interest in ($\theta_{A}$, $\theta_F$) parameter space. Each value has a typical uncertainty of 0.03 bits due to finite sampling statistics.}
\label{tab:HWFVals}
\end{table}

\subsection{Bound of the Entropic Uncertainty Relation}
\label{sec:RHSBound}
We have so far focused on the uncertainty component of our primary entropic uncertainty relation. Let us now turn to the bound. 
cQED allows nearly independent measurements of each component of the bound:
\begin{equation}
\min_{i,j,f} \left\{
- \log_2 \left\{ p_j\, p_{f|i} \right\} 
-\frac{2}{\ln{2}} \Re(g_j A_{\rm wv}) 
\right\}
\label{eqn:bound}
\end{equation}
Let us discuss each of the four non-constant components in turn.

\subsubsection{Measuring Conditional Probabilities}
\begin{figure}[h]
	\centering
	\includegraphics[width=0.7\linewidth]{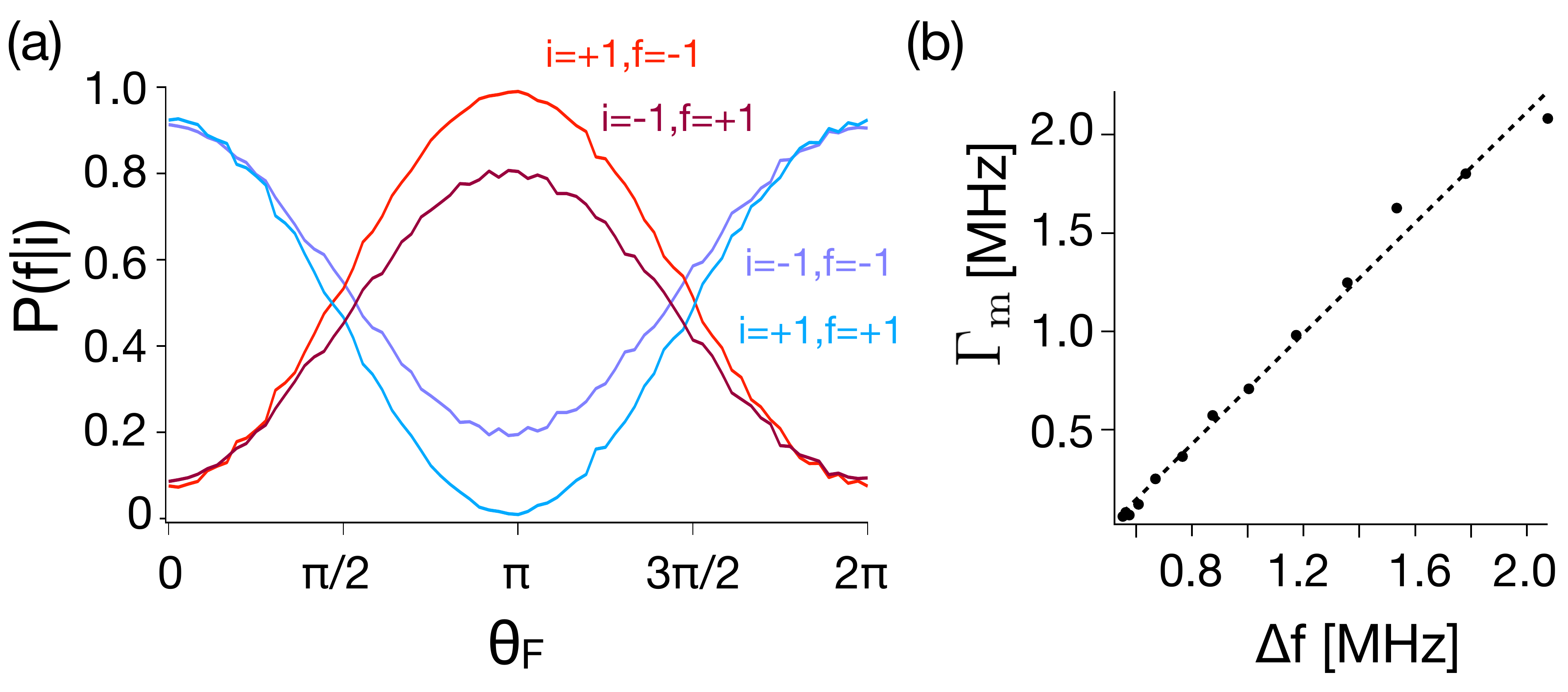}
	\caption[RHS Components]{Components of RHS (a) The conditional probability of measuring outcome $f$ from strongly measuring $\theta_F$ after preparing an $i$ eigenstate. Each choice of $i$ and $f$ is indicated. Red hues correspond to $i \neq f$ and blue hues correspond to $i=f$. Darker shades correspond to the case when $i$ is prepared as an excited state, leading to added $T_1$ decay. (b) Calibrating the measurement strength. By populating the cavity with $\bar{n}$ photons, we shift the qubit frequency by a fixed amount $\Delta f$, causing dephasing $\Gamma$. The slope is $\frac{4\chi}{\kappa}$, providing the measurement strength.}
	\label{fig:RHScomponents}
\end{figure}

We will start with the conditional probability $p_{f|i}$ of obtaining outcome $f$ given a preparation $\ket{i}$.
The measurement procedure consists of preparing either $\ket{0}$ or $\ket{1}$ (eigenstates of $\mathcal{I}$)
and measuring $F$ (with the pre-rotation discussion in Section \ref{sec:EURexp}). 
Thus, the conditional probability depends only on the final measurement angle, $\theta_F$.
The sequence is similar to Rabi oscillations \cite{Naghiloo2019}, modified by the initial $\pi$ pulse.
The results are shown in Figure \ref{fig:RHScomponents}(a).

\subsubsection{Measuring $p_j$ and $g_j$}
Now consider the components associated with the Taylor expansion of the weak measurement Kraus operator, $K_j$. The Taylor expansion produced Equation [\ref{eqn:KrausTaylor}] and provided expressions for $p_j$ and $g_j$ which depend only on the measurement strength.

To calibrate the measurement strength, we populate the cavity with photons and measure the qubit's $T_2^*$ decay time with a Ramsey sequence \cite{Naghiloo2019}.
The measurement  provides the ensemble dephasing rate, $\Gamma_\mathrm{m} = 8\chi^2 \bar{n}/\kappa$ and the ac Stark shift, $\Delta \omega_q = 2\chi \bar{n}$.
$\kappa/(2\pi) = 4.5$ MHz is the cavity linewidth measured with low power transmission via a vector network analyzer.
From these values, we infer the dispersive coupling rate, $\chi/(2\pi) = -1.5$ MHz, and the mean intracavity photon number during the weak measurement, $\bar{n} = 0.5$.
We integrate the weak signal for $\delta t=250$ ns, setting $\delta t/\tau = 0.375$.
From this we calculate 
$\sqrt{p_j} = 	 \left( \frac{ \delta t }{ 2 \pi \tau } \right)^{1/4}  
\exp \left( - \frac{ \delta t }{ 4 \tau }   \left[ j^2 + 1 \right]  \right)$ and
$g_j  =  \frac{ \delta t }{ 2 \tau } j $ for each value of $j$ in the span sampled during the LHS measurement.
See Figure \ref{fig:RHScomponents}(b).

\subsubsection{Measuring weak values}
\begin{figure}
	\centering
	\includegraphics[width=0.8\linewidth]{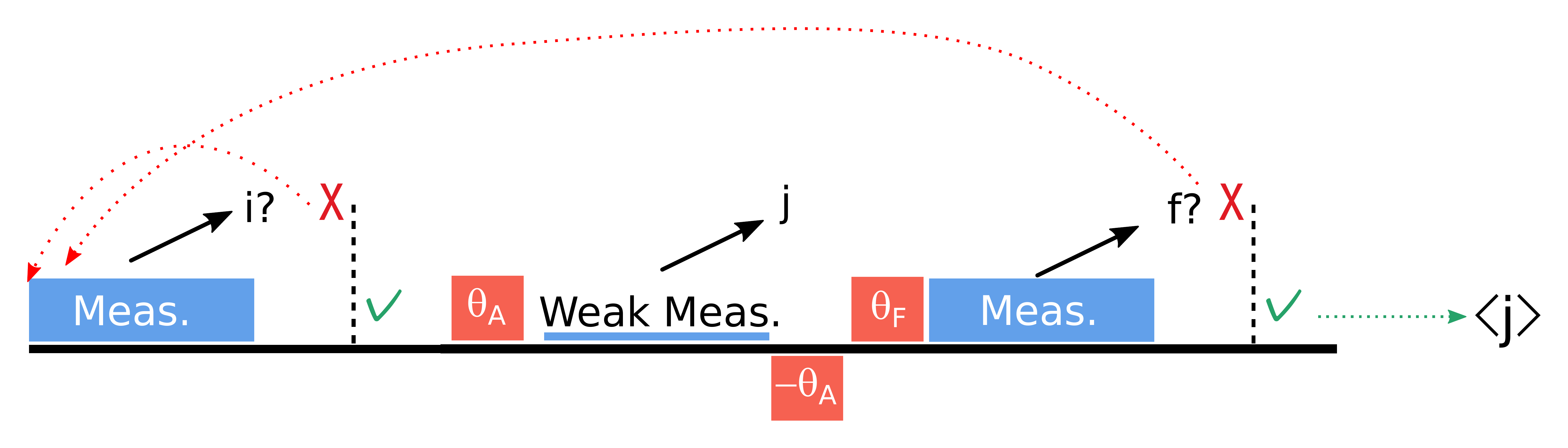}
	\caption[Weak Value Measurement Protocol]{To measure weak value $A_{\rm wv}^{i,f}$, the state $\ket{i}$ is prepared and measured. If the undesired outcome results, the run is discarded and begins again. A weak ${A}$ measurement results in outcome $j$. Then $F$ is strongly measured to obtain outcome $f$. If $f$ is not the desired outcome the run is discarded and begins again. If the desired outcome results, the outcome $j$ is included in a final average which estimates the weak value.}
	\label{fig:EURweakvalueprotocol}
\end{figure}

The weak value is a pre- and post-selected (conditional) expectation value \cite{Aharonov1988,Aharonov1989}:
\begin{equation}
A^{i,f}_\mathrm{wv}  = \frac{ \bra{f} A \ket{i} }{ \brakett{f}{i} },
\end{equation}
where we have explicitly denoted the $\mathcal{I}$ and $F$ outcomes, $i$ and $f$.
We experimentally probe the weak value through the protocol shown in Figure \ref{fig:EURweakvalueprotocol}.
The protocol is similar to measuring the LHS of the entropic uncertainty relation but differs because of post-selection.

Each choice of $i$ and $f$, corresponds to a different weak value.
As an example, consider the process for measuring the weak value for $i=-1$ and $f=+1$.
First, two counters for the numbers of total and successful trials initialize to zero.
After heralding the ground state, a $\pi$-pulse prepares the pre-selected $\mathcal{I}$ eigenstate, $\ket{1}$.
Fast rotations prepare the state for an ${A}$ measurement which results in outcome $j$.
A final set of rotations prepare the state for the strong $F$ measurement. The number of total trials increments.
For the case of measuring $A^{-1,+1}_{\rm wv}$, if the $F$ measurement results in $f=-1$, the outcome is discarded and the sequence repeats.
If the $F$ measurement results in $f=+1$, then the success counter increments and the value of $j$ is appended to a list of successful outcomes.
After a sufficient number of trials, the average value of successful $j$ outcomes is divided by the fraction of successful trials, producing the weak value.


\begin{figure}
	\centering
	\includegraphics[width=0.45\linewidth]{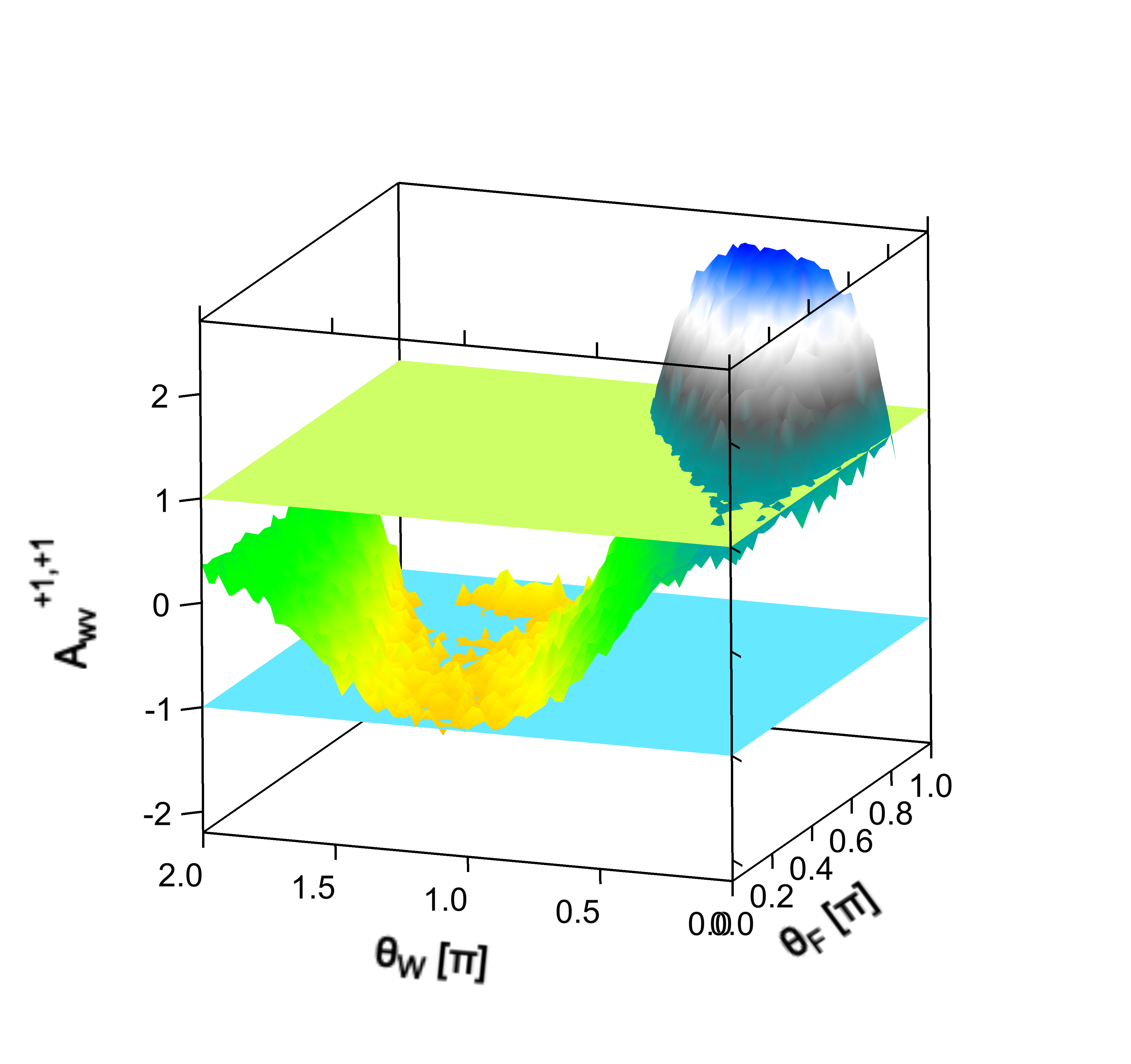}
	\includegraphics[width=0.45\linewidth]{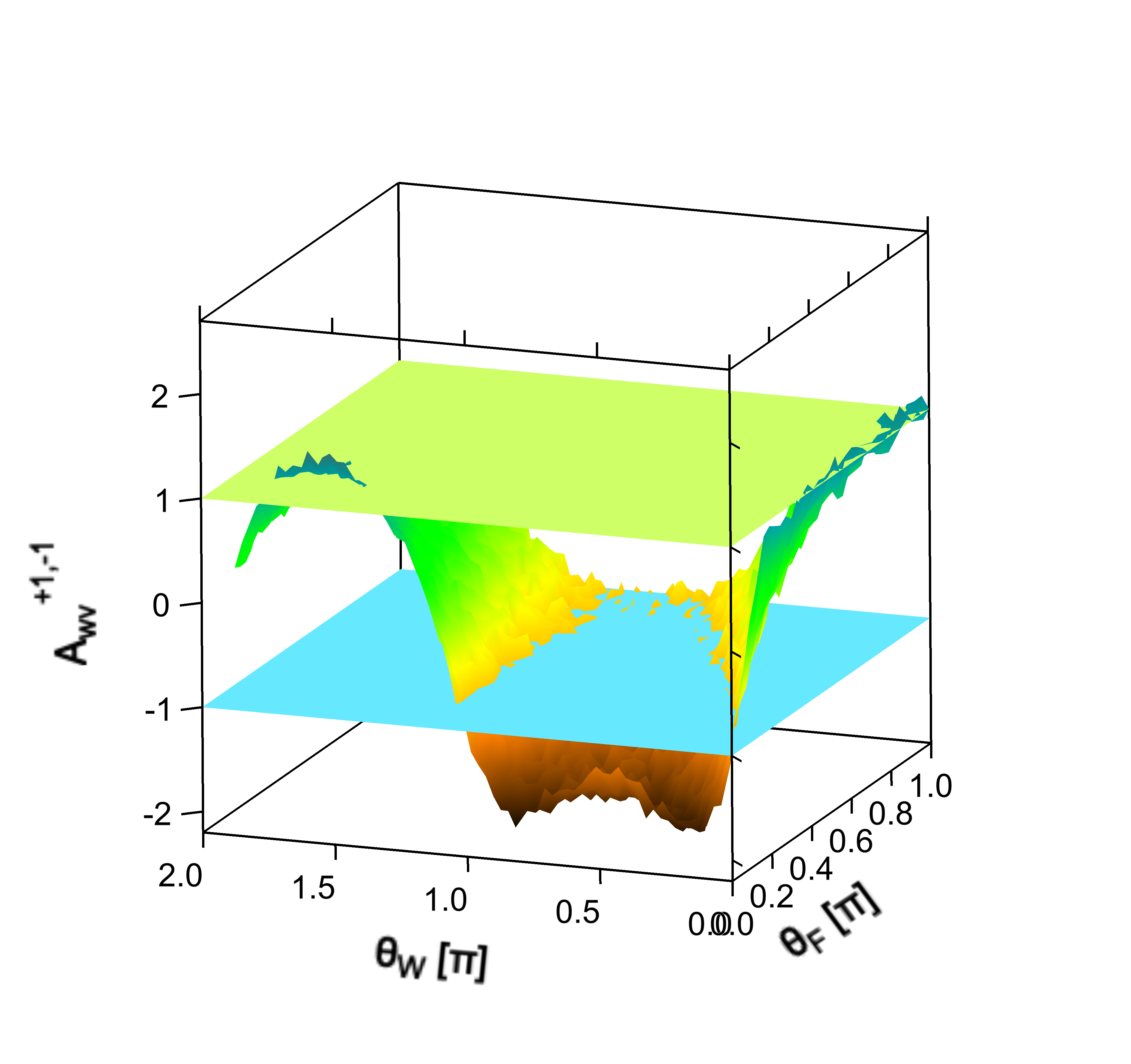}
	\caption[Weak Values]{The weak value, $A_{\rm wv}^{i,f}$, for two choices of $i$ and $f$ across a span of ${A}F$ compositions. Outside of $\theta_F \in \left\{\pi/6, 5\pi/6\right\}$, the sampling success probability for one of  these weak values becomes prohibitively low. The weak value is expected to show similar behavior for $\theta_F>\pi$.
	Pastel blue and green planes indicate the maximum expectation value for Pauli operators. Weak values beyond these are anomalous.
	\label{fig:RHSWeakVal}
}
\end{figure}

Weak value averaging depends on successful measurement outcomes.
The success probability depends on the overlap between $\ket{f}$ and $\ket{i}$ and is equal to $\frac{1}{2}\left(1+\cos\theta_F\right)$. Dephasing due to the ${A}$ measurement can slightly improve this probability, but the $\theta_F$ dependence dominates.
When $\theta_F$ takes on values outside the range $\left[\pi/6,5\pi/6\right]$, the post-selection success probability becomes too small to measure accurately in a reasonable amount of time. Fortunately, this range contains the point $\theta_F=\pi/2$, which is a primary point of interest for the uncertainty relation.
In order to expedite measurements, we only measure $A^{+1,+1}_{\rm wv}$ and $A^{+1,-1}_{\rm wv}$, using the fact that $A^{+1,+1}_{\rm wv} = -A^{-1,-1}_{\rm wv}$. Figure \ref{fig:RHSWeakVal} displays the results of these two measurements.

Many of our weak values are anomalous. The maximally anomalous weak value is 2.7, well beyond 1---the maximal value for an expectation value of a Pauli operator.
More anomalous weak values do not play a significant role in the uncertainty relation's bound because they violate the Taylor approximation assumed in the bound (see Equations [\ref{eqn:deriveRHS}]).

Some of our measured weak values display apparent suppression of anomalous weak values (see especially $A^{+1,-1}_{\rm wv}$ in Figure \ref{fig:RHSWeakVal}).  
We attribute this to $T_1$ effects.
$T_1$ jumps are rare because the measurement time (250 ns) is much shorter than the $T_1$ time (50 $\mu$s).
However, anomalous weak values require sampling rare events, namely preparation of one eigenstate and measurement of a nearly orthogonal eigenstate. 
While the probability of such a nearly-orthogonal projection is low, $T_1$ decay increases the probability.

\subsection{Combing Elements of the RHS}
\begin{figure}
	\centering
	\includegraphics[width=0.7\linewidth]{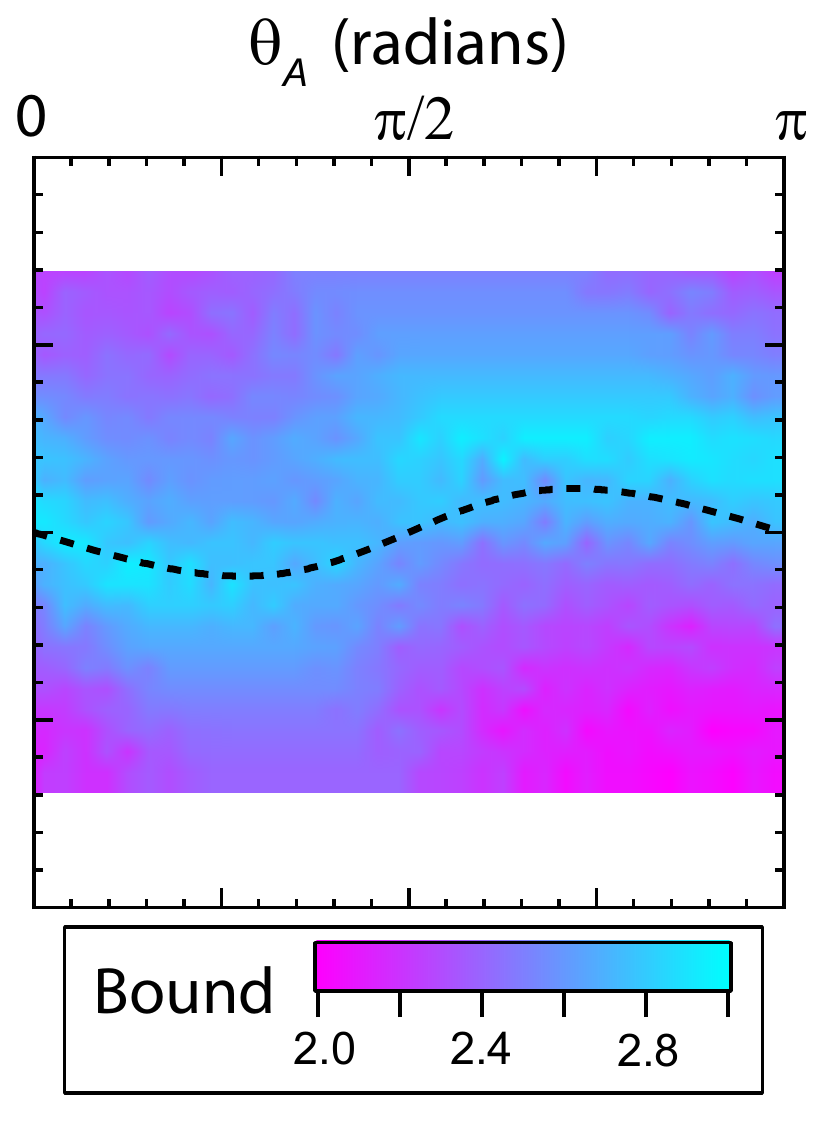}
	\caption[RHS of Entropic Uncertainty Relation]{The bound for the sum of entropies for our two POVMs. The theoretical maximum of the bound is indicated with the black dashed curve.}
	\label{fig:EURRHS}
\end{figure}

We now have each element to construct the argument for the minimization over $i$,$j$, and $f$ in the RHS of Inequality [\ref{eqn:EUR}].
Figure \ref{fig:EURRHS} shows the result of the minimization for each configuration of ${A}F$.
This bound is always lower than the measured entropy, as expected. 

Consider how well the bound does in predicting the entropy.
The bound's tightness, the $\mathrm{LHS - RHS}$, is maximal throughout a set of orientations near $\theta_F = \pi / 2$.
Here, the tightness is $2.45 \pm 0.05$ bits. 
Theoretically, the tightness is 0.7 bits, but inefficient detection ($\eta=10\%$) raises the entropy sum's by 1.66 bits.

The bound follows a similar qualitative shape as the LHS. The maximum of the bound is near $\theta_F=\pi/2$, indicating nearly maximal disagreement between $\mathcal{I}$ and $\sigma_x$. However, the maximum shifts sinusoidally with \tw. 
For example, when $\tw = \pi/4$, the maximally disagreeing ${A}F$ measurement has $\theta_F=0.53\pi$, when the measurement strength is $\delta t/\tau = 0.17$.
When $\theta_F=\pi/2$, setting $\tw$ to $\pi/4$ reconciles disagreeing operators, $\sigma_z$ and $\sigma_x$.
Phrased alternatively, for fixed $\theta_F=\pi/2$, the bound decreases as \tw increases from 0 to $\pi/4$.

The weak value decreases the bound.
The second term of the bound enters with a negative sign. The signs of $A_{\rm wv}$ and $g_j$ correlate so that the sign of their product is positive.
Hence, the weak value term decreases the bound, reconciling incompatible observables.


\chapter{Time-Reversed Feedback}	
\section{Introduction}
In this chapter, we investigate the origin of irreversible dynamics from reversible processes.
This is Loschmidt's paradox. We study single quantum trajectories and compare forward and backward path probabilities to infer a statistical arrow of time quantified by an entropic measure. Moreover, we implement post-selected feedback protocols wherein the extraction and subsequent use of information appear to set a definite direction for the arrow of time. Though causal-order feedback does not affect entropy production, a reversed causal order feedback protocol reverses the flow of entropy, annihilating rather than creating entropy, revealing the essential role of causality in the direction of the arrow of time.

Standard dynamics of quantum systems, as governed by Schr\"odinger's equation, are time-reversal invariant. Taking $t\rightarrow -t$ and taking the complex conjugate leaves the Schr\"odinger equation unaffected:
\begin{equation}\label{eqn:TRSchro}
	\begin{split}
	i\hbar \frac{\partial}{\partial t}\psi(t) &= H\psi(t), \\
	\rightarrow ~(-)i\hbar\frac{\partial}{\partial (-t)}\psi(t) &= H\psi (t),\\
	\rightarrow ~ i\hbar \frac{\partial}{\partial t}\psi(t) &= H\psi(t). 
	\end{split}
\end{equation}  
While under such dynamics the von Neumann entropy 
is constant, many quantum mechanical systems allow for the gain or loss of information corresponding to changes in the von Neumann entropy. For example, measurement is a process that can change a system's entropy.
When a system starts in a mixed state (maximal entropy), measurement projects the system onto a pure eigenstate, resulting in zero entropy.

On the other hand,
consider measuring $\sigma_x$ of the state $\rho=\ket{0}\bra{0}$. 
The state is initially pure, with no entropy. But upon projection onto a $\sigma_x$ eigenvector, the state's information (Shannon) entropy in the original basis has increased.

Each of these phenomena displays entropy production or annihilation. However, systems described by unitary (i.e. norm-preserving) dynamics can have no such entropy change. This is the quantum analog of Loschmidt's paradox wherein phase-space volume conservation (as guaranteed by Liouville's theorem) seemingly cannot produce phase space enlarging processes, as demanded by the laws of thermodynamics.

Changes in entropy arise when subsystems are traced over, yet due to their presence in an inaccessible environment, or as a measurement apparatus \cite{Manikandan2018}.
Such an \textit{open} quantum system allows for entropy-producing evolution \cite{Alicki2012}. 
Thus,  open quantum systems can have stochastic evolution which obeys more nuanced laws of entropy production \cite{Campisi2011a, Camati2018,Mancino2018, Holmes2018}. Rather than satisfying Clausius' maxim that thermodynamic entropy always increases, statistical entropy of open quantum systems can fluctuate between positive and negative values.

A stochastic version of the second law of thermodynamics for classical systems resolved Loschmidt's paradox \cite{Evans1994, Lebowitz1994, Campisi2011}.
The law is expressed as a fluctuation theorem that describes the statistics of time-ordered and time-reversed dynamics \cite{Sevick2008, Deffner2011, Uzdin2016}.
Such fluctuation theorems have been successfully measured in a variety of classical systems \cite{Wang2002, Carberry2004, Risler2015, Kalziqi2018}, as well as quantum systems \cite{Hofmann2017, Brunelli2018,Harrington2019}.
	
\section{Stochastic Thermodynamics}
The expectation of time-irreversible dynamics, especially entropy production, is firmly grounded in thermodynamics.
However, for small systems, such as particles undergoing Brownian motion, standard thermodynamics begin to break down. 
As thermodynamic variables describe more strongly stochastic processes, common notions like positive entropy production are challenged. These quantities fluctuate, and can occasionally take on negative values.

	\subsection{Classical Fluctuation Theorems}
	\label{sec:deriveFT}

	We begin with a classical treatment of stochastic thermodynamics, following the development of Ref.~\cite{VanDenBroeck2013}.
	Consider a system and a reservoir at temperature $T$.
	We wish to describe the dynamics of the system while interacting with the reservoir. The state is indexed by $m$, and the probability of occupying state $m$ follows one of the standard ensembles.
	We suppose that the master equation is Markovian and has the following form:
	\begin{equation}\label{eqn:masterEquation}
	\dot{p}_m = \sum_{m'} H_{m',m} \, p_m'
	\end{equation}
	where $H_{m,m'}$ is the transition matrix element from state $m'$ to $m$.
	In the case of equilibrium with the bath, the transition matrix obeys detailed balance:
		\begin{equation}\label{eqn:detailedBalance}
	H_{m,m'} \, p^{\rm eq}_{m'} = H_{m',m} \, p^{\rm eq}_m \,,
	\end{equation}
	where $p^{\rm eq}_m$ is the equilibrium distribution for the particular ensemble.
	
	We focus on the stochastic dynamics of a single trajectory, emphasizing its thermodynamics.
	The interaction with the reservoir induces jumps in the trajectory (which we will interpret as heat).
	The resulting discontinuities break up the trajectory into sub-intervals of smooth, unitary evolution (during which work is performed).
		We specify the interval after jump $j$ with a single index $m_j(t)$, which depends on time, $t\in[t_j,t_{j+1})$.
	
	Because energy is a state function, the total change in the energy across the trajectory is unaffected by the discontinuities. The total change in energy only depends on the final and initial states.
	We use this fact to define the first law of thermodynamics for the stochastic trajectory.
	The energy is composed of two path-dependent quantities, which we call heat, $\mathcal{Q}$ and work, $\mathcal{W}$:
	\begin{equation}\label{eqn:firstLaw}
		\Delta E  = \mathcal{Q} +\mathcal{W}.
	\end{equation}
	
	Following common approaches in quantum thermodynamics cite{Naghiloo2020}, we associate work with the smooth evolution (due to energy level shifts) and heat with the jump evolution (due to state changes).
	Within the smooth interval, the work becomes a state variable of the trajectory (a microstate) and can be calculated as the energy difference at the interval's endpoints.
	\begin{equation}\label{eqn:work}
		\mathcal{W} = \left[ E_{m_0(t_1) } - E_{m_0(t_0)} \right]
			 + \ldots + \left[ E_{m_N(t_f) } - E_{m_N(t_{N})} \right].
	\end{equation}
	The heat describes the energy related to jumps:
	\begin{equation}\label{eqn:heat}
		\mathcal{Q} = \sum_j \left[E_{m_j(t_j)} - E_{m_{j-1}(t_j)}\right] .
	\end{equation}

	Entropy is often thought of as an ensemble property.
	However, with the ability to calculate the probability of a single trajectory, we can define a microstate entropy via the surprisal, similarly to Section \ref{sec:entropy}.
	This definition of entropy retains its state-variable characteristic \cite{Seifert2005,Seifert2012}, and so it only depends on the initial and final states:
	\begin{equation}\label{eqn:fullEntropy}
		\Delta \mathcal{S} = -k \left( \ln P[{m_f}] - \ln P[{m_i}] \right),
	\end{equation}
	where $k$ is Boltzmann's constant.
	
	We decompose this entropy into two components:
	\begin{equation}\label{eqn:entropyComponents}
	\Delta \mathcal{S} = \Delta \mathcal{S}_r + \Delta \mathcal{S}_i.
	\end{equation}
	The first component, $\Delta \mathcal{S}_r$, describes reversible entropy flow via heat exchange with the reservoir.  
	It comes from the Clausius definition:
	\begin{equation}\label{eqn:entropyE}
	\Delta \mathcal{S}_r = \frac{\mathcal{Q}}{T}
	\end{equation}
	The second component, $\Delta \mathcal{S}_i$ describes irreversible entropy production and will relate to forward and backward path probabilities.

	To see this, consider the probability of the forward path, $P_F$.
	The total probability is given by the product of each independent step.
	The probability of the path to (i) start at the initial state, (ii) evolve for the interval, (iii) jump, (iv) evolve, (v) jump, etc.~is:
	\begin{equation}\label{eqn:chainProb}
		P_F = P[m_0] \times P[m_0(t_0)\rightarrow m_0(t_1)] \times P[m_0(t_1) \rightarrow m_1(t_1)] \times \ldots \times P[m_f].
	\end{equation}
	
	The actual calculation would be incredibly tedious, but we can greatly simplify it by considering groups of terms in turn.
	First, the jump probabilities are simply given by the specified $H_{m_{j+1},m_j}$ of the master equation.
	Second, consider the no-jump probabilities for a single interval.
	The probability of each smooth interval $[t_j,t_{j+1}]$ can be calculated as the probability of having no jumps, for every infinitesimal interval $dt\in[t_{j-1},t_{j}]$, $1-H_{m_{j+dt},m_j}$.
	Integrating the interval results in a path integral \cite{VanDenBroeck2013}. 
	Fortunately, we do not need to perform this calculation because this group of no-jump factors cancels in the forward and reverse path probabilities \cite{Crooks1999}.
	The final group contains the boundary terms, which are given by the ensemble distribution.
	
	Turning to the backward path probability, $P_B$, we follow the same prescription. The jumps now occur with probability $H_{m_j, m_{j+1}}$ (indices inverted). The smooth intervals have the same probability as in the reversed case.
	This treatment produces an expression analogous to Equation [\ref{eqn:chainProb}] for $P_B$.
	
	We can then calculate the log-ratio of the forward-to-backward path probabilities:
	\begin{equation}\label{eqn:pfpbEntropy}
	\ln \frac{P_F}{P_B} = \ln P[m_1] + \sum_j \ln\frac{H_{m_j,m_{j+1}}}{H_{m_{j+1},m_j}} - \ln P[m_N].
	\end{equation}
	We now applied detailed balance to finish the proof.
	From Equation [\ref{eqn:detailedBalance}], we have:
	\begin{equation}\label{eqn:getSrev}
	\sum_j\ln\frac{H_{m_j, m_{j+1}}}{H_{m_{j+1},m_j}}  
	= \sum_j \ln\frac{P^{\rm eq}_{m_j}}{P^{\rm eq}_{m_{j+1}}} 
	=  - \beta \mathcal{Q} 
	 = - \frac{ \Delta \mathcal{S}_r}{ k }.
	\end{equation}
	Substituting expressions for the total and irreversible entropies (Eqns. [\ref{eqn:fullEntropy}] and [\ref{eqn:getSrev}]) into Equation [\ref{eqn:pfpbEntropy}], we can see:
	\begin{equation}\label{eqn:combineEntropies}
		k \ln \frac{P_F}{P_B} =	\Delta \mathcal{S} - \Delta \mathcal{S}_r.
	\end{equation}
	Thus, we identify the irreversible entropy production in terms for forward and backward path probabilities:
	\begin{equation}\label{eqn:logRclassical}
	\boxed{
	\ln \frac{P_F}{P_B} = \frac{\Delta \mathcal{S}_i	}{k}
	}	
	\end{equation}
	From this derivation, especially at Equation [\ref{eqn:pfpbEntropy}], we can see why $\Delta \mathcal{S}_r$ and $\Delta \mathcal{S}_i$ have been called boundary and bulk terms, respectively \cite{Seifert2005,Smerlak2012, Seifert2012}.
	The reversible entropy flow relates to the final and initial state of the trajectory, while the irreversible entropy production relates to the intermediate evolution.
	
	More explicit statistical mechanical calculations based on reversing phase-space trajectories provide the same result \cite{evans2016fundamentals}.
	Moreover, it can be proved based solely on results of measure theory  \cite{Shargel2010}.
	From this expression for the entropy production in terms of forward and reversed path probabilities, we can derive several additional stochastic thermodynamic relations.

	\subsubsection*{Detailed Fluctuation Theorem}
	The detailed fluctuation theorem describes the probability that the entropy production takes on some specific value, $P(\Delta\mathcal{S}_i = \mathcal{P})$.
	The theorem takes the form of a ratio between the probabilities of $\pm \mathcal{P}$. 
	We use the Dirac delta functional%
	\footnote{This object is called a functional because unlike a function which maps a number to a number, this object maps a function to a number (namely the function's value at zero).}
	 to select trajectories that create entropy $\mathcal{P}$:
	\begin{equation}
	\label{eqn:FT}
	\begin{split}
	P(\Delta\mathcal{S}_i=\mathcal{P} ) &= \sum_{\rm traj.} P_F \,\delta\left[ \mathcal{P}- k \ln \frac{P_F}{P_B}\right]\\
			&= e^{\mathcal{P}} \sum_{\rm traj.}  P_B  \,\delta\left[ \mathcal{P} - k \ln \frac{P_F}{P_B}\right]\\	
			&= e^{\mathcal{P}} \sum_{\rm traj.}  P_B  \,\delta\left[ -\mathcal{P} - k \ln \frac{P_B}{P_F}\right]\\	
			&= e^{\mathcal{P}} P(\Delta\mathcal{S}_i=-\mathcal{P}) \\			
	\end{split}
	\end{equation}
	Thus, the ratio of probabilities for trajectories producing an irreversible entropy of $\pm\mathcal{P}$ is equal to $e^\mathcal{P}$.

	This expression represents one of a variety of generalized second laws of thermodynamics \cite{Brandao2013}, applied to stochastic systems.
	It is an exact equality that characterizes much more than the narrow (and sometimes false) statement about the monotonic increase of entropy. It not only allows for negative entropy fluctuations but requires at least small negative entropy fluctuations to balance the ratio. 
	These negative fluctuations can be thought of as a state update procedure becoming less certain about the actual state \cite{Bartolotta2016}.
	The apparent absence of observed entropy annihilation arises from the exponential suppression of backward trajectories. 
	Because of its role in the time-reversibility of dynamics, such adherence to fluctuation theorems is often an important characteristic of arrow-of-time statistics \cite{Lebowitz1994, Campisi2011}.

	\subsubsection*{Integral Fluctuation Theorem}
	We also obtain the integral fluctuation theorem. This theorem returns us to the ensemble level of traditional thermodynamics.
	We calculate an average over an ensemble of trajectories.
	\begin{equation}\label{eqn:IFT}
	\left\langle e^{-\Delta\mathcal{S}_i/k} \right\rangle = \left\langle e^{\ln\frac{P_B}{P_F}} \right\rangle = \left\langle\frac{P_B}{P_F}\right\rangle
	= \sum_{\rm traj.} P_F \frac{P_B}{P_F}=1\,.
	\end{equation}
	
	\subsubsection*{Jarzynski's Equality and the Second Law}
	In the special case that the trajectory begins and ends in equilibrium states, we can also derive Jarzynski's equality.
	In this case, the entropy production obeys $T \Delta\mathcal{S}_i = \mathcal{W} - \Delta F^{\rm eq}$.
	So the integral fluctuation theorem becomes:
	\begin{equation}\label{eqn:Jarzynski}
		\begin{split}
		\left\langle e^{-\Delta\mathcal{S}_i/k} \right\rangle & =1 \\
		\left\langle e^{-\beta \mathcal{W} \, +\beta\Delta F} \right\rangle & = 1\\
		\left\langle e^{-\beta \mathcal{W}}  \right\rangle &= e^{-\beta\Delta F}.
		\end{split}
	\end{equation}
	This seminal equality has also been extended to include measurement and feedback \cite{Sagawa2008, Sagawa2010, Wachtler2016}.

	The more familiar second law about entropy increase also follows from the integral fluctuation theorem:
	\begin{equation}\label{eqn:standardSecondLaw}
	\left\langle\Delta\mathcal{S}_i \right\rangle
	 = \left\langle \ln e^{\Delta\mathcal{S}_i} \right\rangle
	 \geq  \ln \left\langle e^{\Delta\mathcal{S}_i} \right\rangle
	 = \ln 1 =0.
	\end{equation}
	The inequality comes from an application of Jensen's inequality which applies to any convex function $f$: 
	$f(\bar{x}) \leq \bar{f}(x)$, where, for the sake of clarity, averages are indicated with bars.

	\subsection{Thermodynamics of Quantum Trajectories}
	We have shown in Section \textit{1.3.2} how to update a state based on the stochastic outcome of a partial measurement.
	Based on these trajectories, we can calculate forward and backward path probabilities, allowing for a characterization of the entropy production (and annihilation) in open quantum systems.

	Reversibility is a key tenet for the calculation in Section \ref{sec:deriveFT}. The reversibility of quantum measurement dynamics is enabled by the positivity of the POVM mapping \cite{quantumMeasurement}. For a measurement of non-zero strength, some backaction will be imparted on the system. Measuring outcome $j$ changes the initial state $\rho_0$ to $\rho_1 \propto K_{j}\rho_0 K_j^\dagger$ via the Kraus operator, $K_{j}$ (see Sect.~\ref{sec:partialMeasThy}). This update protocol is reversible since we may apply a time-reversed measurement that undoes the backaction. The operator $\Theta$ produces the reversing measurement: $\tilde{K}_{j} = \Theta K_{j} \Theta^\dagger$. The reversed measurement operator reverses a single step in the trajectory:
	\begin{equation}
		\begin{split}
		\tilde{\rho}_0  & \propto \Mtr	 \tilde{\rho}_1 					\MDtr \\
		& \propto K_{-j}  \T   \M \rho_0 \MD   \TD 			K_{-j}^\dagger\\
		&\propto \T\rho_0\TD,
		\end{split}
	\end{equation}
	where we have used the fact that $\Theta\Theta^\dagger=1$.
	We applied the so-called ``passive'' transformation by negating the measurement record measurements \cite{Dressel2016}.
	 This reduces to $\tilde{\rho}_0 = \T\rho_0\TD$, the time-reversed initial state. Thus, because we track the individual record steps, $j$, we can apply time-reversed measurement operators to reverse the dynamics of measurement.
	 
		With individual, reversible quantum trajectories we may proceed with calculating an irreversible entropy production.
		We define:
		\begin{equation}
		\mathcal{Q} \equiv \ln \frac{P_F}{P_B}
		\end{equation}
		where $P_F$ and $P_B$ are the probabilities of a forward or reversed trajectory, respectively. 
		This definition of entropy has offered insights into a statistical arrow of time \cite{Dressel2016,Harrington2019}.
		Similar to Equation [\ref{eqn:logRclassical}], $\mathcal{Q}$ obeys a detailed fluctuation theorem:
		\begin{equation}
		\frac{P(+\mathcal{Q})}{P(-\mathcal{Q})} = e^\mathcal{Q}.
		\label{eqn:quantumFT}
		\end{equation}
		The derivation of this equation is more difficult to the lack of clear definitions of state variables and measurements \cite{Alhambra2016, Aberg2018}.
	
		After introducing the experiment in Section \ref{sec:exp}, we will describe the calculating of probabilities for these forward and reversed trajectories in Section \ref{sec:trajProb} and verify the detailed fluctuation theorem in Section \ref{sec:expFT}

\section{Experiment}
\label{sec:exp}
\subsection{Setup}
	The experiment consists of a qubit coupled to a single mode of the electromagnetic field. 
	The lowest two levels of a superconducting transmon circuit form the qubit, and a 3D aluminum cavity stores the single EM mode.
	A dispersive qubit-cavity interaction allows weak measurements of the qubit.
	 The qubit's state is inferred from coherent states transmitted through the cavity. Transmitted signals receive a qubit-state-dependent frequency shift according to the dispersive Jaynes-Cummings Hamiltonian:
	 \begin{equation}
	 	H_{\rm JC}/\hbar = \frac{\omega_q}{2}\sigma_z + a^\dagger a(\omega_c + \chi \sigma_z).
	 	\label{eqn:JC}
	 \end{equation}
	The dispersive interaction, $H_{\mathrm{int}}/\hbar = \chi a^\dagger a \sigma_z$ 
	has been interpreted as a qubit-state-dependent frequency shift on the cavity. 
	
	\begin{figure}[H]
		\includegraphics[width=0.7 \linewidth]{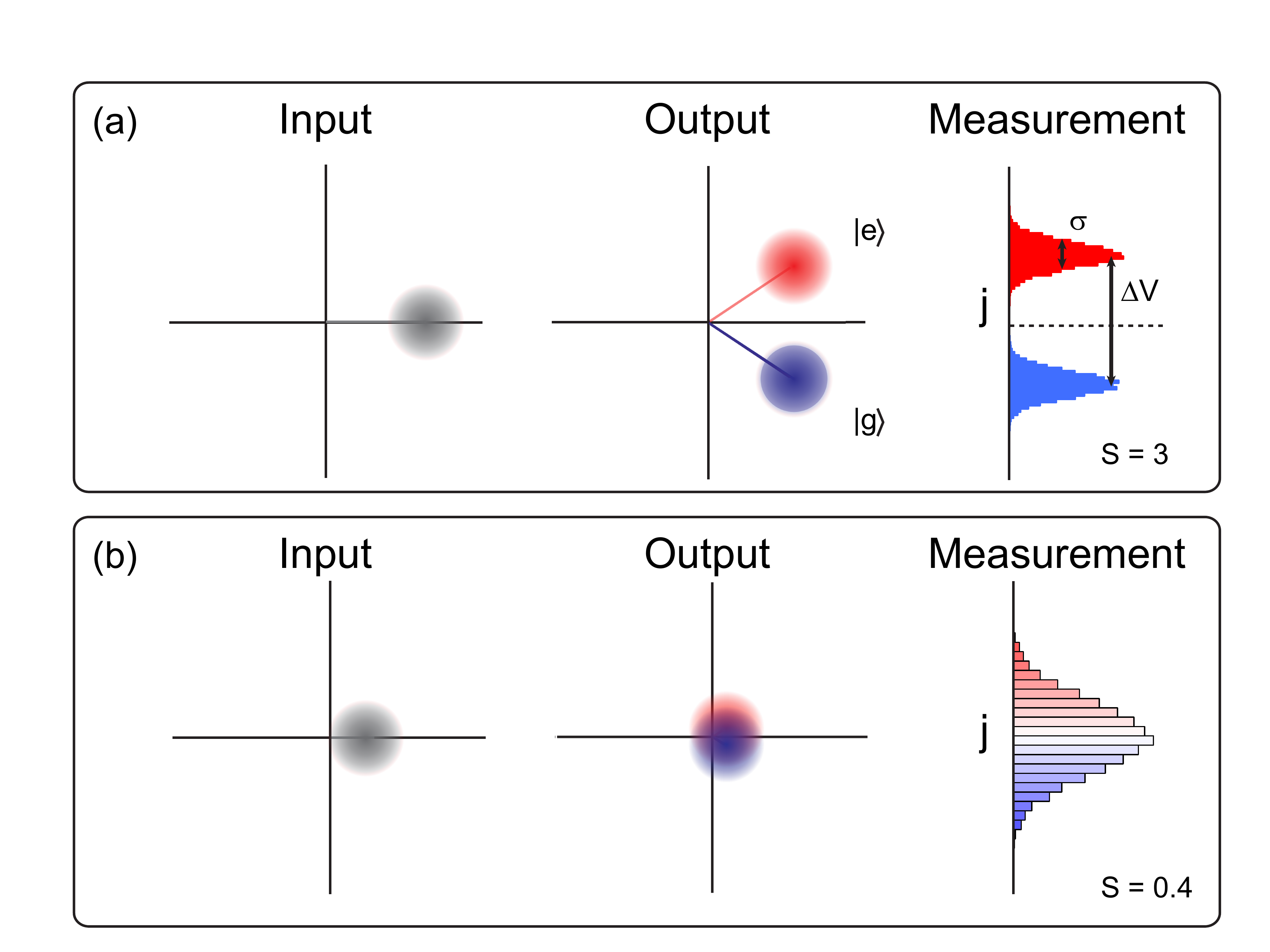}
		\centering
		\caption[Readout Distributions]{Our readout scheme allows both strong and weak measurements. \textbf{(a)} For strong measurements, a coherent tone sent into the cavity receives a qubit-state-dependent phase shift resulting in two well-separated Gaussian distributions ($S$=3). \textbf{(b)} Conversely, we implement weak measurements by decreasing the intra-cavity photon number so that the qubit states do not resolve ($S$=0.4). 
		} 
		\label{fig:readout}
	\end{figure}
	
	Readout consists of homodyne monitoring of the field's quadrature component.
	The outcomes are Gaussian-distributed with a variance of $\sigma^2$ and a mean which shifts by $\Delta V$ depending on the qubit's state, as shown in Figure \ref{fig:readout}.
	The separation parameter, $S\equiv \frac{\Delta V^2}{\sigma^2}$ 
	can be expressed in experimental parameters \cite{Murch2013}:
	\begin{equation}
		S= \frac{8\, \delta t \, \eta }{\tau}.
	\label{eqn:S}
	\end{equation}
	The measurement duration is $\delta t= 145$ ns, the characteristic measurement strength is $\frac{1}{\tau} = \frac{8 \chi^2 \bar{n} \eta}{\kappa} = 2\pi \times 2.54$ MHz, and the quantum efficiency is $\eta$=0.31.
	The measurement strength depends on the dispersive shift, $\chi/2\pi=-0.25$ MHz, the variable intra-cavity photon number, $\bar{n}$, and the cavity linewidth $\kappa/2\pi=3.37$ MHz.
	 When $S=3$, ground- and excited-state preparations result in significantly different homodyne voltages  [Fig.~\ref{fig:readout}(a)]. Decreasing the intra-cavity photon number, so that $S$ = 0.4, implements a weak measurement. The weak measurement does not completely distinguish ground from excited qubit states [Fig.~\ref{fig:readout}(b)], thus inducing small backaction on the state.
	
	The Kraus operator for measurement update (see Sec. 1.3.2) is given by 
	\begin{equation}
	\M = \sqrt[4]{\frac{\delta t}{2\pi \tau}} \exp[-(j-\sigma_z)^2 \delta t/2 \tau].
	\end{equation}
%
	 From the updated state, we can directly calculate expectation values of Pauli operators, $\langle \sigma_{i} \rangle  = \Tr[\rho'\sigma_i]$. Using Bayesian update rules \cite{Korotkov2016}, we attain measurement-update equations for the Bloch coordinates: 
	\begin{subequations}
		\label{eqn:bayesian}
		\begin{align}
		Z_j &= \tanh[\frac{j \delta t}{2 \tau} + C]  \\
		X_j &=\sqrt{1-	\langle \sigma_z \rangle ^2} e^{-\gamma \delta t}
		\end{align}
	\end{subequations}
	where $\gamma$ is the dephasing rate and $C=\frac{\ln[1+Z_0]}{\ln[1-Z_0]}$ depends on the initial coordinate $Z_0$.
	
	\begin{figure}
		\includegraphics[width=0.7 \linewidth]{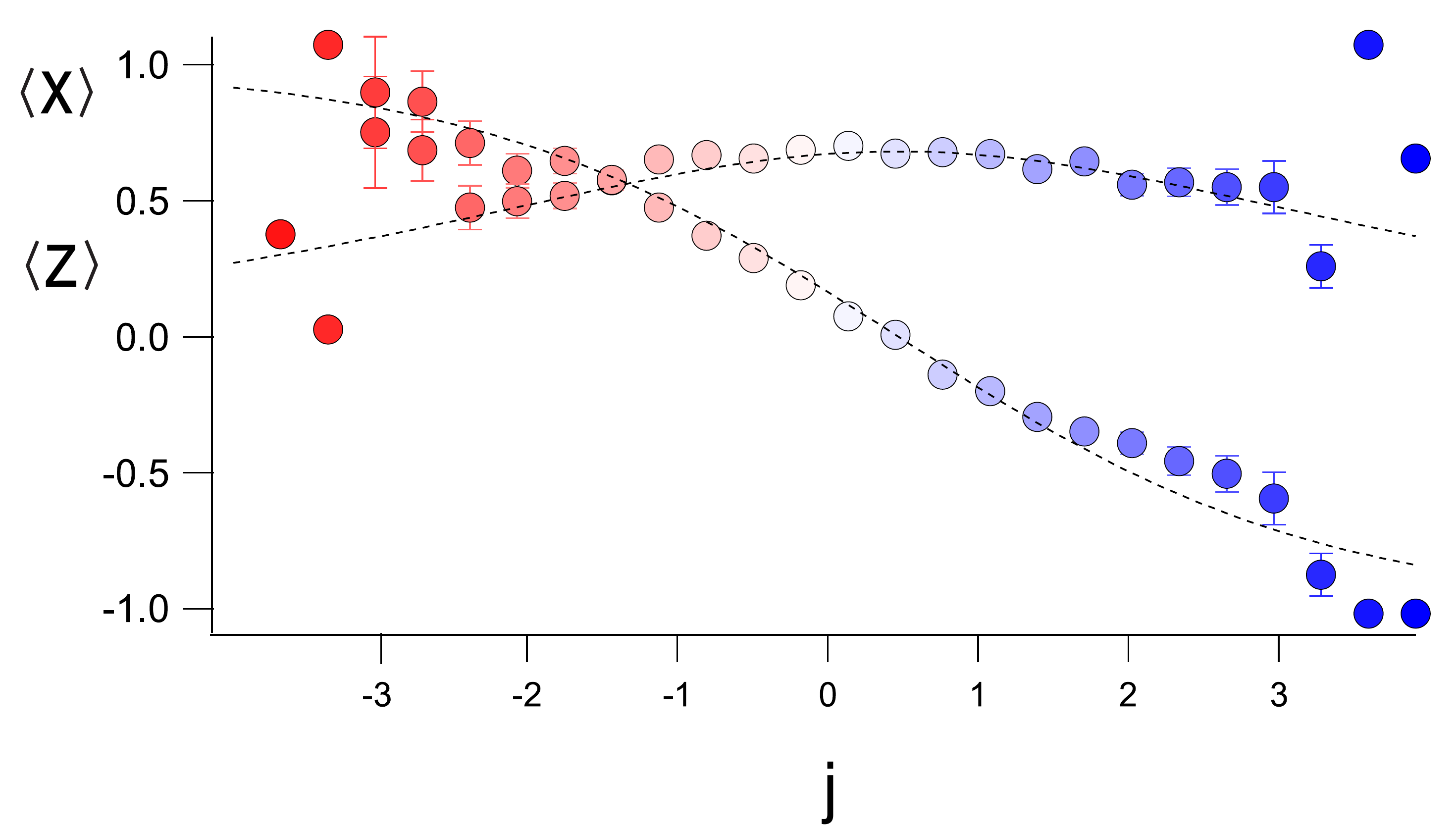}
		\centering
		\caption[Correlated Tomography]{ In the weak measurement regime, Bayesian state update (Eqns.~4) enables faithful state reconstruction via tomographic outcomes conditioned on average homodyne voltages. \SY~is not shown but is consistent with zero, as expected for rotations about $\sigma_y$.
		} 
		\label{fig:corrTomo}
	\end{figure}

	In Figure \ref{fig:corrTomo}, we show that the measurement-update equations (Eqns. [{\ref{eqn:bayesian}]), accurately predict the measured state. For $1.3\times10^6$ experimental runs we collect a weak measurement record, $j$, followed by a tomographic projective measurement. We then bin trajectories based on $j$ and compare the bin’s average tomography value for \SZ~(or \SX~or \SY) with the predicted value. 
	 Good agreement  between this correlated tomography and the predictions of measurement-update equations implies that we accurately track quantum trajectories.

	\subsection{Trajectory probabilities}
	\label{sec:trajProb}
		
		We now turn to calculating the probability of individual trajectories in order to infer the likelihood of the arrow of time. Within the POVM formalism, the probability of measurement outcome $j$ is $P(j)= \Tr[K_j \rho K_j^\dagger]$. In order to directly compare forward and backward probabilities we condition on the appropriate initial condition. We reverse trajectories with an ``active transformation'' \cite{Dressel2016}, where coordinates are inverted and the order of measurement outcomes are reversed, while the sign of the measurement outcome is fixed. 
		We calculate the forward and backward path probabilities via conditioned Bayesian inference:

		\begin{subequations}
			\label{eqn:PfPb}
			\begin{align}
			P_F(j|Z_0) &= \frac{1+Z_0}{2} e^{-\frac{(j-\Vg)^2 \delta t}{2 \tau}} 
					 + \frac{1-Z_0}{2} e^{-\frac{(j-\Ve)^2 \delta t}{2 \tau}}\nonumber \\
			P_B(j|\zf) &= \frac{1+\zf}{2} e^{-\frac{(j-\Vg)^2 \delta t}{2 \tau}}  
				+ \frac{1-\zf}{2} e^{-\frac{(j-\Ve)^2 \delta t}{2 \tau}} \nonumber 
			\end{align}
		\end{subequations}
		where \Vg~ (\Ve) is the mean voltage measured when the qubit is prepared in the ground (excited) state and \zf~ is the Bloch coordinate propagated to time $t$ and then negated.
		
		With forward and backward probabilities for individual quantum trajectories, we estimate the statistical arrow of time from the log ratio of forward and backward probabilities:
		\begin{equation}
		Q \equiv \ln\left[\frac{P_F}{P_B}\right]
		\label{eqn:Q}
		\end{equation}
		which in general is non-zero. The preference for forward-moving trajectories ($Q>0$) comes from the effect of measurement to project towards eigenstates. 
		As the measurement dynamics unfold, previous measurements induce backaction towards an eigenstate, causing the relative weights of Equations \ref{eqn:PfPb} to favor continued progression towards that eigenstate. Moreover, initial conditions may bias these weights to further heighten the movement towards an eigenstate. Thus, despite the reversibility of individual measurements, we nonetheless statistically infer $\bar{Q}>0$, i.e. trajectories tend to move forward in time.

		\subsubsection{Feedback Protocols}	
				\begin{figure}[h!]
			\includegraphics[width=0.7\linewidth]{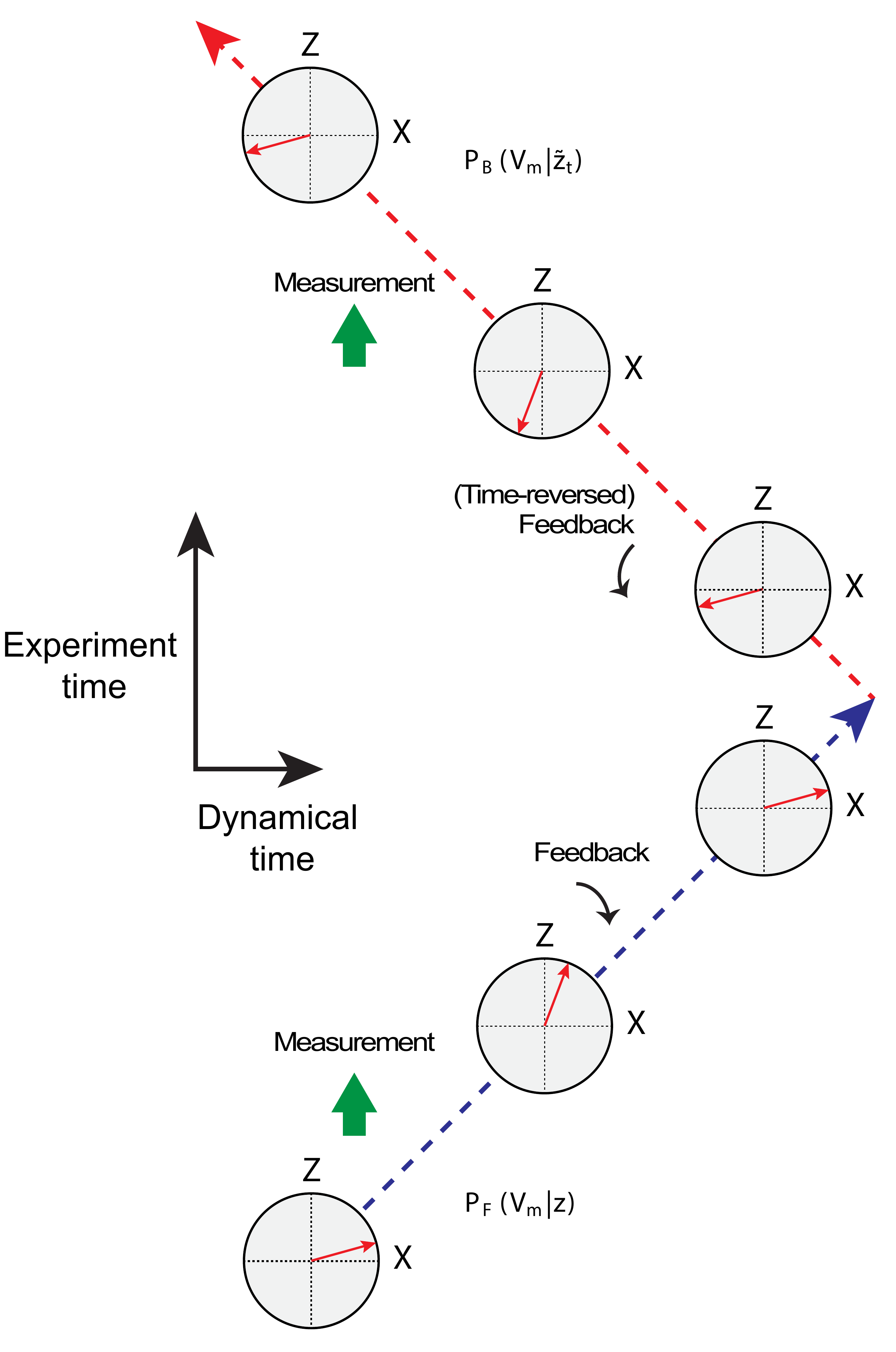}
			\centering
			\caption[Feedback Protocol]{Our feedback experiments entail stabilizing an initial state and comparing forward and backward path probabilities under measurement dynamics. Experimental realizations involve \textbf{(a)} first preparing a state which is then perturbed by measurement-induced backaction. \textbf{(b)} By applying a corrective unitary feedback rotation we \textbf{(c)} return the qubit to its initial state. We make an {equitable} comparison of forward and backward probabilities by \textbf{(d)} flipping the dynamics (in analysis) before \textbf{(e)} applying the time-reversed unitary feedback (negative angle) and \textbf{(f)} calculating the measurement probability according to Equation 5. The backaction of the forward measurement outcome restores the system to the time-reverse of the initial state.}
			\label{fig:protocol}
		\end{figure}
		We now turn to our feedback protocol for state stabilization, visualized in Figure \ref{fig:protocol}. The goal of feedback to maintain an initial state in the face of decoherence and stochastic evolution from environmental interactions.  
		However, in order to consider the role of both causal and anti-causal feedback in the arrow of time, we implement a post-selected feedback scheme.
		After preparing the target state, we allow the qubit to stochastically evolve under the influence of measurement. We then apply a corrective rotation
		pulse, $\theta_{\mathrm{app}}$, about the $Y$-axis with a randomly chosen angle between 
		$-\frac{\pi}{4}$ and $\frac{\pi}{4}$. 
		In post-analysis, we select realizations wherein we chose the ideal feedback angle, $\theta_{\mathrm{ideal}} = \tan^{-1}\left[\frac{Z_{j}}{X_{j}}\right]$, within $\frac{\pi}{20}$.

		
		Under this feedback protocol, we then evaluate forward and backward probabilities of individual trajectories for causally-ordered feedback (COF). Critically, the backward probability, $P_B(j|\zf)$, is based on the time-reversed coordinate \textit{after} the time-reversed feedback (Fig.~\ref{fig:protocol}). At this stage, the state has undergone both feedback and its time-reversal, effectively canceling the effect of feedback [12]. Thus in the case of causal order feedback, the arrow of time ratio, seen in Figure \ref{fig:Q}, is the same as if we had not applied feedback. Our results for forward feedback are consistent with previous studies \cite{Dressel2016,Harrington2019}.
		
		Now consider feedback with anticausally-ordered feedback (ACOF). We utilize the same post-selected feedback protocol, but after state preparation, we proceed with a random rotation succeeded by weak measurement. Interpreted as a feedback protocol, we correct the state and then measure to see what feedback should be applied. With this protocol we observe a backward arrow of time, seen in Figure \ref{fig:Q}, reflecting the anticausal order of the ``feedback''. In one sense, the measurements which correctly ``unmeasure'' the rotation are so exceedingly rare that an ensemble consisting of these trajectories must annihilate entropy. In another sense, because the arrow of time is a fundamentally causal relationship, an ACOF protocol produces a reversed arrow of time. 
		
		\begin{figure}[h]
			\includegraphics[width=0.7\linewidth]{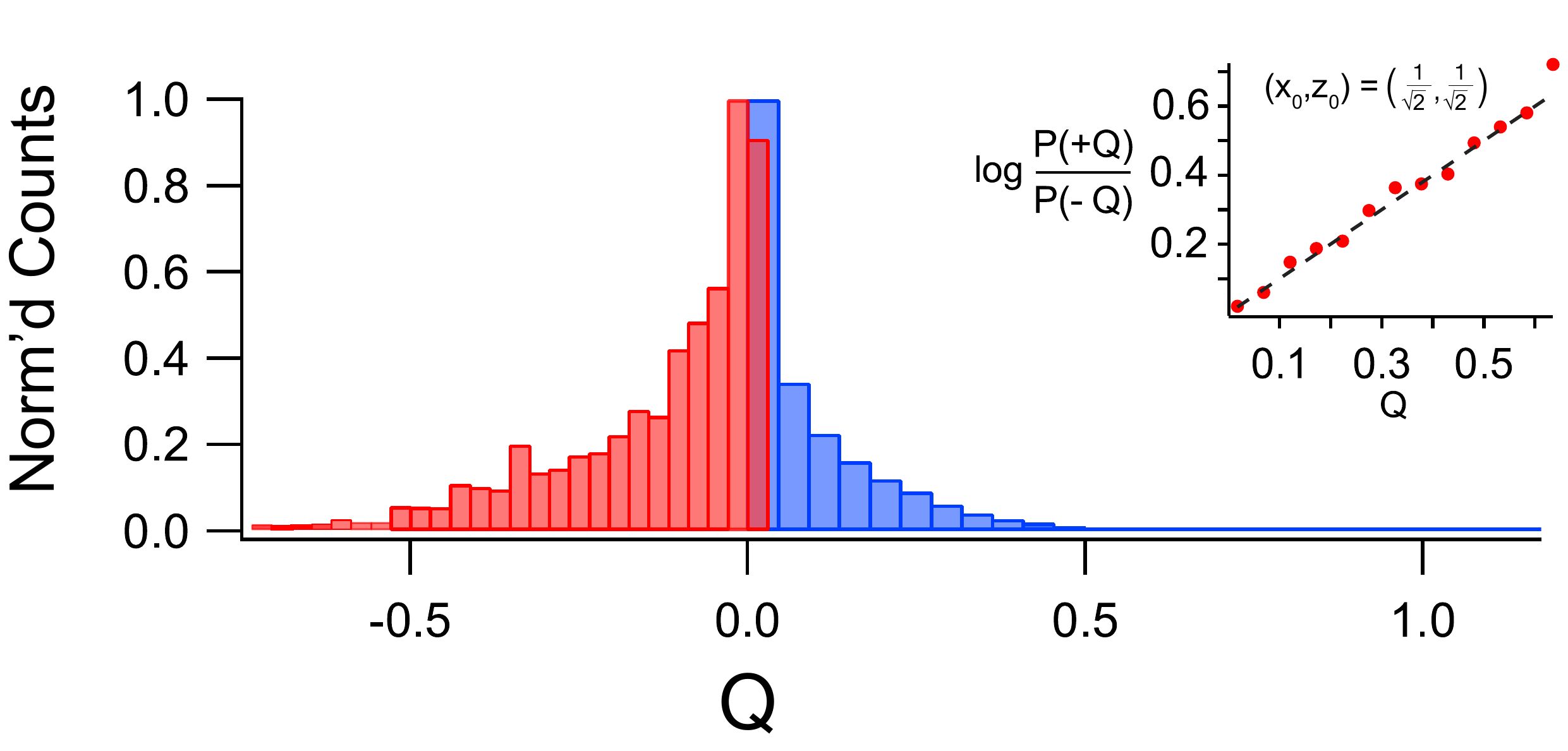}
			\centering
			\caption[Entropy Histograms]{Histogram of the log ratio of forward and backward probabilities, Q (Eqn.~\ref{eqn:Q}), calculated from individual quantum trajectories. In standard COF (blue) the trajectories are essentially all entropy-producing due to the initial state, $\rho = \ket{+x}\bra{+x}$. Conversely, for ACOF (red), time apparently flows backward due to the imposed anticausality. The figure inset shows validation of a fluctuation theorem for a different initial condition, as indicated.}
			\label{fig:Q}
		\end{figure}

	\subsection{Fluctuation Theorems}
		\label{sec:expFT}
		We could, in principle, use this data to verify the detailed fluctuation theorem (Eqn.~\ref{eqn:quantumFT}). However, as Figure \ref{fig:Q} shows, entropy-annihilating trajectories do not occur for states prepared in $\ket{+x}\bra{+x}$.
		
		In this case, the initial state is ``too unlikely'' to produce entropy annihilating trajectories.
		Over an ensemble of many initial states, the measurement process produces entropy in accordance with the detailed fluctuation theorem.
		However, when the ensemble is pre-selected to an initial state which strongly favors entropy production, the resulting distribution of $\mathcal{Q}$ is severely skewed.
		Such a biased initial state deviates from an integral fluctuation theorem by a quantity called the absolute irreversibility \cite{Harrington2019}.
	
		 With preliminary data of a different initial condition, $(X_0, Z_0) = (\frac{1}{\sqrt{2}}, \frac{1}{\sqrt{2}})$, we find that the entropy production and annihilation do balance according to the detailed fluctuation theorem (Figure \ref{fig:Q}, inset).

\section{Conclusion}
In this experiment, we have investigated the nuance of entropy production (and annihilation) in quantum systems. The details of such processes require a microscopic view of individual trajectories, and an ability to evaluate probabilities of time-reversed trajectories.

The foundation of generalized second laws includes an essential connection to causality \cite{Evans1996, Donoghue2020}. We investigated the influence of COF and ACOF protocols on the fluctuation theorem. We found that feedback enforces a causal direction by allowing information gained to then be imparted into the system. We found that this preference had no effect in an already causal world. However, reversing the causality via ACOF flipped the sign of entropy production, annihilating entropy and reversing the arrow of time.

The influence of feedback on the flow of entropy, suggests that these techniques could be used to create information-based engines \cite{Kim2011}, even when causality can be reversed \cite{Felce2020}. These would be quantum versions of Sz\'ilard's classical engine \cite{Szilard1964}.


\begin{spacing}{1.0}

\appto{\bibsetup}{\emergencystretch=1em}

\bibliographystyle{unsrt}
\bibliography{fullBib}

\end{spacing}
\clearpage
%

\end{document}


\section{Brief Introduction to Object-Oriented Programming}
	Script Writing
	
	We'd like a tool to organize better
	
	A class is feature of OOP that allows us to create more structure in code	
		In C++ (which could be called "C+classes"), the first class-like objects were called structs
		
	A class as various attributes, which can be things like functions or variables
		
	use of self (see 9.3.5) 

	\subsection{Minimal Example}
	instance of class Lab ()

\section{Generating Pulse Sequences}
	\subsection{Problem Statement}
		It will be instructive to consider a useful example.
		Let us examine the requirements for performing a Rabi experiment.
		The experiment consists of driving the qubit for a variable duration before measuring the qubit's state.

		The sequence of measurements is composed of a series of steps, where each step contains (i) a pulse of variable duration at the qubit frequency and (ii) a fixed pulse at the cavity frequency.
		In addition, our setup requires (iii) a trigger pulse to be sent to the data acquisition card (named ``Alazar'' after the manufacturer).
		A single step (sometimes called a pattern when referring to an older instrument in our lab) lasts for a fixed interval.
		The number of points in the interval can be varied, but it should be a power of 2 to accommodate the AWG's memory structure. Because we almost always set our AWG's sampling rate to 1 GS/s, one point corresponds to 1 ns. 
		
		Each of these pules in generated by an arbitrary waveform generator (AWG). We seek to program the set of drives into the AWG outputs.
		The outputs consist of channel outputs and marker outputs.
		Channels have variable output amplitudes, but markers only have a single amplitude (on or off).
		We typically use channels to create control pulses because control relies on amplitude control.
		We typically use markers as triggers.
		For a single-qubit, single-cavity Rabi experiment, we typically need two drives for the qubit and one drive for the cavity.
		
		So we'll choose a number of duration points, say 51. For each duration, we'll send a pulse to drive the qubit, and then we'll send a pulse to drive the cavity.
		For each repetition, the cavity readout is interpreted as either a $-1$ or a $+1$. We'll average over many such repetitions to get the expectation value at a single duration value. 
		We have a choice in the order of repetitions. We could repeat a single step in the sequence many times, or we could loop through the full set of durations and then average each duration. The latter choice is supported by the arbitrary waveform generator.
		
		\subsection{Sequence Structure}		
		Thus, we will specify the output of a single channel with a ``Sequence matrix''.
		An example is shown in Figure \ref{fig:sequenceDiagram}.
		It shows the pulse train for the duration of a single step, and shows all of these for all of the steps in the sequence.
		
		Such output needs to be specified for each output port of the AWG. The output ports include both channel outputs and marker outputs.

		\subsection{Pulse Class}
		The pulses we seek to generate act like gates to the microwave generator.
		We'll be programming an arbitrary waveform generator (AWG) whose output goes to the $I$ and $Q$ components of a mixer.
		The mixer's local oscillator (LO) port receives a high frequency drive, which we assume as the form $\cos(\omega_{\rm LO}t)$.
		The radio frequency (RF) port of the mixer outputs a combination of products between the I and Q ports with the LO port:
		\begin{equation}\label{eqn:IQMixer}
			\mathrm{RF}(t) = \cos(\omega_{\rm LO} t) *\mathrm{I}(t)	+\sin(\omega_{\rm LO} t) *\mathrm{Q}(t)
		\end{equation}

		A square pulse is the simplest gate. The square pulse has three parameters: start time, amplitude, and duration.
		A square pulse is appropriate in situations where the frequency distribution of the pulse is irrelevant, such as during cavity probing.
		Suppose the square pulse enters the I port. 
		During the pulse duration, the RF port outputs the drive at the appropriate level.
		
		In other cases, especially the qubit drives, we utilize more complex gates in order to shape the frequency components.
		The the most common pulse shaping we do is single-sideband modulation (SSM).
		SSM consists of using two components of the mixer input. With a sinusoidally-modulated gate amplitude, we can adjust the output frequency of the drive.
		This helps us avoid mixer leakage at the qubit frequency, by moving the carrier frequency (the input to the LO port) away from the qubit transition.
		If our mixer-input pulses are $\mathrm{I}(t) = \cos(\omega_{\rm SSM}t)$ and $\mathrm{Q}(t) = \cos(\omega_{\rm SSM} t + \pi/2)$, then the RF port becomes $\mathrm{RF}(t) = \cos[(\omega_{\rm LO} +\omega_{\rm SSM})t]$. 
		So we can move the LO frequency up by $\omega_{\rm SSM}$, avoiding carrier leakage.
		To allow for SSM, we add two more parameters to our pulses: the SSM frequency and the drive's phase.
		
		We note that in order to maintain drive consistency, we define the phase relative to the time step's initial time.
		We need this so that we maintain coherent control of the qubit.
		The phase of the drive sets the qubit's phase. So in the case when we want to to do multiple pulses on the qubit (e.g.~in a Ramsey sequence), we must maintain phase consistency between the first and subsequent pulses.
		We do so by setting the reference phase to $0$ at $t=0$.
		The vertical stripes which appear in Figure \ref{fig:pusleSequence} correspond to a constant phase of the SSM.

		The Pulse class provides a way to organize the different parameters within a pulse.
		We initialize a pulse instance with the pulse's duration, start time, and amplitude.
		If we want a pulse with SSM, we need to specify the ssm\_freq and the phase (if desired).
		We create an new instance with the command:
		new\_pulse = Pulse(duration=pi\_time, start\_time=4000, amplitude=1).
		The use of the variable names (e.g.~``duration=...'') are not necessary, but they can help prevent entry-order mistakes.
		
		The Pulse class is primarily intended as an organization tool for passing arguments to a Sequence instance.
		However, it also has a few self-contained functions.
		A Pulse has a toString() function which returns a formatted descriptor string of the Pulse object.
		This could be improved by moving to a \_\_repr\_\_() function, which describes how to represent the function, i.e.~as text.
		
		The Pulse object can return an array displaying the programmed amplitudes by calling the make() function.
		And finally, the copy() function returns a new (deep) copy of the Pulse.

	\subsection{Sequence Class}
		A single pulse takes its place in the full structure of the sequence.
		A sequence defines the final performance of the program we've written.
		
		Pulses can enter a sequence in multiple ways.
		A \texttt{Sequence} instance has a function called \texttt{add\_sweep()}. 
		The function takes a \texttt{Pulse }instance and a set of optional sweep parameters.
		The function interprets the parameters according to the variable \texttt{sweep\_type}.
		The function then inserts into each step of the sequence, an updated Pulse object for the given step.
		For example, if the sweep type was set to duration, as in the Rabi experiment described above, then each step in the sequence would receive a Pulse of incremented duration.
		
		Inserting a pulse into a sequence occurs at a single step.
		The step is represented as an array of floats, corresponding to each time point.
		Adding a pulse amounts to adding the Pulse's amplitude to the \fixit{step array}.

		The Sequence class is the primary class for sequence generation.
		Figure \ref{fig:seqDemo} shows the organizational structure of the Sequence class.
		The primary attribute is channel list.
		We mentioned above that there are four channels and four markers.
		For each of these outputs, we define a single sequence matrix, which programs the channel for the particular experiment we need to run (e.g.~Rabi or Ramsey experiment).
		The sequence matrices are stored in a list called \texttt{channel\_list}.
		
		\texttt{channel\_list} has four elements, each corresponding to a single channel.
		Each element of \texttt{channel\_list} has three elements, corresponding to the channel, marker 1 and marker 2.
		Because \texttt{CH1} and \texttt{CH2} share markers, the entry for CH2's markers are redundant and ignored. Likewise, CH4's markers are ignored.
		Thus, the call to \texttt{channel\_list[0][1]} would result in the sequence matrix (a 8142x101 matrix) for CH1's first marker.

		\subsection{Example Sequence Construction}

\section{Data Acquisition}
	\subsection{DAC Specifications}
		records, buffers, ``readout''

\section{Addressing Instruments with SCPI}

	\subsection{Example: Spectroscopy}

\section{Full Code}

\bibliographystyle{unsrt}

\bibliography{fullBib}